\documentclass[prb,twocolumn,showpacs,superscriptaddress,amsmath,amssymb]{revtex4}
\usepackage{graphicx}
\usepackage{dcolumn} 
\usepackage{bm}      
\usepackage{color}   

\begin{document}

\title{Polaronic-Quasiparticle Picture for Generation Dynamics of Coherent Phonons  in Semiconductors: Transient and Non-Linear Fano Resonance}


\author{Yohei Watanabe}
\affiliation{Doctoral Program in Materials Science, Graduate School of Pure and Applied Sciences, University of Tsukuba, Tsukuba, Ibaraki 305-8573, Japan}
\author{Ken-ichi Hino}
\email{hino@ims.tsukuba.ac.jp}
\affiliation{Division of Materials Science, Faculty of Pure and Applied Sciences, University of Tsukuba, Tsukuba 305-8573, Japan}
\affiliation{Center for Computational Sciences, University of Tsukuba, Tsukuba 305-8577, Japan}
\author{Muneaki Hase}
\affiliation{Division of Applied Physics, Faculty of Pure and Applied Sciences, University of Tsukuba, Tsukuba 305-8573, Japan}
\author{Nobuya Maeshima}
\affiliation{Center for Computational Sciences, University of Tsukuba, Tsukuba 305-8577, Japan}
\affiliation{Division of Materials Science, Faculty of Pure and Applied Sciences, University of Tsukuba, Tsukuba 305-8573, Japan}
\date{\today}

\begin{abstract}

We examine generation dynamics of coherent phonons (CPs) in both of polar and non-polar semiconductors -- such as  GaAs and Si -- based on a polaronic-quasiparticle (PQ) model.
In the model concerned, the PQ operator is composed of two kinds of 
operators.
One is a quasiboson operator -- defined as a linear combination of a set of pairs of electron operators -- and the other is  a longitudinal optical (LO) phonon operator.
The problem of transient and non-linear Fano resonance (FR) is tackled in particular; the vestige of this quantum interference effect was observed exclusively in lightly $n$-doped Si immediately after carriers were excited by an ultrashort pulse-laser [M.~Hase, {\it et. al.}, Nature {\bf 426}, 51 (2003)], though not  observed yet in GaAs. 
It is shown that the phonon energy state is embedded in a continuum state formed by a set of {\it adiabatic} eigenstates of the quasiboson.
This result implies the possibility of manifestation of the transient FR in the present optically-non-linear  system.
Moreover, both of a retarded  longitudinal susceptibility and an LO-phonon displacement as a function of time are derived in an analytic manner.
These physical quantities are mainly affected by  an effective interaction between  quasiboson and LO-phonon ($M_{\boldsymbol{q}}$), and a diagonal part of non-adiabatic interaction  integral over time ($\mathfrak{I}_{\boldsymbol{q}}$).
Two important results are as follows.
(i) The longitudinal susceptibility
allows one to calculate photoemission spectra from transient photoexcited-states, which is  linear in a weak probe pulse-laser turned on at time $t_p$. 
An asymmetric spectral profile  characteristic of FR comes into existence in undoped Si just in the region of $T^\prime_{\boldsymbol{q}12} < t_p \ll T^\prime_{12}$, while this becomes almost symmetric in the region of  $t_p \gtrsim T^\prime_{12}$; where $T^\prime_{12}$ and $T^\prime_{\boldsymbol{q}12}$ represent phenomenological damping time-constants of induced carrier-density with isotropic and {\it anisotropic} momentum-distribution, respectively, with $T^\prime_{\boldsymbol{q}12}\ll T^\prime_{12}$.
The asymmetry in spectra is attributed to both phase factors of $\arg{M_{\boldsymbol{q}}}$ and  ${\rm Im}\/\mathfrak{I}_{\boldsymbol{q}}$.
These factors contribute to spectra in GaAs in a different manner.
(ii) The Fourier-transform of LO-phonon displacement  into  frequency domain is examined.
The associated power spectra
show  an asymmetric spectral-profile due to FR exclusively in undoped Si, however, a symmetric one in GaAs
irrespective of ${\rm Im}\/\mathfrak{I}_{\boldsymbol{q}}$.
Such obtained results seem consistent with the reported experimental results.
An initial offset phase is  also examined, which is extracted from an asymptotic sinusoidal form of LO-phonon displacement.
It is found that this  phase is affected  by $\arg{M_{\boldsymbol{q}}}$ and ${\rm Im}\/\mathfrak{I}_{\boldsymbol{q}}$ in addition to a pulse area due to pump pulse-laser.
According to the results of (i) and (ii), some fundamental features pertinent to CP generation are nicely understood by virtue of the PQ picture of concern.

\end{abstract}

\maketitle

\section{Introduction}
\label{sec1}

The recent progress of  laser technology toward the development of ultrashort pulsed laser with high intensity has opened up a new research area  exploring ultrafast transitory phenomena governed by strongly-photoexcited  electronic states in diverse fields of physics and chemistry.~\cite{ultrafastXIX}
The coherent phonon (CP) generation concerned in this study is undoubtedly one of the representative phenomena successfully exposed by virtue of ultrashort pulsed laser; its temporal width is of an order of 10 femtosecond (fs) much shorter  than a period of a longitudinal optical (LO) phonon.~\cite{lightscatteringVIII,kuznetsov1,kuznetsov2}
Here, the LO-phonon mode is driven at one stroke immediately after the pulse exertion on a crystal
to show coherent vibrations in phase each other.
The CPs have been observed in variety of physical systems such as 
semiconductors,\cite{cho1,pfeifer1,dekorsy1,dekorsy2,cho2,chang,misochko1,sabbah1,hase1,riffe1,kato} 
dielectrics,\cite{nelson,bakker1,bakker2,bakker3}
 semimetals/metals,\cite{cheng1,cheng2,dekorsy3,hunsche,hase2,hase6,hase3,misochko3,decamp,hase4,misochko2,murray,misochko6,misochko7,hase5,hase7,li}
high-$T_c$ superconductors,\cite{chwalek1,chwalek2,albrecht,misochko4,misochko5,bozovic} 
and other materials;\cite{mishina,ishioka1,watanabe1,watanabe2} a great number of other references are cited in Refs.~\onlinecite{lightscatteringVIII} and \onlinecite{kuznetsov1}.
The CP generation dynamics is well described in terms of a classical model following  a damped forced-oscillation, and this is assessed by an initial phase built in the harmonic oscillator with an asymptotic sinusoidal form in time.\cite{lightscatteringVIII,kuznetsov1}

There are two models  hitherto presented for the generation dynamics, namely, the impulsive stimulated Raman scattering (ISRS) model\cite{ yan1,yan2,kutt} 
and the displacive excitation of CP (DECP) model,\cite{zeiger,garrett,merlin1}
aside from one more mechanism of ultrafast screening of space-charge/surface-depletion field inherent in polar semiconductors.\cite{pfeifer1,dekorsy1}
In the former of the ISRS model, the stimulated Raman scattering induced by  a pump pulse is considered to govern the generation dynamics, in which the oscillator is driven by an impulsive external force -- a delta-function-like force --  associated with  Raman polarizability, namely, a derivative of electronic susceptibility with respect to lattice displacement.\cite{lightscatteringVIII,kuznetsov1}
The resulting harmonic oscillation becomes of the sine-form with the initial phase equal to zero.
On the other hand, in the latter of the DECP model, optical transitions of electrons into excited states alter an equilibrium position of lattice, leading to vibration toward a new position,
in which an effective external force driving this vibration is considered as a step-function-like force.\cite{lightscatteringVIII,kuznetsov1}
The resulting harmonic oscillation becomes of the cosine-form with the  initial phase equal to $\pi/2$.
In fact, initial phases are observed mostly between $0$ and $\pi/2$, depending on additional physical effects due to pulse intensity, pulse shape, and so on.\cite{hasex}
Therefore, phenomenological models hybridizing the above two models have been devised to be made in consistency with the observed initial phases.\cite{garrett,merlin1,stevens,bragas,riffe1} 

Indeed the classical models mentioned above have successfully revealed overall features of  the CP generation dynamics, but it is the matter of course that a microscopic theory based on non-equilibrium quantum dynamics is required to bring into light not only details of the dynamics of concern -- such as an origin of phenomenological external forces driving  CPs, and the related built-in  initial phases -- but also still unexplored quantum effects.
In fact, theoretical studies toward this direction are scarce until now; to the best of one's knowledge, the density-matrix theory  was applied to understand the external forces from the viewpoint  of the group theory,\cite{pfeifer2,scholz1}
 the time-dependent Schr\"{o}dinger equation was solved numerically to 
understand an experimentally observed quantum effect,\cite{lee}
 the time-dependent density-functional theory was applied to reproduce measured CP signals in Si,\cite{shinohara}
 the Fano-Anderson Hamiltonian\cite{mahan} was applied to classical dynamics to show the dependence of a  Fano resonance (FR) effect\cite{fano1} on an initial phase,\cite{riffe2} 
and a simple two-level model was applied to CP generation to show the initial-phase-dependence on pulse width and detuning.\cite{nakamura}

As regards the quantum effect concomitant to the CP generation, Hase {\it et al.} observed  
the transient FR effect in a lightly $n$-doped Si crystal immediately after carriers were excited by an ultrashort laser pulse.\cite{hase1}
This was considered to result from quantum interference between excited carries  and an LO-phonon, which played the  roles of continuum  and discrete states, respectively.
Moreover, the speculation was made that  the observed FR would show the evidence of  the birth of  a polaronic quasiparticle (PQ), which would be likely formed in a strongly interacting carrier-LO-phonon system in a moment.\cite{Gaal2007}
It is noted that the transient FR of concern has been observed exclusively in a lightly $n$-doped Si crystal and semimetals/metals such as Bi and Zn\cite{hase5,hase7,misochko6,misochko7} till now, however, not observed in $p$-doped Si and GaAs crystals.\cite{misochko1,kato}
In passing, recently, dynamic FR-like interference between Rabi oscillations and CPs has been observed in CuCl semiconductor microcavities.\cite{yoshino}

Thus far, there are a number of  theoretical studies regarding these experimental findings.
By solving the time-dependent Schr\"{o}dinger equation in the system of GaAs, Lee {\it et al}.\cite{lee} calculated a displacement function of  CP, and showed that the associated continuous-wavelet transform of this function became of the asymmetric  shape featuring FR spectra, though apparently opposed to  existing experimental results, as mentioned above.
The authors further discussed the origin of the resulting  FR  in a Feynman-diagrammatic manner to conclude that the FR arose from interference between two types of vibrational Raman-scattering processes.
Riffe\cite{riffe2} derived the classical Fano oscillator model  from the Fano-Anderson Hamiltonian, and showed that Fano's $q$-parameter $q^{(F)}$, which determines the degree of asymmetry of spectra, was incorporated in an initial phase.
Misochko and Lebedeva\cite{misochko7} showed the different dependence of $q^{(F)}$ on the initial phase by taking  the  Fourier-transform of the Fano's spectral formula in a direct manner into a temporal region.
Further, they confirmed that the experimental results of the CP signals of Bi were in harmony with the obtained initial-phase dependence.

The present article is aimed at the following goals. 
To begin with, one constructs a fully-quantum-mechanical model for the CP generation dynamics available for both of polar and non-polar semiconductors such as undoped GaAs and undoped Si.
Next,  by means of this model, one tracks the origin of the manifestation of the transient FR, and further, understands  fundamental quantum processes sprouting a seed of  CPs in the early stage of the whole CP dynamics, which will be termed hereafter as the early time-region.
Here, this region is defined as the temporal region during which a great number of carriers still stay in excited states and the quantum processes govern the CP dynamics.
It is noted that any delayed formation of  LO-phonon-plasmon coupled modes  in polar semiconductors is not taken into account throughout this study, since these coupled modes are not created instantaneously within the early-time region.\cite{kuznetsov1,varga,mooradian,klein,kuznetsov3,hase8,ishioka2,hu,Huber05,Chang2010}
It is desirable to scrutinize the above-mentioned supposition of the introduction of the PQ to the CP dynamics, and  the role of this quasiparticle played for  the formation of the transient FR.\cite{hase1}
Here, the PQ operator is defined as being composed of two kinds of operators.
That is, one is a quasiboson operator, given by a linear combination of a set of pairs of electron operators, and the other is  an LO-phonon operator.
A set of the expansion coefficients of both of the operators is obtained by solving   Fano's resonant-scattering problem.\cite{fano1}
To be more specific, according to the present PQ picture, it is understood that the transient FR of concern is caused by interference between a continuum state of the quasiboson and a discrete state of the LO-phonon, as  shown in detail in Sec.~\ref{sec2}.

\begin{figure}[tb]
\begin{center}
\includegraphics[width=7.0cm,clip]{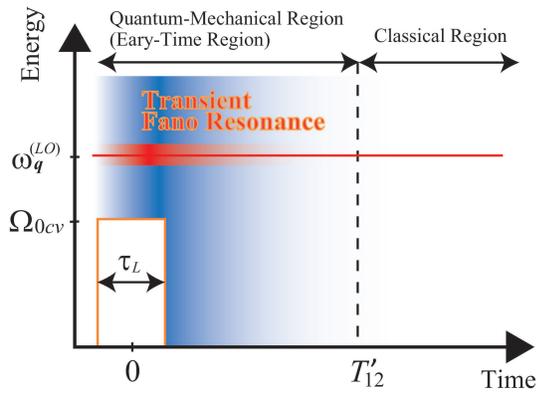}
\caption{(Color online) The schematic diagram of time-evolution of the overall CP generation dynamics. Here, the energy of LO-phonon $\omega^{(LO)}_{\boldsymbol{q}}$ with momentum $\boldsymbol{q}$ is  represented by a red solid-line, and the energy of quasiboson continuum-state is represented by gradation of blue color, where the gradation shows schematic change of excited carrier density, the lowest limit of this is provided by Rabi frequency $\Omega_{0cv}$ due to  squared-pulse laser represented by an orange solid-line with temporal width  $\tau_L$, and the degree of blurring of the LO-phonon energy represents the extent of transient FR-width. Further, $T^\prime_{12}$ represents a phenomenological damping time-constant of induced carrier-density with isotropic momentum distribution.
This is rough estimate of  the border between  the quantum-mechanical  and classical regions. 
}
\label{fig1}
\end{center}
\end{figure}

Together with  this result of the transient Fano dynamics,  time-evolution of the overall CP generation dynamics is depicted in the schematic diagram of Fig.~\ref{fig1}, where a rough border between the quantum-mechanical region, namely, the early-time region, and a classical region are delimited by a phenomenological damping time-constant $T^\prime_{12}$ of induced carrier-density with isotropic momentum distribution; this corresponds to excited-carrier relaxation time and dephasing time.
Moreover, it is seen in this figure that the energy of LO-phonon $\omega^{(LO)}_{\boldsymbol{q}}$ with momentum $\boldsymbol{q}$ -- represented by a red solid-line --  is embedded in the continuum state of quasiboson -- represented by gradation of blue color -- ; where the gradation shows schematic change of excited carrier density, and the lowest limit of this is provided by Rabi frequency $\Omega_{0cv}$ due to  squared laser-pulse -- represented by an orange solid-line -- with temporal width  $\tau_L$. 
In addition, the degree of blurring of the LO-phonon energy given above represents the extent of transient FR-width.

Once the set of the PQ operators is ready in hand,  physical quantities relevant to the present system are provided in terms of a retarded Green functions associated with these operators;
time-evolution of this Green function is described by means of an adiabatic expansion with respect to PQ states with time fixed.
Both of a retarded  dielectric function and an LO-phonon displacement function in the systems of Si and GaAs are examined based on the PQ picture developed here, where
opaque interband transitions accompanying real excited carriers  are exclusively considered.
Transient photoemission spectra from photoexcited states are obtained by  the dielectric function, in which
an asymmetric spectral profile  characteristic of FR comes into existence  in Si (see Sec. \ref{sec3B}).
On the other hand, the power spectra associated with an LO-phonon displacement function show  an asymmetric spectral profile due to FR in Si, however, this is always symmetric in GaAs (see Sec. \ref{sec3D}).
This seems consistent with the reported experimental results.\cite{hase1,misochko1}

It should be pointed out that there is a different type of FR from the present one in semiconductors, which is generated by incoherent Raman scattering driven by irradiation of a continuous-wave (cw) laser.\cite{russell,parker,hart,cerdeira2,cerdeira1,cerdeira0,temple,cerdeira3,
cerdeira4,balkanski,chandrasekhar1,arya,chandrasekhar2,klein,menendez,belitsky,nunes,pusep,jin}
It is well known  that this FR process results from  interference between two contributions of the Raman scattering; one is an electronic Raman process playing a role of a continuum state, and the other is a vibrational Raman process playing another role of a discrete state.
In the system of  heavily $p$-doped Si, the energetically-degenerate final states of these two processes are allowed to be coupled each other by an electron-LO-phonon interaction, leading to the FR.\cite{cerdeira0,cerdeira3,balkanski,arya,klein,belitsky}
Here, this interaction causes the transition from an electronic state in a heavy-hole subband of a valence band to that in a light-hole subband of the valence band, accompanying emission of an LO-phonon.
The similar FR is also observed in the system of heavily $n$-doped Si in the X-valley.\cite{chandrasekhar1}
It is interesting that the manner of doping has an effect on a sign of the resulting Fano's $q$-parameter, that is, $q^{(F)}$ tends to be positive (negative) in the $p(n)$-doping case, showing FR spectra with a dip (peak ), followed by a peak (dip).\cite{belitsky}
In addition, the FR is also discerned in the system of $\delta$-doped GaAs.\cite{nunes,pusep,jin}
It is speculated that little of the above-mentioned interference between the two Raman processes
would affect the transient FR concerned in this system; consult the discussion made in  Sec.~\ref{sec3C}.

As is well known, FR is considered as one of the fundamental concepts common to diverse fields of physics.\cite{miroshnichenko1}
In particular, within the restriction just  to the FR processes triggered by laser irradiation, the FR of concern is regarded as an unusual {\it transient} and optically-{\it non-linear} processes,  considerably distinct from most of FR processes observed in linear optical processes due to cw-laser  irradiation.\cite{non-linearFR}
Actually, the present FR is revealed just in an ultrafast process and is caused by an optically non-linear manner, where carriers are strongly excited by a short pulsed-laser (such as the fs-laser concerned here), followed by an interaction with an LO-phonon, and eventually this process is detected by a weak probe pulse.
It seems that the number of studies pertaining to this type of transient and non-linear FR is  really limited due presumably  to difficulties of experimental measurements and theoretical predictions in advance.
To the best of one's knowledge, one example belonging to this is a transient excitonic-FR manifesting itself in ultrafast optical processes.\cite{meier,siegner1,siegner2,hino2}
Precisely for this reason, the studies directed toward this type of FR would have the potential to explore a new area of research with enriched physics.
Hereafter, the FR of concern is termed  as transient FR just for the sake of simplicity unless otherwise stated, though this should be referred to as transient and non-linear FR, as titled in the present article.
 
Materials parameters of both of undoped Si and undoped GaAs, and laser parameters employed in the present study are listed in Ref.  \onlinecite{SMparameter}.
It is understood that the concerned material of undoped Si/GaAs is referred to just as Si/GaAs hereafter unless otherwise stated.
The remainder of this article is organized as follows. Section~\ref{sec2} describes the theoretical framework,
Sec.~\ref{sec3} presents the results and discussion, and Sec.~\ref{sec4} presents the conclusion. 
Atomic units (a.u.) are used throughout unless otherwise stated.

\section{Theory}
\label{sec2}

The total Hamiltonian $\hat{H}$ of the present system is given as follows:
\begin{equation}
\hat{H}=\hat{H}_e+\hat{H}^\prime(t)+\hat{H}_p+\hat{H}_{e-p},
\label{H}
\end{equation}
where $\hat{H}_e$, $\hat{H}^\prime(t)$, $\hat{H}_p$, and $\hat{H}_{e-p}$ represent an electron Hamiltonian, an interaction Hamiltonian between laser and  electron at time $t$, an LO-phonon Hamiltonian, and an interaction Hamiltonian between electron and phonon, respectively.
$\hat{H}_e$ is provided within a two-band model just taking account of the energetically-lowest conduction band ($c$) and the energetically-highest valence band ($v$) as follows:
\begin{equation}
\hat{H}_e
=\sum_{b(=c,v), \boldsymbol{k}}
\varepsilon^\prime_{b\boldsymbol{k}}\:a^\dagger_{b\boldsymbol{k}}a_{b\boldsymbol{k}}
+{1\over 2}\sum_{\boldsymbol{q}(\not=\boldsymbol{0})}V^{(C)}_{\boldsymbol{q}}\:\hat{\rho}_{\boldsymbol{q}}\:
\hat{\rho}_{-\boldsymbol{q}},
\label{He}
\end{equation}
where
$a^\dagger_{b\boldsymbol{k}}\:(a_{b\boldsymbol{k}})$ represents a creation (annihilation) operator of electron with $b$ and $\boldsymbol{k}$ as a band index and the associated Bloch momentum, respectively.
Hereafter, it is understood that a spin index of electron is suppressed in the notation of $\boldsymbol{k}$.
Further,
$V^{(C)}_{\boldsymbol{q}}$ represents a Coulomb potential given by
\begin{equation}
V^{(C)}_{\boldsymbol{q}}={4\pi \over \epsilon_\infty V}{1\over  \boldsymbol{q}^2}
\label{VC}
\end{equation}
with $\boldsymbol{q}$, $V$, and $\epsilon_\infty$ as momentum, volume of crystal, and a dielectric constant in the high-frequency limit, respectively,
$\hat{\rho}_{\boldsymbol{q}}$ is defined as 
\begin{equation}
\hat{\rho}_{\boldsymbol{q}}=\sum_{b,\boldsymbol{k}} a^\dagger_{b,\boldsymbol{k}+\boldsymbol{q}}
a_{b\boldsymbol{k}},
\label{rho}
\end{equation}
and
$\varepsilon^\prime_{b\boldsymbol{k}}$ is given by
\begin{equation}
\varepsilon^\prime_{b\boldsymbol{k}}
=\varepsilon_{b\boldsymbol{k}}-{1\over 2}\sum_{\boldsymbol{q}(\not=\boldsymbol{0})}V^{(C)}_{\boldsymbol{q}}
\label{varepsilonbar},
\end{equation}
where
$\varepsilon_{b\boldsymbol{k}}$ represents an energy dispersion of $b$-band electron.
$\hat{H}^\prime(t)$ is expressed as
\begin{equation}
\hat{H}^\prime(t)
=-{1\over 2}\sum_{b,b^\prime(\not=b), \boldsymbol{k}}
\left[
\Omega_{bb^\prime}(t)\:a^\dagger_{b\boldsymbol{k}}a_{b^\prime\boldsymbol{k}}
+\Omega_{b^\prime b}^*(t)\:a^\dagger_{b^\prime\boldsymbol{k}}a_{b\boldsymbol{k}}
\right],
\label{H'}
\end{equation}
where
$\Omega_{bb^\prime}(t)=d_{bb^\prime}F(t)$ with $F(t)$ an electric field  of an irradiated pump-laser and 
$d_{bb^\prime}$ an electric dipole moment  between $b$ and $b^\prime$ bands; slight $\boldsymbol{k}$-dependence of the dipole moment is neglected.
Further, a factor of $1/2$ in front of the summation of the right-hand side implies that
the summations with respect to both $b$ and $b^\prime$ are taken over all of bands concerned here.
$\hat{H}_p$ is given by
\begin{equation}
\hat{H}_p=\sum_{\boldsymbol{q}}\omega^{(LO)}_{\boldsymbol{q}}\:c^\dagger_{\boldsymbol{q}}c_{\boldsymbol{q}},
\label{Hp}
\end{equation}
where
$c^\dagger_{\boldsymbol{q}}\:(c_{\boldsymbol{q}})$ represents a creation (annihilation) operator of LO phonon with an energy dispersion $\omega^{(LO)}_{\boldsymbol{q}}$.
The zero-point energy of the phonon is omitted just for the sake of simplicity.
Further, $\hat{H}_{e-p}$ is given by
\begin{equation}
\hat{H}_{e-p}
=\sum_{b,\boldsymbol{q},\boldsymbol{k}}
\left(
g_{b\boldsymbol{q}}c_{\boldsymbol{q}}
a^\dagger_{b\boldsymbol{k}+\boldsymbol{q}}a_{b\boldsymbol{k}}
+g_{b\boldsymbol{q}}^*c_{\boldsymbol{q}}^\dagger
a^\dagger_{b\boldsymbol{k}}a_{b\boldsymbol{k}+\boldsymbol{q}}
\right),
\label{Hep}
\end{equation}
where
$g_{b\boldsymbol{q}}$ is 
a coupling constant of a $b$-band electron with an LO phonon.

The theoretical framework is shown below by dividing the present section into the following two stages of subsections.
In Sec.~\ref{sec2A}, we introduce a set of quasi-boson operators composed of pairs of electrons in a conduction band and a valence band by means of an adiabatic recipe where time $t$ is kept fixed as a parameter. 
Here, these quasi-bosonic particles are considered as a plasmon and an eigen mode of  electron-hole pairs;
for the sake of simplicity, hereafter, this eigen mode is just termed as an electron-hole pair, unless otherwise stated.
Further, introducing an interaction of the quasi-boson with LO-phonon, we derive an equation of motion of  a PQ operator,
and  obtain the associated retarded Green function by solving this time-evolution equation.
In Sec.~\ref{sec2C}, employing the PQ operators thus obtained, we derive a dielectric function relevant to the system of concern, and analytic expressions of power spectra relevant to a CP displacement function.

\subsection{Introduction of Polaronic Quasiparticle Operators}
\label{sec2A}

\subsubsection{Quasi-Boson Operators}
\label{sec2A1}
First, we take into consideration the equation of motion for
a composite operator
$A_{\boldsymbol{q}}^\dagger(\boldsymbol{k}bb^\prime)$
defined by
\begin{equation}
A_{\boldsymbol{q}}^\dagger(\boldsymbol{k}bb^\prime)
=a^\dagger_{b,\boldsymbol{k}+\boldsymbol{q}}a_{b^\prime\boldsymbol{k}},
\label{A+}
\end{equation}
which is provided in terms of  a Heisenberg equation as
\begin{eqnarray}
&&-i\left({d\over dt}+{1\over T_{\boldsymbol{q}\boldsymbol{k}bb^\prime}}\right)A_{\boldsymbol{q}}^\dagger(\boldsymbol{k}bb^\prime)
\nonumber\\
&&=[\hat{\mathcal{H}}_e(t), A_{\boldsymbol{q}}^\dagger(\boldsymbol{k}bb^\prime)]
+[\hat{H}_{e-p}, A_{\boldsymbol{q}}^\dagger(\boldsymbol{k}bb^\prime)],
\label{Heisenberg}
\end{eqnarray}
where an electronic Hamiltonian
$\hat{\mathcal{H}}_e(t)$ is defined as
\begin{equation}
\hat{\mathcal{H}}_e(t)=\hat{H}_e+\hat{H}^\prime(t).
\label{calH}
\end{equation}
Further, $T_{\boldsymbol{q}\boldsymbol{k}bb^\prime}$ represents a phenomenological relaxation-time constant of induced carrier density $A_{\boldsymbol{q}}^\dagger(\boldsymbol{k}bb^\prime)$ with anisotropic momentum distribution.
It is noted that the degree of spatial anisotropy shown by $|\boldsymbol{q}|$ is finite here, even though it is quite small; $\boldsymbol{q}\not=\boldsymbol{0}$. 
Consulting Ref.~\onlinecite{appA} which provides the detail of calculations of the first commutator
in the right-hand side of Eq.~(\ref{Heisenberg}),
the following expression is obtained:
\begin{equation}
[\hat{\mathcal{H}}_e(t), A_{\boldsymbol{q}}^\dagger(\boldsymbol{k}bb^\prime)]
\approx
\sum_{\tilde{\boldsymbol{k}}\tilde{b}\tilde{b}^\prime}
A_{\boldsymbol{q}}^\dagger(\tilde{\boldsymbol{k}}\tilde{b}\tilde{b}^\prime)
Z_{\boldsymbol{q}}(\tilde{\boldsymbol{k}}\tilde{b}\tilde{b}^\prime,\boldsymbol{k}bb^\prime),
\label{comm}
\end{equation}
where  $Z_{\boldsymbol{q}}$ is a $c$-number non-hermitian  matrix, the explicit expression of which is given by 
\begin{eqnarray}
&&Z_{\boldsymbol{q}}(\boldsymbol{k}_1b_1b_1^\prime,\boldsymbol{k}_2b_2b_2^\prime)
\nonumber\\
&&=
w_{b_1b_1^\prime\boldsymbol{k}_1\boldsymbol{q}}\delta_{b_1b_2}\delta_{b_1^\prime b_2^\prime}\delta_{\boldsymbol{k}_1\boldsymbol{k}_2}
+V^{(C)}_{\boldsymbol{q}}\delta_{b_1b_1^\prime}\Delta\rho_{b_2b_2^\prime\boldsymbol{k}_2\boldsymbol{q}}
\nonumber\\
&&-\Omega^{(R)}_{b_1\bar{b}_1\boldsymbol{k}_1}\delta_{b_1\bar{b}_2}\delta_{b_1^\prime b_2^\prime}\delta_{\boldsymbol{k}_1\boldsymbol{k}_2}
+\Omega^{(R)}_{\bar{b}_1^\prime b_1^\prime\boldsymbol{k}_1}\delta_{b_1^\prime\bar{b}_2^\prime}\delta_{b_1 b_2}\delta_{\boldsymbol{k}_1\boldsymbol{k}_2}.
\nonumber\\
\label{Z}
\end{eqnarray}
It is understood that the barred index $\bar{b}$ means the index that is not equal to $b$,
that is, $\bar{c}=v$ and $\bar{v}=c$ in Eq.~(\ref{Z}).
Here,
\begin{equation}
w_{bb^\prime\boldsymbol{k}\boldsymbol{q}}
=\varepsilon^{(r)}_{b\boldsymbol{k}+\boldsymbol{q}}
-\varepsilon^{(r)}_{b^\prime\boldsymbol{k}},
\label{w}
\end{equation}
where
$\varepsilon^{(r)}_{b\boldsymbol{k}}$ represents a renormalized $b$-band electron energy
given by\cite{haug}
\begin{equation}
\varepsilon^{(r)}_{b\boldsymbol{k}}
=\varepsilon^\prime_{b\boldsymbol{k}}-\sum_{\boldsymbol{q}}V^{(C)}_{\boldsymbol{q}}
\rho_{bb\boldsymbol{k}+\boldsymbol{q}},
\label{renrgy}
\end{equation}
$\Omega^{(R)}_{b b^\prime\boldsymbol{k}}$
represents a Rabi frequency given by\cite{haug}
\begin{equation}
\Omega^{(R)}_{b b^\prime\boldsymbol{k}}
=\Omega_{bb^\prime}+\sum_{\boldsymbol{q}}V^{(C)}_{\boldsymbol{q}}
\rho_{bb^\prime\boldsymbol{k}+\boldsymbol{q}},
\label{Renergy}
\end{equation}
and
\begin{equation}
\Delta\rho_{bb^\prime\boldsymbol{k}}
=\rho_{bb^\prime\boldsymbol{k}}
-\rho_{bb^\prime\boldsymbol{k}+\boldsymbol{q}}.
\label{Delrho}
\end{equation}
It is noted that the commutator of Eq.~(\ref{comm}) is evaluated by making a factorization approximation
to split four operator terms such as
$
a^\dagger_{\tilde{b},\tilde{\boldsymbol{k}}+\tilde{\boldsymbol{q}}}a_{\tilde{b}^\prime\tilde{\boldsymbol{k}}}
a^\dagger_{b,\boldsymbol{k}+\boldsymbol{q}}a_{b^\prime\boldsymbol{k}}
$
into a product of the operator $A_{\tilde{\boldsymbol{q}}}^\dagger(\tilde{\boldsymbol{k}}\tilde{b}\tilde{b^\prime})$ and a single-particle density matrix
$
\rho_{bb^\prime\boldsymbol{k}}\equiv \langle
a^\dagger_{b,\boldsymbol{k}}a_{b^\prime\boldsymbol{k}}
\rangle,
$
where $\langle \hat{X} \rangle$ means an expectation value of operator $\hat{X}$ with respect to the ground state.
This implies that the equations of motion of functions $A_{\tilde{\boldsymbol{q}}}^\dagger(\tilde{\boldsymbol{k}}\tilde{b}\tilde{b^\prime})$'s, namely, Eq.~(\ref{Heisenberg}), are coupled to
those of functions $\rho_{bb^\prime\boldsymbol{k}}$'s.
In actual calculations, in place of solving such coupled equations, the latter density matrices -- independent of $\boldsymbol{q}$ -- are provided elsewhere by solving the optical Bloch equations 
 -- rather than the semiconductor Bloch equations -- just for the sake of simplicity.

$F(t)$ is expressed as
\begin{equation}
F(t)=F_0(t)\cos{\omega_0 t}.
\label{F}
\end{equation}
Here,   $\omega_0$  represents center frequency of the laser field, and it is understood that
resonant interband transitions accompanying real excited carriers due to $F(t)$ are considered throughout this study, that is, detuning $\Delta\omega$ is positive.
$F_0(t)$ represents an envelope with narrow temporal width $\tau_L$ satisfying
$\tau_L \ll 2\pi/\omega^{(LO)}_{\boldsymbol{q}}$;
in the physical systems of concern, $\tau_L$ is of the order of a couple of tens femtosecond at most.
Hereafter, let $F_0(t)$ be of squared-shape just for the sake of simplicity, that is,
\begin{equation}
F_0(t)=\mathcal{F}_0 \theta(t+\tau_L/2)\theta(t-\tau_L/2)
\label{F0}
\end{equation}
with $\mathcal{F}_0$ constant; the associated pulse area is given by $A_L=d_{cv}\mathcal{F}_0 \tau_L$.
It would be convenient to remove from Eq.~(\ref{Heisenberg}) high-frequency contributions by means of a well-justified rotational-wave approximation.\cite{meystre}
This is done by replacing
$A_{\boldsymbol{q}}^\dagger(\boldsymbol{k}bb^\prime)$ and 
$\rho_{bb^\prime\boldsymbol{k}}$  in Eq.~(\ref{Heisenberg})
by
$e^{i\bar{\omega}_{bb^\prime}t}\bar{A}_{\boldsymbol{q}}^\dagger(\boldsymbol{k}bb^\prime)$ and 
$e^{i\bar{\omega}_{bb^\prime}t}\bar{\rho}_{bb^\prime\boldsymbol{k}}$,
respectively, where
$\bar{\omega}_{cv}=\omega_0$, $\bar{\omega}_{vc}=-\omega_0$, and $\bar{\omega}_{bb}=0$.
Thus, Eq.~(\ref{Heisenberg}) is cast into the form
\begin{eqnarray}
&&-i\left( {d\over dt}+{1\over T_{\boldsymbol{q}\boldsymbol{k}bb^\prime}}\right)\bar{A}_{\boldsymbol{q}}^\dagger(\boldsymbol{k}bb^\prime)
\nonumber\\
&&=[\hat{\mathcal{H}}_e(t), \bar{A}_{\boldsymbol{q}}^\dagger(\boldsymbol{k}bb^\prime)]
-\bar{A}_{\boldsymbol{q}}^\dagger(\boldsymbol{k}bb^\prime)\bar{\omega}_{bb^\prime}
+[\hat{H}_{e-p}, \bar{A}_{\boldsymbol{q}}^\dagger(\boldsymbol{k}bb^\prime)]
\nonumber\\
&&\approx
\sum_{\tilde{\boldsymbol{k}}\tilde{b}\tilde{b}^\prime}
\bar{A}_{\boldsymbol{q}}^\dagger(\tilde{\boldsymbol{k}}\tilde{b}\tilde{b}^\prime)
\bar{Z}_{\boldsymbol{q}}(\tilde{\boldsymbol{k}}\tilde{b}\tilde{b}^\prime,\boldsymbol{k}bb^\prime)
+[\hat{H}_{e-p}, \bar{A}_{\boldsymbol{q}}^\dagger(\boldsymbol{k}bb^\prime)]
\label{Heisenberg2},
\end{eqnarray}
where
$\bar{Z}_{\boldsymbol{q}}(\tilde{\boldsymbol{k}}\tilde{b}\tilde{b}^\prime,\boldsymbol{k}bb^\prime)$
is provided from $Z_{\boldsymbol{q}}(\tilde{\boldsymbol{k}}\tilde{b}\tilde{b}^\prime,\boldsymbol{k}bb^\prime)$
of Eq.~(\ref{Z}) by replacing 
$\rho_{bb^\prime\boldsymbol{k}}$ by $\bar{\rho}_{bb^\prime\boldsymbol{k}}$,
$\Delta \rho_{bb^\prime\boldsymbol{k}}$ by $\Delta \bar{\rho}_{bb^\prime\boldsymbol{k}}$,
$w_{bb^\prime\boldsymbol{kq}}$ by $\bar{w}_{bb^\prime\boldsymbol{kq}}\equiv w_{bb^\prime\boldsymbol{kq}}-\bar{\omega}_{bb^\prime}$,
and
$\Omega^{(R)}_{bb^\prime\boldsymbol{k}}$
by 
\begin{equation}
\bar{\Omega}^{(R)}_{bb^\prime\boldsymbol{k}}=
{1\over 2}\Omega_{0\:bb^\prime}+\sum_{\boldsymbol{q}}V^{(C)}_{\boldsymbol{q}}
\bar{\rho}_{bb^\prime\boldsymbol{k}+\boldsymbol{q}}
\label{Renergy2}
\end{equation}
with $\Omega_{0\:bb^\prime}(t)=d_{bb^\prime}F_0(t)\delta_{b^\prime\bar{b}}$.

The non-hermitian matrix $\bar{Z}_{\boldsymbol{q}}$ would be considered as a slowly varying function in time, since rapidly time-varying contributions are removed in Eq.~(\ref{Heisenberg2}) owing to the rotational-wave approximation.
However, it is noted that $\bar{Z}_{\boldsymbol{q}}$ is discontinuous at $t=\pm \tau_L/2$ due to Eq.~(\ref{F0}). 
Bearing in mind this situation,
we tackle left and right eigenvalue problems of $\bar{Z}_{\boldsymbol{q}}$.\cite{morse,crossing}
These problems are described by
\begin{equation}
U^{L\dagger}_{\boldsymbol{q}}\bar{Z}_{\boldsymbol{q}}
=\mathcal{E}_{\boldsymbol{q}}U^{L\dagger}_{\boldsymbol{q}},\;\;\;
\label{eigenZbarL}
\end{equation}
and
\begin{equation}
\bar{Z}_{\boldsymbol{q}}U^R_{\boldsymbol{q}}
=U^R_{\boldsymbol{q}}\mathcal{E}_{\boldsymbol{q}},
\label{eigenZbarR}
\end{equation}
respectively,
in terms of an eigenvalue diagonal-matrix $\mathcal{E}_{\boldsymbol{q}}$, and  the associated biorthogonal set of eigenvectors $\left\{ U^{L}_{\boldsymbol{q}}, U^{R}_{\boldsymbol{q}} \right\}$
satisfying the orthogonality relation $U^{L\dagger}_{\boldsymbol{q}}U^{R}_{\boldsymbol{q}}=1$
and the completeness $U^{R}_{\boldsymbol{q}}U^{L\dagger}_{\boldsymbol{q}}=1$.\cite{appB}
Here, matrix notations are employed, that is,
$\bar{Z}_{\boldsymbol{q}}=\left\{\bar{Z}_{\boldsymbol{q}}(\tilde{\boldsymbol{k}}\tilde{b}\tilde{b}^\prime,\boldsymbol{k}bb^\prime)\right\}$,
$\mathcal{E}_{\boldsymbol{q}}=\left\{\mathcal{E}_{\boldsymbol{q}\alpha}\right\}$,
and
$U^{L/R}_{\boldsymbol{q}}=\left\{U^{L/R}_{\boldsymbol{q}\alpha}(\boldsymbol{k}bb^\prime)\right\}$,
where 
$\mathcal{E}_{\boldsymbol{q}\alpha}$
and
$U^{L/R}_{\boldsymbol{q}\alpha}(\boldsymbol{k}bb^\prime)$
are the $\alpha$th eigenvalue and eigenvector, respectively.
It is pointed out that similarly to $\bar{Z}_{\boldsymbol{q}}$, both of $\mathcal{E}_{\boldsymbol{q}}$ and $U^{L/R}_{\boldsymbol{q}}$ are slowly varying functions in time, however,  except at $t=\pm\tau_L/2$ and in the vicinity of crossings relevant to non-adiabatic interactions.\cite{crossing}
Equations~(\ref{eigenZbarL}) and (\ref{eigenZbarR}) can be solved in an analytic manner, as shown in Ref.~\onlinecite{appB}.
Accordingly, the $\alpha$th left and right eigenvectors are expressed as
\begin{equation}
U^{L\dagger}_{\boldsymbol{q}\alpha}=N^L_{\boldsymbol{q}\alpha}V^{(C)}_{\boldsymbol{q}}u^{L\dagger}_{\boldsymbol{q}\alpha},
\label{eigenVL}
\end{equation}
and
\begin{equation}
U^R_{\boldsymbol{q}\alpha}=N^R_{\boldsymbol{q}\alpha}V^{(C)}_{\boldsymbol{q}}u^R_{\boldsymbol{q}\alpha},
\label{eigenVR}
\end{equation}
respectively,
where
$u^{L/R}_{\boldsymbol{q}}=\left\{u^{L/R}_{\boldsymbol{q}\alpha}(\boldsymbol{k}bb^\prime)\right\}$,
and $N^{L/R}_{\boldsymbol{q}\alpha}$ is a normalization constant to be determined 
by
\begin{equation}
N^L_{\boldsymbol{q}\alpha}N^R_{\boldsymbol{q}\alpha}[V^{(C)}_{\boldsymbol{q}}]^2
\left(u^{L\dagger}_{\boldsymbol{q}\alpha}u^{R}_{\boldsymbol{q}\alpha}\right)=1.
\label{normalization}
\end{equation}

Given the relation 
$\bar{Z}_{\boldsymbol{q}}=U^R_{\boldsymbol{q}}\mathcal{E}_{\boldsymbol{q}}U^{L\dagger}_{\boldsymbol{q}}$,
Eq.~(\ref{Heisenberg2}) is recast into the form:
\begin{eqnarray}
-i{d B^\dagger_{\boldsymbol{q}\alpha}\over dt}
&=&
B^\dagger_{\boldsymbol{q}\alpha}\mathcal{E}_{\boldsymbol{q}\alpha}
+i\sum_{\alpha^\prime}B^\dagger_{\boldsymbol{q}\alpha^\prime}\mathcal{W}_{\boldsymbol{q}\alpha^\prime\alpha}
\nonumber\\
&&+[\hat{H}_{e-p}, B^\dagger_{\boldsymbol{q}\alpha}].
\label{Heisenberg3}
\end{eqnarray}
Here, the operator $B^\dagger_{\boldsymbol{q}\alpha}$ is defined as
\begin{equation}
B^\dagger_{\boldsymbol{q}\alpha}
=\sum_{\boldsymbol{k}bb^\prime}
\bar{A}_{\boldsymbol{q}}^\dagger(\boldsymbol{k}bb^\prime)
U^{R}_{\boldsymbol{q}\alpha}(\boldsymbol{k}bb^\prime)
\equiv 
\bar{A}_{\boldsymbol{q}}^\dagger U^{R}_{\boldsymbol{q}\alpha},
\label{B+}
\end{equation}
and
\begin{equation}
\mathcal{W}_{\boldsymbol{q}\alpha^\prime\alpha}
=W_{\boldsymbol{q}\alpha^\prime\alpha}
+{\gamma^{(B)}_{\boldsymbol{q}\alpha^\prime\alpha}\over 2},
\label{mathcalW}
\end{equation}
where
\begin{equation}
W_{\boldsymbol{q}\alpha^\prime\alpha}
=\sum_{\boldsymbol{k}bb^\prime}
{d U^{L\dagger}_{\boldsymbol{q}\alpha^\prime}(\boldsymbol{k}bb^\prime) \over dt}
U^{R}_{\boldsymbol{q}\alpha}(\boldsymbol{k}bb^\prime)
\equiv
{d U^{L\dagger}_{\boldsymbol{q}\alpha^\prime} \over dt}
U^{R}_{\boldsymbol{q}\alpha},
\label{W}
\end{equation}
and
\begin{equation}
{\gamma^{(B)}_{\boldsymbol{q}\alpha^\prime\alpha}\over 2}
=\sum_{\boldsymbol{k}bb^\prime}
U^{L\dagger}_{\boldsymbol{q}\alpha^\prime}(\boldsymbol{k}bb^\prime)
{1\over T_{\boldsymbol{q}\boldsymbol{k}bb^\prime}}
U^{R}_{\boldsymbol{q}\alpha}(\boldsymbol{k}bb^\prime).
\label{gammaB}
\end{equation}

Equation~(\ref{Heisenberg3}) is nothing but adiabatic coupled-equations.
Without the second and third terms of the right-hand side of this equation, which cause couplings of adiabatic quasi-boson  states $\alpha$ and $\alpha^\prime$,
$B^\dagger_{\boldsymbol{q}\alpha}(t)$ would be given by
\begin{equation}
B^\dagger_{\boldsymbol{q}\alpha}(t)
=\exp{\left[i\int^t_{t^\prime} \mathcal{E}_{\boldsymbol{q}\alpha}(\tau)d\tau\right]}
B^\dagger_{\boldsymbol{q}\alpha}(t^\prime),
\label{B+2}
\end{equation}
where time-dependence is explicitly shown.
Therefore, $\mathcal{E}_{\boldsymbol{q}\alpha}(t)$ is considered as adiabatic energy  at time $t$ associated with the operator $B^\dagger_{\boldsymbol{q}\alpha}(t)$ thus introduced.
Hereafter, this operator is termed as a creation operator of  quasi-boson, and the corresponding annihilation operator is defined as\cite{dyson}
\begin{equation}
B_{\boldsymbol{q}\alpha}
=\sum_{\boldsymbol{k}bb^\prime}
U^{R\dagger}_{\boldsymbol{q}\alpha}(\boldsymbol{k}bb^\prime)
\bar{A}_{\boldsymbol{q}}(\boldsymbol{k}bb^\prime)
\equiv U^{R\dagger}_{\boldsymbol{q}\alpha}\bar{A}_{\boldsymbol{q}}.
\label{B-}
\end{equation}
Thus, Eq.~(\ref{B+2}) is read as
\begin{equation}
B_{\boldsymbol{q}\alpha}(t)
=\exp{\left[-i\int^t_{t^\prime} \mathcal{E}^*_{\boldsymbol{q}\alpha}(\tau)d\tau\right]}
B_{\boldsymbol{q}\alpha}(t^\prime).
\label{B-2}
\end{equation}
It is noted that
$B_{\boldsymbol{q}\alpha}(t)$ and $B^\dagger_{\boldsymbol{q}\alpha}(t)$ do not satisfy the equal-time commutation relations for a real boson, that is,
$[B_{\boldsymbol{q}\alpha}(t), B^\dagger_{\boldsymbol{q}^\prime\alpha^\prime}(t)] \not= \delta_{\boldsymbol{q}\boldsymbol{q}^\prime}\delta_{\alpha\alpha^\prime}$,
and that $\mathcal{E}_{\boldsymbol{q}\alpha}$ is a complex number in general,\cite{conventionE} even though
$B^\dagger_{\boldsymbol{q}\alpha}$ is  hermitian-conjugate  of $B_{\boldsymbol{q}\alpha}$.
As regards $W_{\boldsymbol{q}\alpha^\prime\alpha}$ of Eq.~(\ref{W}),
this is considered as a non-adiabatic coupling between $\alpha^\prime$ and $\alpha$.
It is readily seen that this term is read as:
\begin{equation}
W_{\boldsymbol{q}\alpha^\prime\alpha}
=\frac{U^{L\dagger}_{\boldsymbol{q}\alpha^\prime}\frac{d\bar{Z}_{\boldsymbol{q}}}{ dt}U^{R}_{\boldsymbol{q}\alpha}}
{\mathcal{E}_{\boldsymbol{q}\alpha^\prime}-\mathcal{E}_{\boldsymbol{q}\alpha}},
\;\;\;\alpha^\prime \not=\alpha,
\label{W2}
\end{equation}
due to the complex analog of the Hellman-Feynman theorem,\cite{crossing}
and $W_{\boldsymbol{q}\alpha\alpha}\not=0$.

\begin{figure}[tb]
\begin{center}
\includegraphics[width=7.0cm,clip]{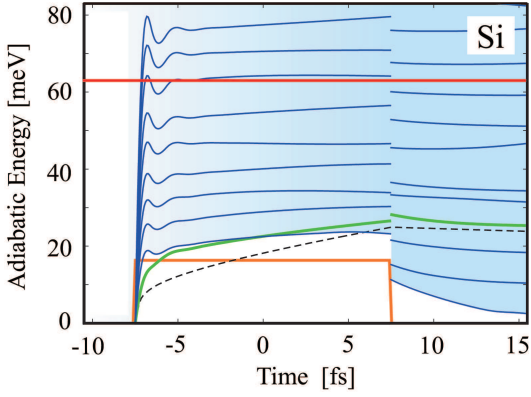}
\caption{Adiabatic energy curves  of Si (in the unit of meV) as a function of time $t$ (in the unit of fs). 
A plasmon-like mode is represented by a green solid line, and a bundle of the electron-hole continua is represented by blue solid lines. 
Further,  the alteration of bare Rabi frequency $\Omega_{0cv}$ as a function of $t$ is represented by an orange solid line; here this is of squared shape [see Eq.~(\ref{F0})].
In addition, the LO-phonon energy, $\omega^{(LO)}_{\boldsymbol{q}}=63$ meV, is represented by a red solid line, and
plasmon energy proportional to total excited electron density $N_{ex}(t)$ is represented by a broken line, just for the purpose of comparison of the plasmon-like mode.
The gradation of  blue color shows schematic change of $N_{ex}(t)$ in $t$, where the lowest limit of this gradation represents the threshold energy of a bundle of the electron-hole continua.
}
\label{fig2}
\end{center}
\end{figure}

Figure~\ref{fig2} shows adiabatic energy curves of $\mathcal{E}_{\boldsymbol{q}\alpha}(t)$ of Si 
as a function of $t$ in the early-time region in the small $\boldsymbol{q}$-limit concerned here;
 to be precise, just positive real parts of $\mathcal{E}_{\boldsymbol{q}\alpha}(t)$ are depicted.
The energetically-lowest ten curves of $\{\mathcal{E}_{\boldsymbol{q}\alpha}(t)\}$ are shown, where these results  are calculated by solving either Eq.~(\ref{eigenZbarL}) or  Eq.~(\ref{eigenZbarR}).
Here, the crystal is assumed to be cubic, and the material parameters given in Ref.~\onlinecite{SMparameter} are employed.
It is considered that the energy indicated by a green solid line is attributed to a plasmon-like mode.
Indeed,  this looks almost proportional to excited electron density,
\(
N_{ex}(t)=(1/V)\sum_{\boldsymbol{k}}\bar\rho_{cc\boldsymbol{k}},
\)
where
the plasmon energy $\omega_{pl}$, given by 
\(
\omega^2_{pl}=4\pi N_{ex}(t)/\epsilon_\infty m_{cv},
\)
 is indicated by a broken line in this figure; $m_{cv}$ is reduced mass of an electron in a joint band composed of $c$ and $v$ bands under an effective mass approximation.
The maximum of $N_{ex}(t)$, represented by $N^0_{ex}$, is  $6.31\times 10^{17}/{\rm cm}^3$.
In fact,
the approximated expression of the adiabatic energy due to this mode is composed of the terms affected by
interband density matrices of $\bar{\rho}_{cv\boldsymbol{k}}(t)$ and $\bar{\rho}_{vc\boldsymbol{k}}(t)$, and
Rabi frequencies of $\bar{\Omega}^{(R)}_{cv\boldsymbol{k}}(t)$ and $\bar{\Omega}^{(R)}_{vc\boldsymbol{k}}(t)$
in a complicated manner, 
as well as of the terms simply proportional to 
intraband density matrices of $\bar{\rho}_{cc\boldsymbol{k}}(t)$ and $\bar{\rho}_{vv\boldsymbol{k}}(t)$
that determines $N_{ex}(t)$.\cite{appB}
Thus, the difference of the functional shapes between the energy of the plasmon-like mode (a green solid line)  and the energy of plasmon mode (a broken line) is
definitely attributed to the transient effect of the interband density matrices and the Rabi frequencies.
On the occasion that  such effects are suppressed in the time region of $t >\tau_L/2$,
the energy of the plasmon-like mode become  identical to a plasmon energy just proportional to
electron density, aside from renormalization effects due to $V^{(C)}_{\boldsymbol{q}}$.
In this context, hereafter it is understood that this mode is termed as plasmon, unless otherwise stated.

Adiabatic-energy levels indicated by  blue solid lines show  a bundle of electron-hole continua, 
the lowest energy of which delineates a threshold of these contributions.
This threshold is approximately given by twice the Rabi frequency of Eq.~(\ref{Renergy2}).\cite{appB}
Just for the purpose of comparison, $\Omega_{0cv}$, corresponding to the first term of Eq.~(\ref{Renergy2}), is represented by an orange solid line,  similarly to Fig.~\ref{fig1}.
Higher-lying {\it discretized} energy levels of this bundle result from 
the incorporation of  the {\it finite} number of sites $N_s$ in the calculations; $V=N_sd^3$ with $d$ as a lattice constant.
It is noted that the contribution of these continua exclusively arises from interband transitions, and no intraband transitions affect the formation of the continua; in fact, the latter effect vanishes in the small $\boldsymbol{q}$-limit.
In addition, the LO-phonon energy, $\omega^{(LO)}_{\boldsymbol{q}}=63$ meV, is also represented by a red solid line
in Fig.~\ref{fig2}.

As is seen in Fig.~\ref{fig2},
the LO-phonon is embedded to the electron-hole continua right after the onset of the laser irradiation, and thus,
the mixing between these two modes likely induces FR.
As evidently seen from this figure, this effect lasts even after the completion of laser irradiation: $t > \tau_L/2$ .
As regards the plasmon mode, this tends to dive into the continua after the completion of laser irradiation;
although it appears that the plasmon forms FR as well, this is not the case without an interaction with a phonon.
It is obvious that the time evolution shown in Fig.~\ref{fig2} corresponds to that in the early-time region already shown in Fig.~\ref{fig1}. 
The time evolution of GaAs is omitted, since this is almost similar to Fig.~\ref{fig2} though the relevant FR dynamics is different each other (see Sec.~\ref{sec3}).

\subsubsection{Quasi-Boson-LO-Phonon Interactions}
\label{sec2A2}

Following Eqs.~(\ref{Hep}), (\ref{B+}), and (\ref{B-}), $\hat{H}_{e-p}$ is rewritten as
\begin{eqnarray}
\hat{H}_{e-p}
&=&\sum_{\boldsymbol{q},\alpha}
\left(
M_{\boldsymbol{q}\alpha}
c_{\boldsymbol{q}} B^\dagger_{\boldsymbol{q}\alpha}
+M^*_{\boldsymbol{q}\alpha}c^\dagger_{\boldsymbol{q}} B_{\boldsymbol{q}\alpha}
\right),
\label{Hep2}
\end{eqnarray}
where an effective coupling constant between a quasi-boson and an LO-phonon becomes
of the form
\begin{equation}
M_{\boldsymbol{q}\alpha}
=
\sum_{\boldsymbol{k}b} g_{b\boldsymbol{q}}
U^{L\dagger}_{\boldsymbol{q}\alpha}(\boldsymbol{k}bb).
\label{M}
\end{equation}
Thus, the commutator of $[\hat{H}_{e-p}, c^\dagger_{\boldsymbol{q}} ]$ is readily evaluated as
\begin{equation}
[\hat{H}_{e-p}, c^\dagger_{\boldsymbol{q}} ]
=\sum_\alpha M_{\boldsymbol{q}\alpha}B^\dagger_{\boldsymbol{q}\alpha}.
\label{comm3}
\end{equation}
Hereafter, $M_{\boldsymbol{q}\alpha}$ is termed as the effective coupling constant throughout this article despite  a function of time.

On the other hand, the evaluation of
the commutator of $[\hat{H}_{e-p}, B^\dagger_{\boldsymbol{q}\alpha} ]$ seems more involved,
because the quasi-boson operators do not satisfy  commutation relations, as mentioned below Eq.~(\ref{B-2}).
By replacing $ [B_{\boldsymbol{q}\alpha}, B^\dagger_{\boldsymbol{q}^\prime\alpha^\prime}] $ and
$ [B^\dagger_{\boldsymbol{q}\alpha}, B^\dagger_{\boldsymbol{q}^\prime\alpha^\prime}] $ by
$\langle [B_{\boldsymbol{q}\alpha}, B^\dagger_{\boldsymbol{q}^\prime\alpha^\prime}] \rangle$
and
$\langle [B^\dagger_{\boldsymbol{q}\alpha}, B^\dagger_{\boldsymbol{q}^\prime\alpha^\prime}] \rangle$,
respectively,
based on the factorization approximation,
it is seen that
\begin{equation}
[\hat{H}_{e-p}, B^\dagger_{\boldsymbol{q}\alpha} ]
\approx
M^{\prime\prime}_{-\boldsymbol{q}\alpha}c_{-\boldsymbol{q}} 
+M^{\prime*}_{\boldsymbol{q}\alpha}c^\dagger_{\boldsymbol{q}} ,
\label{comm4}
\end{equation}
where
\begin{eqnarray}
M^{\prime\prime}_{-\boldsymbol{q}\alpha}
&=&
\sum_{\alpha^\prime}
M_{-\boldsymbol{q}\alpha^\prime}
\langle [B^\dagger_{-\boldsymbol{q}\alpha^\prime},B^\dagger_{\boldsymbol{q}\alpha}] \rangle
\nonumber\\
&=&
\sum_{\boldsymbol{k}bb^\prime} 
\left(
g_{b-\boldsymbol{q}}\bar{\rho}_{bb^\prime\boldsymbol{k}}
-
g_{b^\prime-\boldsymbol{q}}\bar{\rho}_{bb^\prime\boldsymbol{k}+\boldsymbol{q}}
\right)U^{R}_{\boldsymbol{q}\alpha}(\boldsymbol{k}bb^\prime),
\nonumber
\\
\label{M"}
\end{eqnarray}
and
\begin{eqnarray}
M^{\prime*}_{\boldsymbol{q}\alpha}
&=&
\sum_{\alpha^\prime}
M^*_{\boldsymbol{q}\alpha^\prime}
\langle [B_{\boldsymbol{q}\alpha^\prime},B^\dagger_{\boldsymbol{q}\alpha}] \rangle
\nonumber\\
&=&
\sum_{\boldsymbol{k}bb^\prime} 
\left(
g^*_{b\boldsymbol{q}}\bar{\rho}_{bb^\prime\boldsymbol{k}}
-
g^*_{b^\prime\boldsymbol{q}}\bar{\rho}_{bb^\prime\boldsymbol{k}+\boldsymbol{q}}
\right)U^{R}_{\boldsymbol{q}\alpha}(\boldsymbol{k}bb^\prime).
\label{M'}
\end{eqnarray}

It is noted that $M_{\boldsymbol{q}\alpha}$, $M^\prime_{\boldsymbol{q}\alpha}$, and $M^{\prime\prime}_{-\boldsymbol{q}\alpha}$ are slowly varying function in time except at $t=\pm\tau_L/2$ and in the vicinity of crossings, since all of these functions are represented as Eqs.~(\ref{M}), (\ref{M"}), and (\ref{M'})
in terms of the adiabatic eigenvectors, $U^{L\dagger}_{\boldsymbol{q}}$ and $U^R_{\boldsymbol{q}}$, and the slowly time-varying density-matrices, $\bar{\rho}_{bb^\prime\boldsymbol{k}}$ [see also the remark below Eq.~(\ref{eigenZbarR})].
This fact is a cardinal point that gives a theoretical basis to the introduction of  the PQ picture aimed at in the present study.

Since the present framework is given within the two-band model,
the coupling constants $g_{c\boldsymbol{q}}$ and $g_{v\boldsymbol{q}}$ are rewritten as
\begin{equation}
g_{c\boldsymbol{q}}
=g^0_{\boldsymbol{q}}+{1\over2}\Delta g_{\boldsymbol{q}},
\label{gc}
\end{equation}
and
\begin{equation}
g_{v\boldsymbol{q}}
=g^0_{\boldsymbol{q}}-{1\over2}\Delta g_{\boldsymbol{q}},
\label{gv}
\end{equation}
respectively,
where
\begin{equation}
g^0_{\boldsymbol{q}}={1\over2}\left(g_{c\boldsymbol{q}}+g_{v\boldsymbol{q}}\right),\;\;\;
\Delta g_{\boldsymbol{q}}=g_{c\boldsymbol{q}}-g_{v\boldsymbol{q}}.
\label{g2}
\end{equation}
Putting Eqs.~(\ref{gc}) and (\ref{gv}) into Eqs.~(\ref{M}), (\ref{M"}), and (\ref{M'}),
the effective coupling constants, $M_{\boldsymbol{q}\alpha}$, $M^{\prime\prime}_{-\boldsymbol{q}\alpha}$, and
$M^{\prime*}_{\boldsymbol{q}\alpha}$, become\cite{appB}
\begin{equation}
M_{\boldsymbol{q}\alpha}
=g^0_{\boldsymbol{q}}N^L_{\boldsymbol{q}\alpha}+\Delta g_{\boldsymbol{q}}\Delta N^L_{\boldsymbol{q}\alpha},
\label{M2}
\end{equation}
\begin{equation}
M^{\prime\prime}_{-\boldsymbol{q}\alpha}
=g^0_{-\boldsymbol{q}}N^R_{\boldsymbol{q}\alpha}+\Delta g_{\boldsymbol{q}}\Delta N^R_{\boldsymbol{q}\alpha},
\label{M"2}
\end{equation}
and
\begin{equation}
M^{\prime*}_{\boldsymbol{q}\alpha}
=g^{0*}_{\boldsymbol{q}}N^R_{\boldsymbol{q}\alpha}+\Delta g^*_{\boldsymbol{q}}\Delta N^R_{\boldsymbol{q}\alpha},
\label{M'2}
\end{equation}
respectively,
where 
$\Delta N^{L/R}_{\boldsymbol{q}\alpha}$ is provided by
\begin{equation}
\Delta N^L_{\boldsymbol{q}\alpha}
={1\over2}\sum_{\boldsymbol{k}} 
\left[
U^{L\dagger}_{\boldsymbol{q}\alpha}(\boldsymbol{k}cc)-U^{L\dagger}_{\boldsymbol{q}\alpha}(\boldsymbol{k}vv)
\right],
\label{NL}
\end{equation}
and
\begin{eqnarray}
\Delta N^R_{\boldsymbol{q}\alpha}
&=&{1\over2}\sum_{\boldsymbol{k}} 
\left[
\Delta\bar{\rho}_{cc\boldsymbol{k}\boldsymbol{q}}
U^{R}_{\boldsymbol{q}\alpha}(\boldsymbol{k}cc)
-\Delta\bar{\rho}_{vv\boldsymbol{k}\boldsymbol{q}}
U^{R}_{\boldsymbol{q}\alpha}(\boldsymbol{k}vv)
\right.
\nonumber\\
&&+
(\bar{\rho}_{cv\boldsymbol{k}}+\bar{\rho}_{cv\boldsymbol{k}+\boldsymbol{q}})
U^{R}_{\boldsymbol{q}\alpha}(\boldsymbol{k}cv)
\nonumber\\
&&-
\left.
(\bar{\rho}_{vc\boldsymbol{k}}+\bar{\rho}_{vc\boldsymbol{k}+\boldsymbol{q}})
U^{R}_{\boldsymbol{q}\alpha}(\boldsymbol{k}vc)
\right].
\label{NR}
\end{eqnarray}
It is remarked that in the small-$\boldsymbol{q}$ limit,
$N_{\boldsymbol{q}\alpha} \propto |\boldsymbol{q}|$, and
$\Delta N^{L/R}_{\boldsymbol{q}\alpha} $ is independent of $\boldsymbol{q}$.

In a non-polar crystal --- such as Si --- with two or more atoms per unit cell, 
an electron-LO-phonon interaction is described by a phenomenological optical-phonon deformation potential.
Here, the associated coupling constant, represented by $g^D_{b\boldsymbol{q}}$, is real and is considered approximately as independent of $\boldsymbol{q}$.\cite{yu}
On the other hand, in a polar or partially ionic crystal --- such as GaAs --- with two or more atoms per unit cell, 
a long-range interaction of an electron and a scalar field induced by atomic displacements is known as a Fr\"{o}hlich
interaction.
Here, the associated coupling constant, represented by $g^F_{b\boldsymbol{q}}$,  is pure imaginary and $|g^F_{b\boldsymbol{q}}| \propto |\boldsymbol{q}|^{-1}$.\cite{yu}
In addition, $g^{0F}_{\boldsymbol{q}} \approx g^F_{b\boldsymbol{q}}$,  and  $\Delta g^F_{\boldsymbol{q}} \approx 0$ because of approximate independence of a band index $b$; where $g^{0j}_{\boldsymbol{q}}$ and $\Delta g^j_{\boldsymbol{q}}$ with $j=D, F$ correspond to Eq.~(\ref{g2}).
Bearing these observations in mind, 
as regards $M_{\boldsymbol{q}\alpha}$ of Eq.~(\ref{M2}) in the small-$\boldsymbol{q}$ limit,
 the leading term of it is independent of $\boldsymbol{q}$, and is given by
\begin{equation}
M^D_{\boldsymbol{q}\alpha}\simeq \Delta g^D_{\boldsymbol{q}}\Delta N^L_{\boldsymbol{q}\alpha} 
\label{M2D}
\end{equation}
for the optical-phonon deformation potential, 
and
\begin{equation}
M^F_{\boldsymbol{q}\alpha}\simeq g^{0F}_{\boldsymbol{q}}N^L_{\boldsymbol{q}\alpha}
\label{M2F}
\end{equation}
for the Fr\"{o}hlich interaction.
The net effective coupling constants are the sum of both contributions, that is,
\begin{equation}
M_{\boldsymbol{q}\alpha}=M^D_{\boldsymbol{q}\alpha}+M^F_{\boldsymbol{q}\alpha}.
\label{M2DF}
\end{equation}
The similar results hold correctly for both $M^{\prime\prime}_{-\boldsymbol{q}\alpha}$ and $M^{\prime*}_{\boldsymbol{q}\alpha}$ as well.

\subsubsection{Equations of Motion of Polaronic Quasiparticle Operators}
\label{sec2A3}

Consulting Eq.~(\ref{comm4}), the adiabatic coupled equations given by Eq.~(\ref{Heisenberg3}) become of the form:
\begin{eqnarray}
-i{d B^\dagger_{\boldsymbol{q}\alpha}\over dt}
&=&
B^\dagger_{\boldsymbol{q}\alpha}
\mathcal{E}_{\boldsymbol{q}\alpha}
+c^\dagger_{\boldsymbol{q}} M^{\prime*}_{\boldsymbol{q}\alpha}
\nonumber\\
&&
+i\sum_{\alpha^\prime}B^\dagger_{\boldsymbol{q}\alpha^\prime}\mathcal{W}_{\boldsymbol{q}\alpha^\prime\alpha}
+M^{\prime\prime}_{-\boldsymbol{q}\alpha}c_{-\boldsymbol{q}} 
\label{HeisenbergB}.
\end{eqnarray}
On the other hand, the equation of motion of the LO phonon is described by
\begin{eqnarray}
-i{d c^\dagger_{\boldsymbol{q}}\over dt}
&=&
\sum_\alpha B^\dagger_{\boldsymbol{q}\alpha}M_{\boldsymbol{q}\alpha}
+c^\dagger_{\boldsymbol{q}}
\omega^{(LO)}_{\boldsymbol{q}}
\label{Heisenbergc}.
\end{eqnarray}
Both of Eqs.~(\ref{HeisenbergB}) and (\ref{Heisenbergc}) are integrated into a single equation in terms of matrix notations as follows:
\begin{equation}
-i{d\over dt}[B^\dagger_{\boldsymbol{q}},c^\dagger_{\boldsymbol{q}}]
=
[B^\dagger_{\boldsymbol{q}},c^\dagger_{\boldsymbol{q}}]h_{\boldsymbol{q}}
+[iB^\dagger_{\boldsymbol{q}}\mathcal{W}_{\boldsymbol{q}}+ M^{\prime\prime}_{-\boldsymbol{q}}c_{-\boldsymbol{q}},0].
\label{Heisenberg4}
\end{equation}
Here, $h_{\boldsymbol{q}}$ is a $[(N+2) \times (N+2)]$ non-hermitian matrix given by
\begin{equation}
h_{\boldsymbol{q}}=
\left(
\begin{array}{cc}
\mathcal{E}_{\boldsymbol{q}}&M_{\boldsymbol{q}}\\
M^{\prime\dagger}_{\boldsymbol{q}}&\omega^{(LO)}_{\boldsymbol{q}}
\end{array}
\right),
\label{h}
\end{equation}
where $N$ represents the number of  electron-hole (discretized) continua aside from two discrete state attributed to
a plasmon and a phonon [see Fig.~\ref{fig2}].
Hereafter, the index $\alpha$ is exclusively represented as $\alpha\equiv(\bar{\alpha},\alpha_1)$, which is composed of
$N$ continua designated by  $\bar{\alpha}=1\sim N$ and a single discrete level (the plasmon) 
designated by  $\alpha_1$.
It is noted that the index $\alpha$ also includes a state with ${\rm Re}\/\mathcal{E}_{\boldsymbol{q}\alpha} <0$
[consult Ref.~\onlinecite{appB}].
Further, another discrete level (the phonon) is designated by  $\alpha_2$, if necessary.
In addition, a matrix element of $h_{\boldsymbol{q}}$ is designated by use of an index $\gamma
=1\sim (N+2)$.
It is understood that
the matrix-indices of $\alpha^\prime$, $\alpha^{\prime\prime}$, and so on, those of $\bar{\alpha}^\prime$, $\bar{\alpha}^{\prime\prime}$, and so on, and those of $\gamma^\prime$, $\gamma^{\prime\prime}$, and so on, will be adopted with the same meanings as $\alpha$ ranging from 1 to $(N+1)$, $\bar{\alpha}$ ranging from 1 to $N$, and $\gamma$ ranging from 1 to $(N+2)$, respectively.
The same rule holds for $\alpha_i^\prime$, $\alpha_i^{\prime\prime}$, and so on $(i=1,2)$.

Now we take account of the following matrix equations:
\begin{equation}
h_{\boldsymbol{q}}V^R_{\boldsymbol{q}}
=V^R_{\boldsymbol{q}}E_{\boldsymbol{q}},
\label{eigenh}
\end{equation}
where
$V^{R}_{\boldsymbol{q}}$ is a $[(N+2) \times N_0]$-matrix, and
$E_{\boldsymbol{q}}$ is a $(N_0\times N_0)$ diagonal matrix
with $N_0$ as the number of independent solutions of Eq.~(\ref{eigenh}).
Equation~(\ref{eigenh}) provides a theoretical basis on which both of 
LO-phonon and plasmon are brought  on an equal footing into connection 
with the CP dynamics concerned here.
Hereafter, the solution is exclusively
labeled as $\beta$, $\beta^\prime$, and so on, and
$N_d\equiv N_0-N$, implying the number of discrete solutions.
The problem posed here is classified  into the following three cases dependent on the number of the independent solutions.\cite{seaton}

The first case is that both discrete levels of $\alpha_1$ (plasmon) and $\alpha_2$ (phonon) are embedded into the continuum
$\bar{\alpha}$, and $N_0$ is equal to $N$ with $N_d=0$.
Thus, the above matrix equations belong to a Fano problem; aside from $h_{\boldsymbol{q}}$ being non-hermitian.
To be specific, this is a multichannel scattering problem with one open channel and two closed channels.
Here, $E_{\boldsymbol{q}}$ coincides with given energy $\mathcal{E}_{\boldsymbol{q}}$:
$E_{\boldsymbol{q}\bar{\alpha}\bar{\alpha}}\equiv E_{\boldsymbol{q}\bar{\alpha}}=\mathcal{E}_{\boldsymbol{q}\bar{\alpha}}$.
This case corresponds to the situation of Fig.~\ref{fig2}  that both of the plasmon and phonon dive into the continuum after the completion of pulse excitation.
On the other hand, the second case is that one of the two levels of  $\alpha_1$ and $\alpha_2$ is embedded into the continuum ($N_d=1$), and the third case is that neither $\alpha_1$ nor $\alpha_2$ is embedded into the continuum ($N_d=2$). The last two cases likely take place just during the very short period of pulse irradiation, as seen in Fig.~\ref{fig2}. 
Hereafter, the first case is exclusively examined, unless otherwise stated, since this case lasts for approximately one hundred fs till $T^\prime_{12}$, during which  neither population relaxation of excited carriers nor  dephasing process are dominant yet.

In the case of concern, the index of $\beta$ is equivalent to that of $\bar{\alpha}$, since there are $N$ independent solutions of Eq.~(\ref{eigenh}): $N_0=N$.
Thus, in terms of the $[(N+2) \times N]$-rectangular matrix $V^{R}_{\boldsymbol{q}}$,
a set of $N$ operators,
$F^\dagger_{\boldsymbol{q}\beta}\;(\beta=1\sim N)$, is defined 
as
\begin{equation}
F^\dagger_{\boldsymbol{q}}
=[B^\dagger_{\boldsymbol{q}},c^\dagger_{\boldsymbol{q}}]V^R_{\boldsymbol{q}}.
\label{F+}
\end{equation}
Further, an $[N \times (N+2)]$-rectangular matrix $\bar{V}^{R}_{\boldsymbol{q}}$ is introduced to ensure the inverse relation of it, that is,
\begin{equation}
[B^\dagger_{\boldsymbol{q}},c^\dagger_{\boldsymbol{q}}]
=F^\dagger_{\boldsymbol{q}}\bar{V}^R_{\boldsymbol{q}}.
\label{F+2}
\end{equation}
To make both equations compatible each other, the following relation is readily required:
\(
\bar{V}^{R}_{\boldsymbol{q}}V^{R}_{\boldsymbol{q}}=V^{R}_{\boldsymbol{q}}\bar{V}^{R}_{\boldsymbol{q}}=1
\).

Given Eq.~(\ref{F+}), 
Eq.~(\ref{Heisenberg4}) becomes of the form:
\begin{eqnarray}
-i{d\over dt}F^\dagger_{\boldsymbol{q}\beta}
&=&
F^\dagger_{\boldsymbol{q}\beta}E_{\boldsymbol{q}\beta}
+i\sum_{\beta^\prime}F^\dagger_{\boldsymbol{q}\beta^\prime}\mathcal{I}_{\boldsymbol{q}\beta^\prime\beta}
\nonumber\\
&&+\sum_{\beta^\prime}\mathcal{M}^{\prime\prime}_{-\boldsymbol{q}\beta\beta^\prime}F_{-\boldsymbol{q}\beta^\prime}.
\label{HeisenbergF}
\end{eqnarray}
This  is the adiabatic coupled equations for $F^\dagger_{\boldsymbol{q}}$.
In Eq.~(\ref{HeisenbergF}),
$F_{\boldsymbol{q}}$ is  hermitian-conjugate of $F^\dagger_{\boldsymbol{q}}$, namely,
\begin{equation}
F_{\boldsymbol{q}}
=V^{R\dagger}_{\boldsymbol{q}}
\left[
\begin{array}{c}
B_{\boldsymbol{q}}\\
c_{\boldsymbol{q}}
\end{array}
\right],
\label{F-}
\end{equation}
\begin{equation}
\mathcal{M}^{\prime\prime}_{-\boldsymbol{q}\beta\beta^\prime}
=\left(\sum_{\alpha}M^{\prime\prime}_{-\boldsymbol{q}\alpha}V^R_{\boldsymbol{q}\alpha\beta}
\right)
\bar{V}^{R\dagger}_{-\boldsymbol{q}\alpha_2\beta^\prime},
\label{calM"}
\end{equation}
and
\begin{equation}
\mathcal{I}_{\boldsymbol{q}}=
I_{\boldsymbol{q}}+{\gamma^{(0)}_{\boldsymbol{q}}\over 2}.
\label{mathcalgamma}
\end{equation}
$I_{\boldsymbol{q}}$ is a non-adiabatic interaction given by
\begin{eqnarray}
I_{\boldsymbol{q}\beta^\prime\beta}
&=&\sum_{\alpha\alpha^\prime\alpha^{\prime\prime}}
\frac{d \left( 
\bar{V}^{R}_{\boldsymbol{q}\beta^\prime\alpha^\prime}U^{L\dagger}_{\boldsymbol{q}\alpha^\prime\alpha^{\prime\prime}}
\right)}{dt}
\left(
U^{R}_{\boldsymbol{q}\alpha^{\prime\prime}\alpha}V^{R}_{\boldsymbol{q}\alpha\beta}
\right)
\nonumber\\
&=&
\sum_{\alpha\alpha^\prime}
\bar{V}^{R}_{\boldsymbol{q}\beta^\prime\alpha}W_{\boldsymbol{q}\alpha\alpha^\prime}
V^{R}_{\boldsymbol{q}\alpha^\prime\beta}
\nonumber\\
&&
+
\sum_{\alpha}
\frac{d
\bar{V}^{R}_{\boldsymbol{q}\beta^\prime\alpha}
}{dt}
V^{R}_{\boldsymbol{q}\alpha\beta},
\label{I}
\end{eqnarray}
where
$I_{\boldsymbol{q}}\not = -I^\dagger_{\boldsymbol{q}}$
because of
$\bar{V}^{R}_{\boldsymbol{q}}\not=V^{R\dagger}_{\boldsymbol{q}}$
and 
$U^{L\dagger}_{\boldsymbol{q}}\not=U^{R\dagger}_{\boldsymbol{q}}$, and 
further,
a phenomenological damping factor 
$\gamma^{(0)}_{\boldsymbol{q}}$
pertinent to $F^\dagger_{\boldsymbol{q}}$
is given by
\begin{equation}
{\gamma^{(0)}_{\boldsymbol{q}\beta^\prime\beta}\over 2}
=\sum_{\alpha\alpha^\prime}
\bar{V}^{R}_{\boldsymbol{q}\beta^\prime\alpha^\prime}
{\gamma^{(B)}_{\boldsymbol{q}\alpha^\prime\alpha}\over 2}
V^{R}_{\boldsymbol{q}\alpha\beta}.
\label{gamma0}
\end{equation}
In comparison of  Eq.~({\ref{HeisenbergF}}) with Eq.~({\ref{Heisenberg3}) in the  light of the discussion below 
Eq.~({\ref{B+2}}), it is readily seen that
$E_{\boldsymbol{q}\beta}(t)$ is a non-interacting adiabatic energy at time $t$
associated with the operator $F^\dagger_{\boldsymbol{q}\beta}(t)$ thus introduced.
Hereafter, this operator is termed as a creation operator of  PQ, and
then the corresponding annihilation operator corresponds to $F_{\boldsymbol{q}\beta}(t)$;
it is remarked that these are not bosonic operators.

The coupling constant $\mathcal{M}^{\prime\prime}_{-\boldsymbol{q}\beta\beta^\prime}$
is attributed to $M^{\prime\prime}_{-\boldsymbol{q}\alpha}$ relevant to
non-vanishing commutator between different quasi-boson operators [see Eq.~(\ref{M"})].
This couples the PQ of state $\beta^\prime$ with that  of $\beta$ accompanying
momentum transfer from $-\boldsymbol{q}$ to $\boldsymbol{q}$.
Hereafter, such an effect of momentum exchange is considered as negligibly small.
Further, the approximation of
$\gamma^{(0)}_{\boldsymbol{q}\beta^\prime\beta}
\approx \gamma^{(0)}_{\boldsymbol{q}\beta}\delta_{\beta^\prime\beta}$
is made with
\begin{equation}
\gamma^{(0)}_{\boldsymbol{q}\beta}(t)\equiv {\rm Re}[\gamma^{(0)}_{\boldsymbol{q}\beta\beta}(t)] \theta(t+\tau_L/2) \ge 0,
\label{gamma00}
\end{equation}
just for the sake of simplicity.
Under these approximations, Eq.~(\ref{HeisenbergF}) is read as
\begin{eqnarray}
-i{d\over dt}F^\dagger_{\boldsymbol{q}\beta}
&\approx&
F^\dagger_{\boldsymbol{q}\beta}
\left(E_{\boldsymbol{q}\beta}
+i{\gamma^{(0)}_{\boldsymbol{q}\beta}\over 2}
\right)
+i\sum_{\beta^\prime}F^\dagger_{\boldsymbol{q}\beta^\prime}I_{\boldsymbol{q}\beta^\prime\beta},
\nonumber\\
\label{HeisenbergF2}
\end{eqnarray}
where  the indexes $\beta$ and $\beta^\prime$ are  equivalent to those of  $\bar{\alpha}$ and $\bar{\alpha}^\prime$, respectively, and
$E_{\boldsymbol{q}\beta}=\mathcal{E}_{\boldsymbol{q}\bar{\alpha}}$.
It is noted that
the major approximation made  thus far is the factorization approximation implemented in Eqs.~(\ref{comm}) and (\ref{comm4}),  in addition to the rotational-wave approximation implemented in Eq.~(\ref{Heisenberg2}) though it seems to matter less than the former approximation;
aside from other unimportant assumptions such as the employment of the two-band model and
the above-mentioned approximation of Eq.~(\ref{gamma00}).

\subsubsection{A Retarded Green Function}
\label{sec2B2}

One solves Eq.~(\ref{HeisenbergF2})  in an approximate manner to obtain the closed analytic forms of $F^\dagger_{\boldsymbol{q}\beta}$ and $F_{\boldsymbol{q}\beta}$; the detail of the derivation is given in Ref.~\onlinecite{appC}.
The point of this derivation is to approximate the non-adiabatic interaction $I_{\boldsymbol{q}}(t)$ under the following assumption.

As seen in the second equality of Eq.~(\ref{I}), $I_{\boldsymbol{q}}(t)$
is affected by the two contributions, namely,  $W_{\boldsymbol{q}}(t)$ and
a time-derivative of $\bar{V}^R_{\boldsymbol{q}}(t)$.
There are two pronounced effects on $W_{\boldsymbol{q}}(t)$.
One is an effect of a crossing between adiabatic states of $\alpha^\prime$ and $\alpha$.
Here, the adiabatic energy curves of $\mathcal{E}_{\boldsymbol{q}\alpha^\prime}(t)$ and $\mathcal{E}_{\boldsymbol{q}\alpha}(t)$ tend to cross at $t=t_j$ 
--- termed as exceptional point\cite{crossing} ---,
 resulting in spike-like change of $W_{\boldsymbol{q}\alpha^\prime\alpha}(t)$;
it is seen that this effect arises from the energy denominator  in Eq.~(\ref{W2}) reminiscent of a Landau-Zener coupling.\cite{nikitin}
The other is an effect of abrupt change of $\Omega^{(R)}_{bb^\prime\boldsymbol{k}}(t)$ of Eq.~(\ref{Renergy}),
 namely, $d\Omega^{(R)}_{bb^\prime\boldsymbol{k}}(t)/dt$,
 at $t=\pm\tau_L/2$
because of the impulsive temporal-shape of $F_0(t)$ of Eq.~(\ref{F0}).
These two effects are also incorporated in the time-derivative of $\bar{V}^R_{\boldsymbol{q}}(t)$\textbf{\emph{\textbf{}}} in Eq.~(\ref{I}).
Accordingly, $I_{\boldsymbol{q}}(t)$ would be well described by 
\begin{equation}
I_{\boldsymbol{q}}(t) 
\approx \sum_j \mathfrak{I}_{\boldsymbol{q}}^{(j)\dagger}\delta(t-t_j), 
\label{I4-1}
\end{equation}
where it is understood that a set of 
$t_j$'s also includes $\pm \tau_L/2$, and $\mathfrak{I}_{\boldsymbol{q}}^{(j)}$ is a constant matrix
with
$
\mathfrak{I}^{{(j)}\dagger}_{\boldsymbol{q}}\not=-\mathfrak{I}^{(j)}_{\boldsymbol{q}}.
$
From among a set of the off-diagonal elements $\{\mathfrak{I}^{(j)}_{\boldsymbol{q}\beta^\prime(\not=\beta)\beta}\}$, 
just the single leading contribution, namely, $\mathfrak{I}^{(D)}_{\boldsymbol{q}\beta^\prime(\not=\beta)\beta}$ at $t=t_D$, is retained.
In practical calculations, $t_D$ is set equal to $\tau_L/2$, that is,
\begin{equation}
t_D={\tau_L \over 2}.
\label{tD}
\end{equation}

In terms of $F_{\boldsymbol{q}}$ and $F^\dagger_{\boldsymbol{q}}$ thus obtained,\cite{appC}
the associated  retarded Green function defined as\cite{schafer} 
\begin{equation}
G^R_{\boldsymbol{q}\beta\beta^\prime}(t,t^\prime)
=-i\theta(t-t^\prime)\Bigl\langle\left[
F_{\boldsymbol{q}\beta}(t), F^\dagger_{\boldsymbol{q}\beta^\prime}(t^\prime)
\right]\Bigr\rangle
\label{RGfct}
\end{equation}
ends up with
\begin{eqnarray}
&&G^R_{\boldsymbol{q}\beta\beta^\prime}(t,t^\prime)
\nonumber\\
&&
=
-i\theta(t-t^\prime)e^{-i\Theta_{\boldsymbol{q}\beta}(t,t_D)}
\sum_{\gamma\gamma^\prime}
V^{R\dagger}_{\boldsymbol{q}\beta\gamma}(t_D)
\nonumber\\
&&
\times
T_{\boldsymbol{q}\gamma\gamma^\prime}(t,t^\prime)
V^R_{\boldsymbol{q}\gamma^\prime\beta^\prime}(t_D)
e^{i\Theta^*_{\boldsymbol{q}\beta^\prime}(t^\prime,t_D)},
\label{RGfct2}
\end{eqnarray}
where 
$\Theta_{\boldsymbol{q}\beta}(t,\tilde{t})$ is an adiabatic energy phase given by
\begin{equation}
\Theta_{\boldsymbol{q}\beta}(t,\tilde{t})
= \int^t_{\tilde{t}} dt^\prime 
E^*_{\boldsymbol{q}\beta}(t^\prime)
-i\mathfrak{Z}_{\boldsymbol{q}\beta}(t,\tilde{t}),
\label{adphase0}
\end{equation}
and
both effects of the phenomenological damping and the non-adiabatic correction attributed to diagonal components $\{\mathfrak{I}^{(j)}_{\boldsymbol{q}\beta\beta}\}$ are incorporated in 
\begin{equation}
\mathfrak{Z}_{\boldsymbol{q}\beta}(t,\tilde{t})
=
\int^t_{\tilde{t}} dt^\prime
\left[
{\gamma^{(0)}_{\boldsymbol{q}\beta}(t^\prime)\over 2}
+I^*_{\boldsymbol{q}\beta\beta}(t^\prime)
\right].
\label{Upsilon}
\end{equation}
Further, the matrix $T_{\boldsymbol{q}}$ is introduced in Eq.~(\ref{RGfct2}) as
\begin{equation}
T_{\boldsymbol{q}}(t,t^\prime)
=\mathcal{T}^\dagger_{\boldsymbol{q}}(t)\mathcal{T}_{\boldsymbol{q}}(t^\prime),
\label{T1}
\end{equation}
where all effects of the off-diagonal components $\{\mathfrak{I}^{(D)}_{\boldsymbol{q}\beta^\prime(\not=\beta)\beta}\}$ are
incorporated  to
$\mathcal{T}_{\boldsymbol{q}}(t)$; for more detail, see Ref.~\onlinecite{appC}.
It is supposed that the moduli of these components are much smaller than unity, 
that is, $|\mathfrak{I}^{(D)}_{\boldsymbol{q}\beta^\prime(\not=\beta)\beta}| \ll 1$.
Then, it is shown that  the leading terms of $\mathcal{T}_{\boldsymbol{q}}(t)$ and $T_{\boldsymbol{q}}(t,t^\prime)$ become constant diagonal-matrices given by
\begin{eqnarray}
\mathcal{T}_{\boldsymbol{q}\beta^\prime\gamma}(t)
\approx \delta_{\beta^\prime\gamma}
e^{-\mathfrak{Z}^*_{\boldsymbol{q}\gamma}(t_D,t_0)},
\label{mathcalTapp2}
\end{eqnarray}
and
\begin{eqnarray}
T_{\boldsymbol{q}\gamma^\prime\gamma}(t,t^\prime)
\approx \delta_{\gamma^\prime\gamma}
e^{-2{\rm Re}\/\mathfrak{Z}_{\boldsymbol{q}\gamma}(t_D,t_0)},
\label{T1app}
\end{eqnarray}
respectively,
where $t_0$ is initial time when an initial condition is imposed before laser irradiation:
$t_0<-\tau_L/2$.
It should be noted that $T_{\boldsymbol{q}}(t,t^\prime)$ is a {\it real} diagonal matrix.

Similarly to Eq.~(\ref{RGfct2}), the expectation values of equal-time commutation relations for the PQ operators are evaluated as
\begin{eqnarray}
\langle [F_{\boldsymbol{q}\beta}(t), F^{\dagger}_{\boldsymbol{q}^\prime\beta^\prime}(t)] \rangle
&=&e^{-i\Theta_{\boldsymbol{q}\beta}(t,t_D)}
\sum_{\gamma\gamma^\prime}
V^{R\dagger}_{\boldsymbol{q}\beta\gamma}(t_D)
T_{\boldsymbol{q}\gamma\gamma^\prime}(t,t)
\nonumber\\
&&
\times
V^R_{\boldsymbol{q}\gamma^\prime\beta^\prime}(t_D)
e^{i\Theta^*_{\boldsymbol{q}\beta^\prime}(t,t_D)},
\label{commF-1}
\end{eqnarray}
and
\begin{equation}
\langle [F_{\boldsymbol{q}\beta}(t), F_{\boldsymbol{q}^\prime\beta^\prime}(t)] \rangle
=
\langle [F^{\dagger}_{\boldsymbol{q}\beta}(t), F^{\dagger}_{\boldsymbol{q}^\prime\beta^\prime}(t)] \rangle
=0.
\label{commF-0}
\end{equation}

\subsection{Analytic Expression of Physical Quantities}
\label{sec2C}

\subsubsection{Longitudinal Susceptibility}
\label{sec2C1}

An effective potential of electrons in the present system consists of two contributions:
one arises from a dynamically screened Coulomb interaction, and the other arises from
an LO-phonon-induced interaction.
Thus, the associated total retarded longitudinal susceptibility, $\chi^{(t)}_{\boldsymbol{q}}(t,t^\prime)$, is represented as a sum of 
a retarded susceptibility  due to the electron-induced interaction, $\chi_{\boldsymbol{q}}(t,t^\prime)$,
and that of an LO-phonon-induced  interaction, $\chi^\prime_{\boldsymbol{q}}(t,t^\prime)$, namely,\cite{schafer}
\begin{equation}
\chi^{(t)}_{\boldsymbol{q}}(t,t^\prime)=\chi_{\boldsymbol{q}}(t,t^\prime)+\chi^\prime_{\boldsymbol{q}}(t,t^\prime).
\label{chitot}
\end{equation}
Both of 
$\chi_{\boldsymbol{q}}(t,t^\prime)$ and $\chi^\prime_{\boldsymbol{q}}(t,t^\prime)$
are defined below
in Eq.~(\ref{chi}) and Eq.~(\ref{chiprime}),
respectively.

One begins with a retarded density-density correlation function of electrons
defined as\cite{schafer,fetter}}
\begin{equation}
iD^R_{\boldsymbol{q}}(t,t^\prime)
=
\theta(t-t^\prime)
\langle
\left[\hat{n}_{\boldsymbol{q}}(t),\hat{n}_{-\boldsymbol{q}}(t^\prime) \right]
\rangle,
\label{ddcfQ2}
\end{equation}
where
\begin{equation}
\hat{n}_{\boldsymbol{q}}(t)
={\hat{\rho}_{\boldsymbol{q}}(t)\over V}
={1\over V}
\sum_{b\boldsymbol{k}}\:a^\dagger_{b\boldsymbol{k}+\boldsymbol{q}}a_{b\boldsymbol{k}},
\label{nq}
\end{equation}
in terms of Eq.~(\ref{rho}).
It is remarked that
\begin{equation}
\hat{n}_{-\boldsymbol{q}}(t)=\hat{n}^\dagger_{\boldsymbol{q}}(t).
\end{equation}
and hence,
\begin{equation}
D^{R\dagger}_{\boldsymbol{q}}(t,t^\prime)=
D^{R}_{-\boldsymbol{q}}(t,t^\prime).
\label{ddcfQ2-1}
\end{equation}
The retarded susceptibility of $\chi_{\boldsymbol{q}}(t,t^\prime)$ is relevant  to 
$D^R_{\boldsymbol{q}}(t,t^\prime)$, namely,
\begin{equation}
\chi_{\boldsymbol{q}}(t,t^\prime)\equiv
4\pi VD^R_{\boldsymbol{q}}(t,t^\prime)
=\chi^*_{-\boldsymbol{q}}(t,t^\prime).
\label{chi}
\end{equation}

By employing the relations of  Eq.~(\ref{B+}),
Eq.~(\ref{nq}) is given by
\begin{equation}
\hat{n}_{\boldsymbol{q}}(t)
={1\over V}
\sum_{\alpha}\:B^\dagger_{\boldsymbol{q}\alpha}N^L_{\boldsymbol{q}\alpha}.
\label{n1}
\end{equation}
According to this expression, it is readily seen that $B^\dagger_{\boldsymbol{q}\alpha}$ represents a fraction of electron density at the state of $\alpha$ weighted with $N^L_{\boldsymbol{q}\alpha}/V$.
Following  Eq.~(\ref{F+2}),
this operator is given by
\begin{equation}
B^\dagger_{\boldsymbol{q}\alpha}
=\sum_{\beta}\:
F^\dagger_{\boldsymbol{q}\beta}\bar{V}^R_{\boldsymbol{q}\beta\alpha},
\label{B2F}
\end{equation}
and thus,
Eq.~(\ref{nq}) is recast into
\begin{equation}
\hat{n}_{\boldsymbol{q}}(t)
={1\over V}
\sum_{\beta\alpha}\:
F^\dagger_{\boldsymbol{q}\beta}\bar{V}^R_{\boldsymbol{q}\beta\alpha}
N^L_{\boldsymbol{q}\alpha}.
\label{n2}
\end{equation}
Then, Eq.~(\ref{ddcfQ2}) becomes of the form:
\begin{eqnarray}
&&iD^R_{-\boldsymbol{q}}(t,t^\prime)
\nonumber\\
&&=
{i\over V^2}
\theta(t-t^\prime)
\sum_{\alpha\alpha^\prime\beta\beta^\prime}
N^{L*}_{\boldsymbol{q}\alpha}(t)
\bar{V}^{R\dagger}_{\boldsymbol{q}\alpha\beta}(t)
G^R_{\boldsymbol{q}\beta\beta^\prime}(t,t^\prime)
\nonumber\\
&&
\times
\bar{V}^R_{\boldsymbol{q}\beta^\prime\alpha^\prime}(t^\prime)
N^L_{\boldsymbol{q}\alpha^\prime}(t^\prime).
\label{ddcfQ3}
\end{eqnarray}

Applying Eq.~(\ref{RGfct2})  to Eq.~(\ref{ddcfQ3}),
Eq.~(\ref{chi} ) ends up with
\begin{eqnarray}
&&i\chi_{-\boldsymbol{q}}(t^\prime+\tau,t^\prime)=i\chi^*_{\boldsymbol{q}}(t^\prime+\tau,t^\prime)
\nonumber\\
&&=
{4\pi\over V}
\theta(\tau)
\left[
\sum_{\bar{\alpha}} e^{-i\mathcal{E}_{\boldsymbol{q}\bar{\alpha}}(t_D)\tau}
\Xi^{(c)}_{\boldsymbol{q}\bar{\alpha}}(t^\prime+\tau,t^\prime)
\right.
\nonumber\\
&&+
\left.
\sum_{p=1,2} e^{-i[\mathcal{E}^{(r)}_{\boldsymbol{q}\alpha_p}(t_D)-i\Gamma_{\boldsymbol{q}\alpha_p}(t_D)/2]\tau}
\Xi^{(r)}_{\boldsymbol{q}\alpha_p}(t^\prime+\tau,t^\prime)
\right],
\nonumber\\
\label{chi-I}
\end{eqnarray}
with $\tau \equiv t-t^\prime \ge 0$.
In Ref.~\onlinecite{appC}, the Fano problem of Eq.~(\ref{eigenh}) is approximately solved for obtaining 
$\{\bar{V}^R_{\boldsymbol{q}\beta\alpha}\}$.
Moreover, 
in Ref.~\onlinecite{SMchi}, the detail of  derivation of Eq.~(\ref{chi-I})  is made, and also
the explicit expressions of 
$\mathcal{E}^{(r)}_{\boldsymbol{q}\alpha_p}(t)$,  $\Gamma_{\boldsymbol{q}\alpha_p}(t)$, 
$\Xi^{(c)}_{\boldsymbol{q}\bar{\alpha}}(t^\prime+\tau,t^\prime)$,
and
$\Xi^{(r)}_{\boldsymbol{q}\alpha_p}(t^\prime+\tau,t^\prime)$
are given, where these expressions are used in actual calculations.
Here, just the leading contributions of these quantities are shown in order to avoid unimportant complexity as follows:
\begin{eqnarray}
&&\mathcal{E}^{(r)}_{\boldsymbol{q}\alpha_p}(t)
\nonumber\\
&&\approx
{1\over 2}\left\{
\mathcal{E}_{\boldsymbol{q}\alpha_1}(t)+\omega^{(LO)}_{\boldsymbol{q}}
+(-1)^{p}
\right.
\nonumber\\
&&
\left.
\times
\left[
(\mathcal{E}_{\boldsymbol{q}\alpha_1}(t)-\omega^{(LO)}_{\boldsymbol{q}})^2
+|M_{\boldsymbol{q}\alpha_1}(t)|^2
\right]^{1/2}
\right\},
\end{eqnarray}
where
$\mathcal{E}_{\boldsymbol{q}\alpha_1}(t) < \omega^{(LO)}_{\boldsymbol{q}}$, 
and
\begin{equation}
\Gamma_{\boldsymbol{q}\alpha_p}(t)
\approx
\delta_{p2}
2\pi\rho_{\boldsymbol{q}\bar{\alpha}_p}(t)
|M_{\boldsymbol{q}\bar{\alpha}_p}(t)|^2,
\label{Gamma}
\end{equation}
where a new index $\bar{\alpha}_p$ is introduced, implying
the index of continuum $\bar{\alpha}$ on the occasion that $\mathcal{E}^{(r)}_{\boldsymbol{q}\alpha_p}(t_D)$ resonantly coincides with
$\mathcal{E}_{\boldsymbol{q}\bar{\alpha}}(t_D)$,
and
$\rho_{\boldsymbol{q}\bar{\alpha}_p}(t)$ represents 
a density of state of $\bar{\alpha}_p$ at time $t$.
Further,
\begin{eqnarray}
&&\Xi^{(c)}_{\boldsymbol{q}\bar{\alpha}}(t^\prime+\tau,t^\prime)
\nonumber\\
&&=
N^{L*}_{\boldsymbol{q}\bar{\alpha}}(t^\prime+\tau)
T_{\boldsymbol{q}\bar{\alpha}\bar{\alpha}}(t^\prime+\tau,t^\prime)
N^L_{\boldsymbol{q}\bar{\alpha}}(t^\prime)
\nonumber
\\
&&
\times
e^{-\mathfrak{Z}_{\boldsymbol{q}\bar{\alpha}}(t^\prime+\tau,t^\prime)}
e^{-2{\rm Re}\mathfrak{Z}_{\boldsymbol{q}\bar{\alpha}}(t^\prime,t_D)},
\label{Xc}
\end{eqnarray}
and
\begin{eqnarray}
&&\Xi^{(r)}_{\boldsymbol{q}\alpha_p}(t^\prime+\tau,t^\prime)
\nonumber\\
&&
\approx 
\delta_{p1}
N_{\boldsymbol{q}\alpha_p}^{L*}(t^\prime+\tau)
\mathcal{D}_{\boldsymbol{q}\alpha_p}(t^\prime+\tau,t_D)
T_{\boldsymbol{q}\alpha_p\alpha_p}(t^\prime+\tau,t^\prime)
\nonumber\\
&&
\times
\mathcal{D}^*_{\boldsymbol{q}\alpha_p}(t^\prime,t_D)
N^{L}_{\boldsymbol{q}\alpha_p}(t^\prime)
e^{-\mathfrak{Z}_{\boldsymbol{q}\bar{\alpha}_p}(t^\prime+\tau,t^\prime)}
\nonumber\\
&&
\times
e^{-2{\rm Re}\mathfrak{Z}_{\boldsymbol{q}\bar{\alpha}_p}(t^\prime,t_D)}
\exp{ \left\{ -\int^{t^\prime}_{t_D}
\Gamma_{\boldsymbol{q}\alpha_p}(t^{\prime\prime})
dt^{\prime\prime}\right\}},
\label{Xr}
\end{eqnarray}
where
$\mathcal{D}_{\boldsymbol{q}\alpha_p}(t,t_D)$ is given by
\begin{eqnarray}
&&
\mathcal{D}_{\boldsymbol{q}\alpha_p}(t,t_D)
\nonumber\\
&&\approx
\frac{
\pi\rho_{\boldsymbol{q}\bar{\alpha}_p}(t_D)
}
{
\bar{\Gamma}_{\boldsymbol{q}\alpha_p}(t,t_D)/2
}
\left\{
\left[
{\Delta\Gamma}_{\boldsymbol{q}\alpha_p}(t;t)
-iM_{\boldsymbol{q}\bar{\alpha}_p}(t)
\right]
\right.
\nonumber\\
&&
\times
\left.
\left[
{\Delta\Gamma}_{\boldsymbol{q}\alpha_p}(t;t_D)
+iM^*_{\boldsymbol{q}\bar{\alpha}_p}(t_D)
\right]
\right.
\nonumber
\\
&&
-
\left[
{\Delta\Gamma}_{\boldsymbol{q}\alpha_p}(t;t)
-\pi\rho_{\boldsymbol{q}\bar{\alpha}_p}(t_D)|M_{\boldsymbol{q}\bar{\alpha}_p}(t)|^2
\right]
\nonumber\\
&&
\times
\left[
{\Delta\Gamma}_{\boldsymbol{q}\alpha_p}(t;t_D)
-\pi\rho_{\boldsymbol{q}\bar{\alpha}_p}(t_D)|M_{\boldsymbol{q}\bar{\alpha}_p}(t_D)|^2
\right]
\nonumber\\
&&
+
\left.
[\pi\rho_{\boldsymbol{q}\bar{\alpha}_p}(t_D)]^2
|M_{\boldsymbol{q}\bar{\alpha}_p}(t)|^2
|M_{\boldsymbol{q}\bar{\alpha}_p}(t_D)|^2
\right\},
\label{calD}
\end{eqnarray}
where
\begin{eqnarray}
\Delta\Gamma_{\boldsymbol{q}\alpha_p}(t;t^\prime)
&\approx&
{1\over 2}\bar{\Gamma}_{\boldsymbol{q}\alpha_p}(t,t_D)
\nonumber\\
&&
-\pi\rho_{\boldsymbol{q}\bar{\alpha}_p}(t_D)
|M_{\boldsymbol{q}\bar{\alpha}_p}(t^\prime)|^2,
\label{Gamma-}
\end{eqnarray}
and
\begin{equation}
\bar{\Gamma}_{\boldsymbol{q}\alpha_p}(t,t_D)
={1\over 2}\left[
\Gamma_{\boldsymbol{q}\alpha_p}(t)
+\Gamma_{\boldsymbol{q}\alpha_p}(t_D)
\right].
\label{Gamma+}
\end{equation}
It is remarked that
$\Delta\Gamma_{\boldsymbol{q}\alpha_p}(t_D;t_D)=0$
at equal time, and thus,
$\mathcal{D}_{\boldsymbol{q}\alpha_p}(t_D,t_D)=1$.

On the other hand, the retarded susceptibility of $\chi^\prime_{\boldsymbol{q}}(t^\prime+\tau,t^\prime)$ introduced in Eq.~(\ref{chitot}) is expressed as
\begin{eqnarray}
\chi^\prime_{\boldsymbol{q}}(t^\prime+\tau,t^\prime)
&=&{4\pi\over V}
|g^\prime_{\boldsymbol{q}}|^2
D^{\prime R}_{\boldsymbol{q}}(t^\prime+\tau,t^\prime),
\label{chiprime}
\end{eqnarray}
where
\begin{equation}
|g^\prime_{\boldsymbol{q}}|^2=\left|{g^0_{\boldsymbol{q}}\over v^{(C)}_{\boldsymbol{q}}}\right|^2
\label{Me-ph}
\end{equation}
with 
$g^0_{\boldsymbol{q}}$ defined in Eq.~(\ref{g2}),
and
\begin{equation}
v^{(C)}_{\boldsymbol{q}}=\epsilon_\infty V^{(C)}_{\boldsymbol{q}}={4\pi\over V}{1\over {\boldsymbol{q}^2}}
\label{vC}
\end{equation}
with
$\epsilon_\infty$ as a background dielectric constant given in Eq.~(\ref{VC}).
Further, 
$D^{\prime R}_{\boldsymbol{q}}(t,t^\prime)$ is a retarded phonon  Green function given by\cite{schafer}
\begin{equation}
D^{\prime R}_{\boldsymbol{q}}(t,t^\prime)
\equiv
\bar{D}^{\prime R}_{\boldsymbol{q}}(t,t^\prime)
+\left[\bar{D}^{\prime R }_{-\boldsymbol{q}}(t,t^\prime)\right]^*,
\label{phononGfct}
\end{equation}
where
\begin{equation}
\bar{D}^{\prime R}_{\boldsymbol{q}}(t,t^\prime)
=-i\Bigl\langle
\left[
c_{\boldsymbol{q}}(t),c^\dagger_{\boldsymbol{q}}(t^\prime)
\right]
\Bigl\rangle
\theta(t-t^\prime).
\label{phononGfct2}
\end{equation}
It is remarked that in the small-$\boldsymbol{q}$ limit,  because of Eq.~(\ref{Me-ph}),
$\chi^\prime_{\boldsymbol{q}}(t^\prime+\tau,t^\prime) \propto |\boldsymbol{q}|^2$ for the Fr\"{o}hlich interaction, while
$\chi^\prime_{\boldsymbol{q}}(t^\prime+\tau,t^\prime) \propto |\boldsymbol{q}|^4$ for the deformation potential interaction.

Employing Eqs.~(\ref{F+}), (\ref{F-}), and (\ref{RGfct2}),
Eq.(\ref{phononGfct2})
is recast into 
\begin{eqnarray}
&&
i\bar{D}^{\prime R}_{\boldsymbol{q}}(t,t^\prime)
=
i\sum_{\beta\beta^\prime}
\bar{V}^{R\dagger}_{\boldsymbol{q}\alpha_2\beta}(t)
G^R_{\boldsymbol{q}\beta\beta^\prime}(t,t^\prime)
\bar{V}^R_{\boldsymbol{q}\beta^\prime\alpha_2}(t^\prime).
\nonumber\\
\label{phcomm2}
\end{eqnarray}
Thus,
$\chi^\prime_{\boldsymbol{q}}(t^\prime+\tau,t^\prime)$ of Eq.~(\ref{chiprime}) becomes
\begin{eqnarray}
&&i\chi^\prime_{\boldsymbol{q}}(t^\prime+\tau,t^\prime)
\nonumber\\
&&
=
{4\pi\over V}
\sum_{p=1,2} 
\left\{
e^{-i[\mathcal{E}^{(r)}_{\boldsymbol{q}\alpha_p}(t_D)-i\Gamma_{\boldsymbol{q}\alpha_p}(t_D)/2]\tau}
\Pi^{(r)}_{\boldsymbol{q}\alpha_p}(t^\prime+\tau,t^\prime)
\right.
\nonumber\\
&&-
\left.
e^{i[\mathcal{E}^{(r)}_{-\boldsymbol{q}\alpha_p}(t_D)+i\Gamma_{-\boldsymbol{q}\alpha_p}(t_D)/2]\tau}
\Pi^{(r)*}_{-\boldsymbol{q}\alpha_p}(t^\prime+\tau,t^\prime)
\right\}.
\label{chiprime2}
\end{eqnarray}
Similarly to Eq.~(\ref{chi-I}), in Ref.~\onlinecite{SMchi}, the detail of this derivation is made, and also
the explicit expression of 
$\Pi^{(r)}_{\boldsymbol{q}\alpha_p}(t^\prime+\tau,t^\prime)$
is given, where this expression is used in actual calculations.
Here, just the leading contribution of this quantity is shown in order to avoid unimportant complexity as follows:
\begin{eqnarray}
&&\Pi ^{(r)}_{\boldsymbol{q}\alpha_p}(t^\prime+\tau,t^\prime)
\nonumber\\
&&\approx
\delta_{p2}
\left|g^\prime_{\boldsymbol{q}}\right|^2
\mathcal{D}_{\boldsymbol{q}\alpha_p}(t^\prime+\tau,t_D)
T_{\boldsymbol{q}\alpha_p\alpha_p}(t^\prime+\tau,t^\prime)
\nonumber\\
&&
\times
\mathcal{D}^*_{\boldsymbol{q}\alpha_p}(t^\prime,t_D)
e^{-\mathfrak{Z}_{\boldsymbol{q}\bar{\alpha}_p}(t^\prime+\tau,t^\prime)}
e^{-2{\rm Re}\mathfrak{Z}_{\boldsymbol{q}\bar{\alpha}_p}(t^\prime,t_D)}
\nonumber\\
&&
\times
\exp{ \left\{ \int^{t^\prime+\tau}_{t_D}
\Gamma_{\boldsymbol{q}\alpha_p}(t^{\prime\prime})
dt^{\prime\prime}\right\}}.
\label{Pr}
\end{eqnarray}

\subsubsection{Transient Induced Photoemission Spectra}
\label{sec2C2}

The total retarded susceptibility
$\chi^{(t)}_{\boldsymbol{q}}(t,t^\prime)$   depends on  passage of $t$ as well as 
the relative time $\tau=t-t^\prime$ 
 in the present non-equilibrium system, differing from equilibrium systems  depending solely on $\tau$
due to temporal translational-invariance being conserved.
According to the linear response theory, an induced charge density $n^{(ind)}_{\boldsymbol{q}}(t)$ introduced by a weak external optical field 
$f(t)$ is provided by\cite{schafer,fetter}
\begin{equation}
n^{(ind)}_{\boldsymbol{q}}(t)=-{i\over 4\pi}\int^t_{t_0} dt^\prime\: \chi^{(t)}_{\boldsymbol{q}}(t,t^\prime)f(t^\prime).
\label{linear1}
\end{equation}
Temporal width of the probe pulse $f(t^\prime)$ is almost of the same order as that of the pump pulse given by Eq.~(\ref{F}) in a usual pump-probe experiments for ultrafast optical processes.
Thus, just for the sake of simplicity,  $f(t^\prime)$  is given by a delta-pulse as
$f(t^\prime)=f_0\delta(t^\prime-t_p)$, where
$f_0$ is a constant, and $t_p$ represents the time at which $f(t^\prime)$ probes transient dynamics of concern.
Then, Eq.~(\ref{linear1}) becomes
\begin{eqnarray}
n^{(ind)}_{\boldsymbol{q}}(t)&\approx&
-{i\over 4\pi}f_0 \chi^{(t)}_{\boldsymbol{q}}(t,t_p).
\label{linear2}
\end{eqnarray}
Defining the Fourier-transform of $\chi ^{(t)}_{\boldsymbol{q}}(t_p+\tau,t_p)$ as\cite{comment1}
\begin{equation}
\tilde{\chi}^{(t)} _{\boldsymbol{q}}(t_p;\omega)
=\int^\infty_0 d\tau\: e^{-i\omega\tau}\chi ^{(t)}_{\boldsymbol{q}}(t_p+\tau,t_p),
\label{tildechitot}
\end{equation}
a frequency distribution function of the induced electron density
at the time $t_p$ is represented as
\begin{eqnarray}
\tilde{n}^{(ind)}_{\boldsymbol{q}}(t_p;\omega)=
-{i\over 4\pi}f_0 \tilde{\chi}^{(t)}_{\boldsymbol{q}}(t_p;\omega),
\label{linear3}
\end{eqnarray}
where $\tilde{n}^{(ind)}_{\boldsymbol{q}}(t_p;\omega)$ is the Fourier-transform of 
$n^{(ind)}_{\boldsymbol{q}}(t_p+\tau)$ given in a similar fashion to 
Eq.~(\ref{tildechitot}).
Therefore, it is understood that $\tilde{\chi}^{(t)} _{\boldsymbol{q}}(t_p;\omega)$ is considered as an optical susceptibility in the $\omega$-domain at the time $t_p$.
There are two remarks regarding $\tilde{n}^{(ind)}_{\boldsymbol{q}}(t_p;\omega)$.
One is that $\tilde{n}^{(ind)}_{\boldsymbol{q}}(t_p;\omega)$ is non-linear in the pump field, however, linear in the probe field.
The other is that contrary to convention, in Eq.~(\ref{tildechitot}), the sign of $\omega$ is defined such that 
 transient photoemission spectra peak at positive $\omega$, while  transient photoabsorption spectra peak at negative $\omega$.

The inverse dielectric function $\epsilon^{-1}_{\boldsymbol{q}}(t_p+\tau,t_p)$ is given as
\begin{eqnarray}
&&\epsilon^{-1}_{\boldsymbol{q}}(t_p+\tau,t_p)
\nonumber\\
&&=\epsilon_\infty^{-1}\left[\delta(\tau)
+{V\over 4\pi}V^{(C)}_{\boldsymbol{q}}
\chi ^{(t)}_{\boldsymbol{q}}(t_p+\tau,t_p)\theta(\tau)\right].
\label{epsilon-1}
\end{eqnarray}
Employing the relation
\(\displaystyle
\int^{\infty}_{-\infty} dt^{\prime\prime}
\epsilon^{-1}_{\boldsymbol{q}}(t_p+\tau,t^{\prime\prime})
\epsilon^{}_{\boldsymbol{q}}(t^{\prime\prime},t_p)
=\delta(\tau),
\)
followed by the Fourier-transform of the associated dielectric function $\epsilon_{\boldsymbol{q}}(t_p+\tau,t_p)$:
\begin{equation}
\tilde{\epsilon}^{} _{\boldsymbol{q}}(t_p;\omega)
=\int^\infty_0 d\tau\: e^{-i\omega\tau}\epsilon ^{}_{\boldsymbol{q}}(t_p+\tau,t_p),
\label{tildeepsilontot}
\end{equation}
one obtains
\begin{eqnarray}
&&\tilde{\epsilon}_{\boldsymbol{q}}(t_p;\omega)
\nonumber\\
&&
\approx
\epsilon_\infty
\left[1-
\frac{
{V\over 4\pi}V^{(C)}_{\boldsymbol{q}}
\tilde{\chi}^{(t)} _{\boldsymbol{q}}(t_p;\omega)
}{
1+{V\over 4\pi}V^{(C)}_{\boldsymbol{q}}
\tilde{\chi}^{(t)} _{\boldsymbol{q}}(t_p;\omega)
}\right],
\label{epsilon-2}
\end{eqnarray}
where
$t_p\gg\tau$ is supposed with Eq.~(\ref{tildechitot}).
Therefore,  a transient absorption coefficient $\alpha_{\boldsymbol{q}}(t_p;\omega)$ at time $t_p$ is given by the following expression:
\begin{equation}
\alpha_{\boldsymbol{q}}(t_p;\omega)
=
{\omega\over n(t_p;\omega)\/ C}A_{\boldsymbol{q}}(t_p;\omega),
\label{alpha}
\end{equation}
where
\begin{equation}
A_{\boldsymbol{q}}(t_p;\omega)
={\rm Im}\/\tilde{\epsilon}^{}_{\boldsymbol{q}}(t_p;\omega).
\label{Aomega0}
\end{equation}
Here, $n(t_p;\omega)$ represents the index of refraction, which is approximately given by
$n(t_p;\omega)\approx\sqrt{\epsilon_\infty}$, and $C$ represents the speed of light.
In the occasion that $A_{\boldsymbol{q}}(t_p;\omega)$ is negative in a certain region of $\omega$, this implies transient induced photoemission spectra, namely, transient negative absorption spectra, at time $t_p$.
The transient photoemission spectra are defined as $\bar{A}_{\boldsymbol{q}}(t_p;\omega) =-A_{\boldsymbol{q}}(t_p;\omega)$.

It seems worth deriving an analytic expression of  $\bar{A}_{\boldsymbol{q}}(t;\omega)$ by introducing some approximations in order to deepen an insight into underlying physics of CP generation.
In particular, behavior of $\bar{A}_{\boldsymbol{q}}(t;\omega)$ in the proximity of phonon energy  is concerned.
In Ref.~\onlinecite{SMappF},  it is shown that Eq.~(\ref{Aomega0}) is reduced to Shore's spectral profile, equivalent to well-known Fano's formula.\cite{shore}
This is given by
\begin{eqnarray}
&&\bar{A}_{\boldsymbol{q}}(t_p;\omega)
=
-A_{\boldsymbol{q}}(t_p;\omega)
\nonumber\\
&&\approx
\mathcal{C}_{\boldsymbol{q}}(t_p)
\nonumber\\
&&
+
\frac
{
\mathcal{A}_{\boldsymbol{q}\alpha_2}(t_p)\left[\omega-\mathcal{E}^{(r)}_{\boldsymbol{q}\alpha_2}(t_p)\right]
+\mathcal{B}_{\boldsymbol{q}\alpha_2}(t_p)\Gamma_{\boldsymbol{q}\alpha_2}(t_p)/2
}
{
\left[\omega-\mathcal{E}^{(r)}_{\boldsymbol{q}\alpha_2}(t_p)\right]^2+\left[\Gamma_{\boldsymbol{q}\alpha_2}(t_p)/2\right]^2
}
\nonumber\\
\label{shore0}
\end{eqnarray}
in the vicinity of $\omega =\mathcal{E}^{(r)}_{\boldsymbol{q}\alpha_2}(t_p)$, where
the three parameters of  $\mathcal{A}_{\boldsymbol{q}\alpha_2}(t_p)$, $\mathcal{B}_{\boldsymbol{q}\alpha_2}(t_p)$, and $\mathcal{C}_{\boldsymbol{q}}(t_p)$ are considered as 
Shore's spectral parameters.
$\bar{A}_{\boldsymbol{q}}(t_p;\omega)$
shows 
excited electronic structure formed by a non-linear optical process due to  pump field at time $t_p$.
Here, an asymmetric spectral profile is exclusively determined by $\mathcal{A}_{\boldsymbol{q}\alpha_2}(t_p)$ in the second term, and this is superimposed with continuum background governed by $\mathcal{C}_{\boldsymbol{q}}(t_p)$ in the first term.
$\bar{A}_{\boldsymbol{q}}(t_p;\omega)$ seems to be a crucial observable to understand manifestation of  transient and non-linear FR that would be
accompanied by CP generation.

The associated
Fano's $q$-parameter is determined in terms of Shore's parameters as
\begin{equation}
q_{\boldsymbol{q}\alpha_2}(t_p)=r_{\boldsymbol{q}\alpha_2}(t_p) + \sigma_{\boldsymbol{q}\alpha_2}(t_p)
\sqrt{[r_{\boldsymbol{q}\alpha_2}(t_p)]^2+1},
\label{q0}
\end{equation}
where
\begin{equation}
r_{\boldsymbol{q}\alpha_2}(t_p)
={\mathcal{B}_{\boldsymbol{q}\alpha_2}(t_p) \over
\mathcal{A}_{\boldsymbol{q}\alpha_2}(t_p)},\;\;\;
\sigma_{\boldsymbol{q}\alpha_2}(t_p)
={\mathcal{A}_{\boldsymbol{q}\alpha_2}(t_p) \over
|\mathcal{A}_{\boldsymbol{q}\alpha_2}(t_p)|}.
\label{rsigma0}
\end{equation}
The origin of the asymmetry can be traced back to the explicit form of 
$\mathcal{A}_{\boldsymbol{q}\alpha_2}(t_p)$ determined by
\begin{equation}
\mathcal{A}_{\boldsymbol{q}\alpha_2}(t_p)
={2{\rm Im}\/\Psi^{(r)\prime}_{\boldsymbol{q}\bar{\alpha}_2}(t_p,t_p)\over
\Gamma_{\boldsymbol{q}\alpha_p}(t_p)/2}v^{(C)}_{\boldsymbol{q}},
\label{RemathcalA0}
\end{equation}
where
\begin{equation}
\Psi^{(r)\prime}_{\boldsymbol{q}\bar{\alpha}_2}(t_p,t_p)
=\left.{d \Psi^{(r)}_{\boldsymbol{q}\bar{\alpha}_2}(t_p+\tau,t_p) \over d\tau}\right|_{\tau=0}.
\label{DPsi}
\end{equation}
Here,
$\Psi^{(r)}_{\boldsymbol{q}\bar{\alpha}_2}(t_p+\tau,t_p)$
is given by
\begin{equation}
\Psi^{(r)}_{\boldsymbol{q}\bar{\alpha}_2}(t_p+\tau,t_p)
=\Xi^{(r)*}_{\boldsymbol{q}\bar{\alpha}_2}(t_p+\tau,t_p)
+\Pi^{(r)*}_{\boldsymbol{q}\bar{\alpha}_2}(t_p+\tau,t_p).
\label{Psir0}
\end{equation}
In Sec.~\ref{sec3B}, the condition that $\mathcal{A}_{\boldsymbol{q}\alpha_2}(t_p)$ vanishes is discussed
in comparison with the results obtained from Eq.~(\ref{Aomega0}) in a numerical manner.

\subsubsection{Power Spectra  of Phonon Displacement Function}
\label{sec2C3}

An expectation value of 
$[c_{\boldsymbol{q}}^\dagger(t)+c_{\boldsymbol{q}}(t)]/2$ with respect to the ground state, namely, 
\begin{equation}
Q_{\boldsymbol{q}}(t)={1\over 2}\bigl\langle
 c_{\boldsymbol{q}}(t)+ c^\dagger_{-\boldsymbol{q}}(t)
\bigr\rangle.
\label{Q}
\end{equation}
is examined.
This is regarded as a classical phonon displacement function, if this expectation value is taken with respect to a coherent state of phonon.
In addition, this is also considered as transition probability of Raman scattering process.\cite{lee}
As a consequence of the Wiener-Khintchine's theorem,\cite{meystre} the Fourier-transform of the two-time correlation
\begin{equation}
C_{\boldsymbol{q}}(t)
=\int^\infty_{-\infty} Q_{\boldsymbol{q}}(t+t^\prime) Q_{\boldsymbol{q}}(t^\prime) dt^\prime
\label{correlation}
\end{equation}
provides the spectral function
\begin{equation}
S_{\boldsymbol{q}}(\omega) 
=\int^\infty_{-\infty} e^{-i\omega t} C_{\boldsymbol{q}}(t) dt.
\label{spectraS}
\end{equation}
Employing the Fourier-transform of  $Q_{\boldsymbol{q}}(t)$:\cite{comment2}
\begin{equation}
\tilde{Q}_{\boldsymbol{q}}(\omega)
=\int^\infty_0 e^{-i\omega t}  Q_{\boldsymbol{q}}(t) dt,
\label{tildeQ}
\end{equation}
it is readily seen that
$S_{\boldsymbol{q}}(\omega)$ is proportional to the power spectra, that is,
\begin{equation}
S_{\boldsymbol{q}}(\omega) \propto \left|\tilde{Q}_{\boldsymbol{q}}(\omega)\right|^2.
\label{spectraS2}
\end{equation}

The phonon operator $c_{\boldsymbol{q}}$ is extracted from  a PQ operator by means of the projection similar to Eq.~(\ref{B2F}), that is,
\begin{equation}
c_{\boldsymbol{q}}(t)
=\sum_{\beta}\:
\bar{V}^{R\dagger}_{\boldsymbol{q}\alpha_2\beta}(t)
F_{\boldsymbol{q}\beta}(t).
\label{c2F}
\end{equation}
Therefore, in view of Eq.~(\ref{Q}), $Q_{\boldsymbol{q}}(t)$ becomes
\begin{eqnarray}
Q_{\boldsymbol{q}}(t)
&=&
X_{\boldsymbol{q}}(t,t_D)
\exp{\left[-\int^t_{t_D}  dt^\prime\: {\Gamma_{\boldsymbol{q}\alpha_2}(t^\prime)\over 2}
\right]}
\nonumber\\
&&\times
\sin{ \left[ \omega_{\boldsymbol{q}}^{(LO)}t
+\theta_{\boldsymbol{q}}(t)
\right]}
+\Delta Q_{\boldsymbol{q}}(t).
\label{Q2}
\end{eqnarray}
In Ref.~\onlinecite{SMQ}, the detail of this derivation is made, and also
the explicit expressions of amplitude
$X_{\boldsymbol{q}}(t,t_D)$,  plasmonic contribution  $\Delta Q_{\boldsymbol{q}}(t)$ to the displacement function, and built-in phase $\theta_{\boldsymbol{q}}(t)$
are given, where these expressions are used in actual calculations.
Here, just the leading contributions of these quantities are shown in order to avoid unimportant complexity as follows:
\(
X_{\boldsymbol{q}}(t,t_D)=
\left|P_{\boldsymbol{q}\alpha_2}(t,t_D)\right|,
\Delta Q_{\boldsymbol{q}}(t) \approx 0
\), and
\begin{equation}
\theta_{\boldsymbol{q}}(t)={\pi\over 2}+
\Delta\alpha^\prime_{\boldsymbol{q}}(t)-\xi_{\boldsymbol{q}}(t)
+\upsilon_{\boldsymbol{q}\bar{\alpha}_2}(t,t_D)
+\theta^0_{\boldsymbol{q}}(t).
\label{thetat}
\end{equation}
Here, $P_{\boldsymbol{q}\alpha_2}(t,t_D)$ is provided by
\begin{eqnarray}
P_{\boldsymbol{q}\alpha_2}(t,t_D)
&\approx&
\mathcal{D}_{\boldsymbol{q}\alpha_2}(t,t_D)
\mathcal{O}^*_{\boldsymbol{q}\alpha_2}(t,t_D)
\nonumber\\
&&
\times
e^{-\mathfrak{Z}_{\boldsymbol{q}\bar{\alpha}_2}(t,t_D)},
\label{P}
\end{eqnarray}
where $\mathcal{O}^*_{\boldsymbol{q}\alpha_2}(t,t_D)$ is given by 
\begin{eqnarray}
\mathcal{O}^*_{\boldsymbol{q}\alpha_2}(t,t_D)
&\approx&\left[\mathcal{T}^\dagger_{\boldsymbol{q}}(t)\langle F_{\boldsymbol{q}}(t_0)\rangle
\right]_{\bar{\alpha}_2}
\nonumber\\
&&\times
\exp{\left[-i\int^{0}_{t_0}
\mathcal{E}^{(r)}_{\boldsymbol{q}\alpha_2}(t^{\prime\prime})
dt^{\prime\prime}\right]}.
\label{mathcalO}
\end{eqnarray}
In Eq.~(\ref{thetat}), 
$\Delta\alpha^\prime_{\boldsymbol{q}}(t)$ is an integrated adiabatic energy-phase at $t$, given by
\begin{equation}
\Delta\alpha^\prime_{\boldsymbol{q}}(t)
=\int^{t}_{0}
\left[
\mathcal{E}^{(r)}_{\boldsymbol{q}\alpha_2}(t^{\prime\prime})
-
\omega_{\boldsymbol{q}}^{(LO)}
\right]
dt^{\prime\prime},
\label{delalphaprime}
\end{equation}
$\xi_{\boldsymbol{q}}(t)$ is a phase associated with
FR dynamics, given by
\begin{equation}
\mathcal{D}_{\boldsymbol{q}\alpha_2}(t,t_D)
=|\mathcal{D}_{\boldsymbol{q}\alpha_2}(t,t_D)|
e^{i\xi_{\boldsymbol{q}}(t)},
\label{xit}
\end{equation}
$\upsilon_{\boldsymbol{q}\bar{\alpha}_2}(t,t_D)$ is defined as
an imaginary part of a diagonal contribution of non-adiabatic coupling
of Eq.~(\ref{I4-1}) integral over time, that is,
\begin{eqnarray}
\upsilon_{\boldsymbol{q}\bar{\alpha}}(t,\tilde{t})
&=&
{\rm Im} \:\mathfrak{Z}_{\boldsymbol{q}\bar{\alpha}}(t,\tilde{t}) 
\nonumber\\
&=&\sum_j{\rm Im}\/\mathfrak{I}^{(j)}_{\boldsymbol{q}\bar{\alpha}\bar{\alpha}}\:\theta(t-t_j)\theta(t_j-\tilde{t})
\label{upsilon}
\end{eqnarray}
with $\tilde{t}=t_D$, 
and
an additional phase  $\theta^0_{\boldsymbol{q}}(t)$ is defined as 
an argument of  
$\mathcal{O}^*_{\boldsymbol{q}\alpha_2}(t,t_D)$, that is,
\(\displaystyle
\mathcal{O}_{\boldsymbol{q}\alpha_2}(t,t_D)
=|\mathcal{O}_{\boldsymbol{q}\alpha_2}(t,t_D)|
e^{i\theta^0_{\boldsymbol{q}}(t)}.
\)

Now, the behavior of $\theta_{\boldsymbol{q}}(t)$ of Eq.~(\ref{thetat}) in the  limit of long 
elapse-time $t_L (\gg t_D)$  is examined.
It is remarked that $\Gamma_{\boldsymbol{q}\alpha_2}(t^\prime)$
diminishes rapidly as $t^\prime$ passes toward $t_L$; hence, the exponential function of
Eq.~(\ref{Q2}) becomes just constant at the time $t_L$.
As regards $\xi_{\boldsymbol{q}}(t_L)$,
it is seen that
$\mathcal{D}_{\boldsymbol{q}\alpha_2}(t_L,t_D)$ becomes
\begin{equation}
\mathcal{D}_{\boldsymbol{q}\alpha_2}(t_L,t_D)
=\left|\zeta_{\boldsymbol{q}}(t_D)\right|^2+i\left[\zeta_{\boldsymbol{q}}(t_D)\right]^*,
\label{calDtL}
\end{equation}
since  $M_{\boldsymbol{q}\bar{\alpha}_2}(t_L)$
vanishes in Eq.~(\ref{calD}), 
where
\(\displaystyle
\zeta_{\boldsymbol{q}}(t_D)=\pi\rho_{\boldsymbol{q}\bar{\alpha}_2}(t_D)
M_{\boldsymbol{q}\bar{\alpha}_2}(t_D).
\) 
Thus, 
$\xi_{\boldsymbol{q}}(t_L)$ is given by
\begin{equation}
\xi_{\boldsymbol{q}}(t_L)
=\tan^{-1}\left({\cos{\varphi_{\boldsymbol{q}}} \over
|\zeta_{\boldsymbol{q}}(t_D)|+\sin\varphi_{\boldsymbol{q}}} \right)
\label{xitL}
\end{equation}
modulus $\pi$,
where $\varphi_{\boldsymbol{q}}$ is defined as
\(\displaystyle
\zeta_{\boldsymbol{q}}(t_D)
=\left|\zeta_{\boldsymbol{q}}(t_D)\right|^{i\varphi_{\boldsymbol{q}}}.
\) 
Further, $\theta_{\boldsymbol{q}}^0(t_L)$ is approximated as
\begin{equation}
\theta^0_{\boldsymbol{q}}(t_L)
\approx
\upsilon_{\boldsymbol{q}\bar{\alpha}_p}(t_D,t_0)
+
\int^{0}_{t_0}
\mathcal{E}^{(r)}_{\boldsymbol{q}\alpha_2}(t^{\prime\prime})
dt^{\prime\prime},
\label{theta0}
\end{equation}
where  Eq.~(\ref{mathcalTapp2}) is employed:
\(\displaystyle
\mathcal{T}^\dagger_{\boldsymbol{q}\bar{\alpha}_2\bar{\alpha}}(t_L)
\approx \delta_{\bar{\alpha}\bar{\alpha}_2}e^{-\mathfrak{Z}_{\boldsymbol{q}\bar{\alpha}_2}(t_D,t_0)}
\). 
The adiabatic energy,
$\mathcal{E}_{\boldsymbol{q}\bar{\alpha}_2}(t)=\mathcal{E}^{(r)}_{\boldsymbol{q}\alpha_2}(t)$, is approximately given by\cite{appC}
\begin{eqnarray}
\mathcal{E}^{(r)}_{\boldsymbol{q}\alpha_2}(t)
&\approx&
\sqrt{(\bar{w}_{cv\boldsymbol{kq}})^2
+\Omega^2_{0cv}(t)}
\nonumber\\
&=&
\sqrt{(\bar{w}_{cv\boldsymbol{kq}})^2
+\left({A_L\over \tau_L}\right)^2}\:
\theta(t+\tau_L/2)\theta(\tau_L/2-t)
\nonumber\\
&&+
\omega^{(LO)}_{\boldsymbol{q}}\:
\theta(t-\tau_L/2),
\label{mathcalEapp}
\end{eqnarray}
where  $\bar{w}_{cv\boldsymbol{kq}}$ is defined right above Eq.~(\ref{Renergy2}), and $\Omega_{0cv}$ and $A_L$ are defined as Rabi frequency and pulse area right below Eqs.~(\ref{Renergy2}) and (\ref{F0}), respectively.
Therefore, $\theta_{\boldsymbol{q}}(t_L)$ is cast into
\begin{equation}
\theta_{\boldsymbol{q}}(t_L)
\approx
{\pi\over 2}-\xi_{\boldsymbol{q}}(t_L)
+\Delta\alpha_{\boldsymbol{q}}
+\upsilon_{\boldsymbol{q}\bar{\alpha}_2}(t_L,t_0)
\label{thetatL}
\end{equation}
modulus $\pi$,
where 
according to Eq.~(\ref{upsilon}), one obtains
\begin{equation}
\upsilon_{\boldsymbol{q}\bar{\alpha}_2}(t_L,t_0)
={\rm Im} \:\mathfrak{I}_{\boldsymbol{q}\bar{\alpha}_2\bar{\alpha}_2}(t_L,t_0)
=\sum_j{\rm Im} \:\mathfrak{I}^{(j)}_{\boldsymbol{q}\bar{\alpha}_2\bar{\alpha}_2}.
\label{Upsilonsum}
\end{equation}
Further, $\Delta\alpha_{\boldsymbol{q}}$ is given by
\begin{equation}
\Delta\alpha_{\boldsymbol{q}}
=\left[
\sqrt{\left(\omega^{(LO)}_{\boldsymbol{q}}\right)^2
+\left({A_L\over \tau_L}\right)^2}
-
{\omega^{(LO)}_{\boldsymbol{q}}\over 2}
\right]\tau_L
\approx A_L,
\label{Deltaalpha}
\end{equation}
where the limit of $\tau_L \rightarrow 0$ is taken in the second equality.\cite{comment3}

For the Fr\"{o}lich interaction, $\varphi_{\boldsymbol{q}}=\pm\pi/2$, and thus,
$\xi_{\boldsymbol{q}}(t_L)=0$, leading to 
\begin{equation}
\theta^F_{\boldsymbol{q}}(t_L)\approx \pi/2
+\upsilon_{\boldsymbol{q}\bar{\alpha}_2}(t_L,t_0)+A_L
\label{thetatL1}
\end{equation}
modulus $\pi$.
On the other hand, for the deformation-potential interaction,
due to $\varphi_{\boldsymbol{q}}=0,\pi$, one obtains
$\xi_{\boldsymbol{q}}(t_L)=\pm\tan^{-1}|\zeta_{\boldsymbol{q}}(t_D)|^{-1}$, leading to 
\begin{eqnarray}
\theta^D_{\boldsymbol{q}}(t_L)
&\approx&
\pm\tan^{-1}{|\zeta_{\boldsymbol{q}}(t_D)|}
+\upsilon_{\boldsymbol{q}\bar{\alpha}_2}(t_L,t_0)+\Delta\alpha_{\boldsymbol{q}}
\nonumber\\
&\approx&
\upsilon_{\boldsymbol{q}\bar{\alpha}_2}(t_L,t_0)+A_L
\label{thetatL2}
\end{eqnarray}
modulus $\pi$,
where in the first equality, the signs correspond, and in the second equality, 
the fact that $|\zeta_{\boldsymbol{q}}(t_D)| \approx 0$ in actual calculations is taken account of.

\section{Results and Discussion}
\label{sec3}

In actual calculations, the single-particle density matrices $\rho_{bb^\prime\boldsymbol{k}}(t)$ are evaluated in advance by solving  optical Bloch equations within the two-band model composed of $c$- and $v$-bands, as mentioned  below Eq.~(\ref{Delrho}).
For the sake of simplicity, carrier-density relaxation time -- labeled as $T^\prime_1$ -- and dephasing time  -- labeled as  $T^\prime_2$ --  are set identical  each other, that is,
$T^\prime_{12}\equiv T^\prime_1=T^\prime_2$.
Further, the  phenomenological  damping constant $T_{\boldsymbol{qk}bb^\prime}$  given in Eq.~(\ref{Heisenberg}) is considered independent of $\boldsymbol{k}$, $b$, and $b^\prime$, and is commonly  labeled as $T_{\boldsymbol{q}12}$.
This leads to 
$\gamma^{(0)}_{\boldsymbol{q}\bar{\alpha}(\not=\bar{\alpha}_2)}(t) =2/T_{\boldsymbol{q}12}$
and
$ \gamma^{(0)}_{\boldsymbol{q}\bar{\alpha}_2}(t)\approx 0 $
because of  Eq.~(\ref{gamma00}) and
$V^R_{\boldsymbol{q}\alpha\bar{\alpha}_2} \approx 0$.
For the sake of later convenience, $T^\prime_{\boldsymbol{q}12}$ is defined as
$
T^\prime_{\boldsymbol{q}12}=T_{\boldsymbol{q}12}+t_D.
$
It is understood that $T^\prime_{12}$ and $T^\prime_{\boldsymbol{q}12}$ represent phenomenological damping time-constants of induced carrier-density with isotropic and anisotropic momentum distributions, respectively.

\begin{figure}[tb]
\begin{center}
\includegraphics[width=7cm,clip]{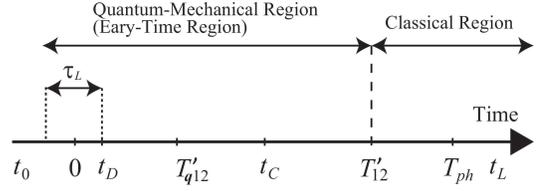}
\caption{Schematic allocation of various time constants employed in actual calculations. For more detail, consult the text and Table~\ref{tab4}.
}
\label{fig8}
\end{center}
\end{figure}

According to Eq.~(\ref{chi-I}), $\chi_{\boldsymbol{q}}(t_p+\tau,t_p)$ is composed of the contributions from  $\Xi^{(r)}_{\boldsymbol{q}\alpha_p}\;(p=1,2)$ and $\Xi^{(c)}_{\boldsymbol{q}\bar{\alpha}}$.
In the temporal region of $t_p< T^\prime_{\boldsymbol{q}12}$,  $\chi_{\boldsymbol{q}}(t_p+\tau,t_p)$ is governed by the terms of  $\Xi^{(r)}_{\boldsymbol{q}\alpha_1}$ and $\Xi^{(c)}_{\boldsymbol{q}\bar{\alpha}}$ attributed to  excitations of plasmon and continuum states, respectively.
On the other hand, in the region of $t_p> T^\prime_{\boldsymbol{q}12}$, $\chi_{\boldsymbol{q}}(t_p+\tau,t_p)$ vanishes, since the contributions of   $\Xi^{(r)}_{\boldsymbol{q}\alpha_1}$ and $\Xi^{(c)}_{\boldsymbol{q}\bar{\alpha}}$ diminish rapidly due to the damping effect relevant to $T^\prime_{\boldsymbol{q}12}$.
In the evaluation of  the transient photoemission spectra $\bar{A}_{\boldsymbol{q}}(t_p;\omega) $ and  the power spectra $S_{\boldsymbol{q}}(\omega)$, the non-adiabatic coupling $\upsilon_{\boldsymbol{q}\bar{\alpha}_2}(t,t^\prime)$ given by  Eq.~(\ref{upsilon})
is approximated by retaining  just two-terms with $\mathfrak{I}^{(D)}_{\boldsymbol{q}\bar{\alpha}_2\bar{\alpha}_2}$ and 
$\mathfrak{I}^{(C)}_{\boldsymbol{q}\bar{\alpha}_2\bar{\alpha}_2}$.
That is, 
\begin{eqnarray}
\upsilon_{\boldsymbol{q}\bar{\alpha}_2}(t,t^\prime)
&\approx&
{\rm Im}\/\mathfrak{I}^{(D)}_{\boldsymbol{q}\bar{\alpha}_2\bar{\alpha}_2}\theta(t-t_D)\theta(t_D-t^\prime)
\nonumber\\
&&+
{\rm Im}\/\mathfrak{I}^{(C)}_{\boldsymbol{q}\bar{\alpha}_2\bar{\alpha}_2}\:\theta(t-t_C)\theta(t_C-t^\prime)
\label{upsilonapp}
\end{eqnarray}
with $t_C > t_D$.

Figure~\ref{fig8} shows schematic allocation of the various time constants mentioned above, namely, 
$t_D$, $T^\prime_{\boldsymbol{q}12}$, $t_C$,  and $T^\prime_{12}$,  in addition to $T_{ph}$, $t_0(=-\infty)$ and $t_L(=+\infty)$;
$T_{ph}$ represents  a phenomenological damping time-constant of LO-phonon due to lattice anharmonicity.\cite{comment1,comment2}
In fact, the numerical values of the first four constants cannot be determined a priori within the scope of the present theoretical framework.
Thus these values are considered as adjustable parameters;
the numerical values  given in Table~\ref{tab4}   are employed in actual calculations.
Time scales for different stages of relaxation phenomena in photoexcited experiments are summarized in Ref.~\onlinecite{kuznetsov1}.
As is shown below, it is certain that in the chronological order given 
in Fig.~\ref{fig8}, asymmetry spectral-profiles of  transient FR of concern are observed both in $\bar{A}_{\boldsymbol{q}}(t_p;\omega) $ and  $S_{\boldsymbol{q}}(\omega)$.

\begin{table}[t]
\caption{Time constants employed in actual calculations of the present study.
}
\begin{tabular}{cr}
\hline\hline
\multicolumn{1}{c}{Time constants} &
\multicolumn{1}{r}{}\\\hline
$t_D$&7.5 fs\\
$T^\prime_{\boldsymbol{q}12}$&27.5 fs\\
$t_C$&\;\;\;\;72.5 fs\\
$T^\prime_{12}$&90 fs\\
$T_{ph}$&5000 fs\\\hline\hline
\end{tabular}
\label{tab4}
\end{table}

Discussion  of the calculated results of $\bar{A}_{\boldsymbol{q}}(t_p;\omega)$ is made in Sec.~\ref{sec3B} with a possible interpretation of CP generation dynamics in Sec.~\ref{sec3C}.
Discussion of $S_{\boldsymbol{q}}(\omega)$ is made in Sec.~\ref{sec3D}.
The comparison of the present results with the existing other studies is given in Sec.~\ref{sec3E}.

\subsection{Transient induced photoemission spectra}
\label{sec3B}

\begin{figure}[tb]
\begin{center}
\includegraphics[width=6.6cm,clip]{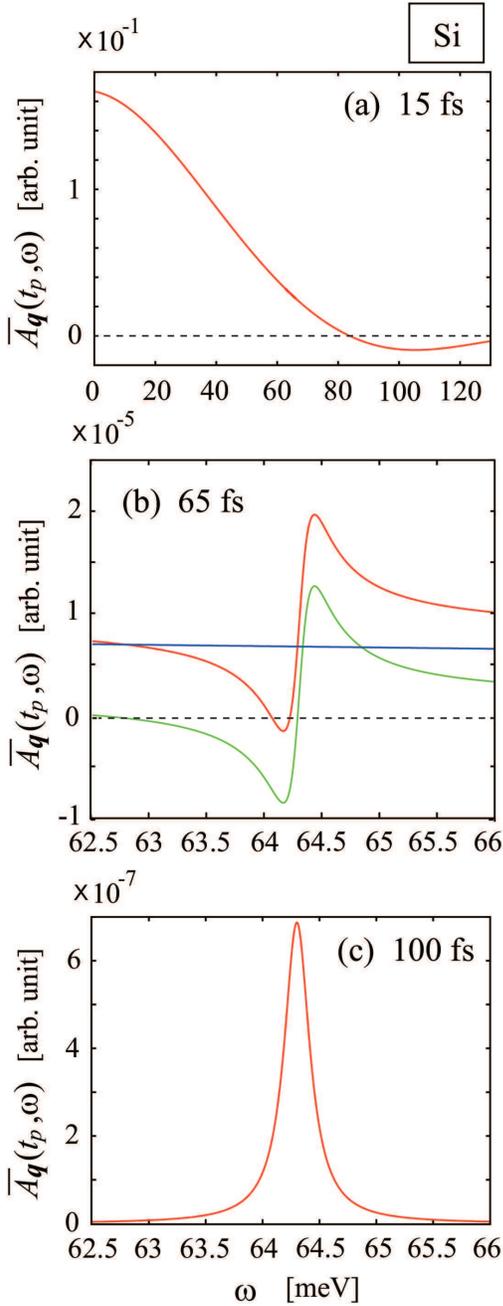}
\caption{(Color online) Transient photoemission spectra $\bar{A}_{\boldsymbol{q}}(t_p,\omega)$ of Si (red line) as a function of frequency $\omega$ (in the unit of meV) at probe time $t_p$ of  (a) 15 fs, (b) 65 fs, and (c) 100 fs. 
Separate contributions to the spectra from 
$\tilde{\chi}_{\boldsymbol{q}}(t_p;\omega)$ and $\tilde{\chi}^\prime_{\boldsymbol{q}}(t_p;\omega)$ are also shown by blue and green lines, respectively. 
}
\label{fig9}
\end{center}
\end{figure}

Transient induced photoemission spectra $\bar{A}_{\boldsymbol{q}}(t_p,\omega)$ of Si  and GaAs as a function of frequency $\omega$ are taken into account with probe time $t_p$ fixed.
As seen from  Eq.~(\ref{chitot}), the two interactions -- the dynamically-screened Coulomb interaction induced by electron and the LO-phonon-induced interaction -- contribute to the total retarded longitudinal susceptibility, 
that is,
\begin{equation}
\tilde{\chi}^{(t)}_{\boldsymbol{q}}(t_p;\omega)=\tilde{\chi}_{\boldsymbol{q}}(t_p;\omega)+\tilde{\chi}^\prime_{\boldsymbol{q}}(t_p;\omega),
\label{tildechitot2}
\end{equation}
where $\tilde{\chi}_{\boldsymbol{q}}(t_p;\omega)$ and $\tilde{\chi}^\prime_{\boldsymbol{q}}(t_p;\omega)$ are Fourier-transforms of  $\chi_{\boldsymbol{q}}(t_p+\tau,t_p)$ and $\chi^\prime_{\boldsymbol{q}}(t_p+\tau,t_p)$ with respect to $\tau$, respectively, similarly to Eq.~(\ref{tildechitot}).
In the small-$\boldsymbol{q}$ limit, $\tilde{\chi}_{\boldsymbol{q}}(t_p;\omega)$ is in proportion to $|\boldsymbol{q}|^2$, while $\tilde{\chi}^\prime_{\boldsymbol{q}}(t_p;\omega)$ is in proportion to $|\boldsymbol{q}|^2$ for the Fr\"{o}hlich interaction and  to $|\boldsymbol{q}|^4$ for the deformation potential interaction, as remarked below Eq.~(\ref{phononGfct2}).
This difference arises from the fact that the former interaction is of long-range, and the latter one is of short-range, which reflects on $\bar{A}_{\boldsymbol{q}}(t_p,\omega)$ through Eq.~(\ref{epsilon-2}), as it should be;
in a non-polar crystal such as Si, due to the presence of spatial inversion-symmetry, lattice absorption vanishes in the limit of a dipole transition with transferred momentum $\boldsymbol{q}$ being  zero.\cite{yu}
The effect of  $\boldsymbol{q}$ on $\bar{A}_{\boldsymbol{q}}(t_p,\omega)$ will be  examined later again from the viewpoint of group theory.
In passing, $|\boldsymbol{q}|$ is set equal to 0.015 a.u. for the evaluation of $\tilde{\chi}^\prime_{\boldsymbol{q}}(t_p;\omega)$ due to the deformation-potential interaction; the edge point of the first Brillouin zone is 0.3 a.u. for reference.

Moreover, as seen from Eqs.~(\ref{Xc}), (\ref{Xr}), and (\ref{Pr}),
an imaginary part of a diagonal contribution of non-adiabatic coupling of Eq.~(\ref{upsilonapp})  is incorporated in 
the total retarded longitudinal susceptibility, $\chi^{(t)}_{\boldsymbol{q}}(t_p+\tau,t_p)$, in the form of
$\upsilon_{\boldsymbol{q}\bar{\alpha}_2}(t_p+\tau,t_p)$ with $\tau >0$.
Thus, in the actual calculations of $\bar{A}_{\boldsymbol{q}}(t_p,\omega)$ in the region of $t_p > t_D$, 
the first term of Eq.~(\ref{upsilonapp}) is left out, and just the second term contributes to $\bar{A}_{\boldsymbol{q}}(t_p,\omega)$ as long as $t_p < t_C$.

Figure~\ref{fig9} shows $\bar{A}_{\boldsymbol{q}}(t_p;\omega)$ of Si by a red line, where
both of the contributions from  $\tilde{\chi}_{\boldsymbol{q}}(t_p;\omega)$ and  $\tilde{\chi}^\prime_{\boldsymbol{q}}(t_p;\omega)$ are shown by blue and green lines, respectively.
At first,  no contributions from non-adiabatic couplings are assumed  throughout the calculations of this figure; ${\rm Im}\/\mathfrak{I}^{(C)}_{\boldsymbol{q}\bar{\alpha}_2\bar{\alpha}_2}$ of Eq.~(\ref{upsilonapp}) is set equal to zero. 
Further, it is understood that $\bar{A}_{\boldsymbol{q}}(t_p;\omega)$ is reckoned from  a background contribution ascribable to $\Xi^{(c)}_{\boldsymbol{q}\bar{\alpha}}$  in Eq.~(\ref{chi-I}); this choice of the baseline, namely, the line of $\bar{A}_{\boldsymbol{q}}(t_p;\omega)=0$, corresponds to removal of this background contribution from $\tilde{\chi}_{\boldsymbol{q}}(t_p;\omega)$.

\begin{figure}[tb]
\begin{center}
\includegraphics[width=7cm,clip]{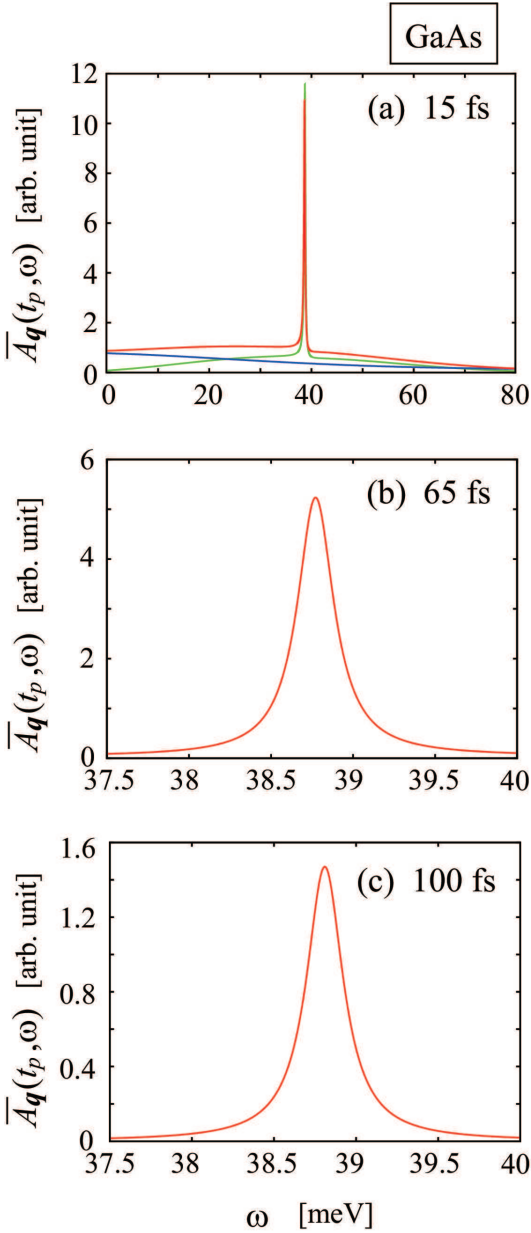}
\caption{(Color online) The same as Fig.~\ref{fig9} but for GaAs.
}
\label{fig10}
\end{center}
\end{figure}

As seen from Fig.~\ref{fig9}(a), $\bar{A}_{\boldsymbol{q}}(t_p,\omega)$ at $t_p=15$ fs is governed by the contribution from $\tilde{\chi}_{\boldsymbol{q}}(t_p;\omega)$ though the associated blue line is overwritten; almost structureless continuum-spectra are formed over a wide range of $\omega$ due to electronic excitation mediated by optical interband-transitions. 
To be specific, the spectra of concern  are attributed to  the greater contribution from the term $\Xi^{(r)}_{\boldsymbol{q}\alpha_1}$  rather than from the term $\Xi^{(r)}_{\boldsymbol{q}\alpha_2}$  in Eq.~(\ref{chi-I}) immediately after the completion of laser irradiation at $t_D$.
On the contrary, the contribution from $\tilde{\chi}^\prime_{\boldsymbol{q}}(t_p;\omega)$ is negligibly small due to the proportion of it to $|\boldsymbol{q}|^4$.
As  is shown in Fig.~\ref{fig9}(b) at $t_p=65$ fs, the contributions from $\tilde{\chi}_{\boldsymbol{q}}(t_p;\omega)$ are damped to be comparable to those from $\tilde{\chi}^\prime_{\boldsymbol{q}}(t_p;\omega)$.
It should be noted that asymmetric spectra with a dip followed by a peak  -- reminiscent of FR -- are manifested, where $\tilde{\chi}^\prime_{\boldsymbol{q}}(t_p;\omega)$ is dominated by the contribution from the term $\Pi^{(r)}_{\boldsymbol{q}\alpha_2}$  in Eq.~(\ref{chiprime2}).
The resulting spectral profile is in sharp contrast with a symmetric Lorentzian-profile of the spectra of
 Fig.~\ref{fig9}(c) at $t_p=100$ fs.
 
The difference between the two spectra is \ out by consulting the behavior of
\begin{equation}
\mathfrak{D}_{\boldsymbol{q}\alpha_2}(\tau,t_D)=
 \mathcal{D}_{\boldsymbol{q}\alpha_2}(t_p+\tau,t_D)\mathcal{D}^*_{\boldsymbol{q}\alpha_2}(t_p,t_D)
 \label{mathfrakD}
\end{equation}
 as a function of $\tau$.
This function is incorporated in $\Pi^{(r)}_{\boldsymbol{q}\alpha_2}(t_p+\tau,t_p)$ given by Eq.~(\ref{Pr}), and for the trace of $\mathcal{D}_{\boldsymbol{q}\alpha_2}(t,t_D)$ of Si as a function of $t$, see Ref.~\onlinecite{SMmathcalD}.
$\Pi^{(r)}_{\boldsymbol{q}\alpha_2}(t_p+\tau,t_p)$ is  considered to be marked
mostly with the $\tau$-dependent argument of 
$\mathfrak{D}_{\boldsymbol{q}\alpha_2}(\tau,t_D)=|\mathfrak{D}_{\boldsymbol{q}\alpha_2}(\tau,t_D)|e^{i\vartheta_{\boldsymbol{q}}(\tau)}$, 
that is,
\begin{equation}
\vartheta_{\boldsymbol{q}}(\tau)=
\xi_{\boldsymbol{q}}(t_p+\tau)-\xi_{\boldsymbol{q}}(t_p),
\label{vartheta}
\end{equation}
where $\xi_{\boldsymbol{q}}(t)$ is given by Eq.~(\ref{xit}), and also shown in Ref.~\onlinecite{SMmathcalD}.
It is seen that with the increase of $\tau$ from zero, $\vartheta_{\boldsymbol{q}}(\tau)$ varies from zero to $-\pi/2$ when $t_p=65$ fs, whereas $\vartheta_{\boldsymbol{q}}(\tau)$  remains almost unaltered with $\vartheta_{\boldsymbol{q}}(\tau)\approx 0$ when $t_p=100$ fs, where $\xi_{\boldsymbol{q}}(t_p=65\:{\rm fs})\approx 0$ and $\xi_{\boldsymbol{q}}(t_p=100\:{\rm fs})\approx -\pi/2$.
Therefore, the $\tau$-dependent phase-change results in the asymmetry in $\bar{A}_{\boldsymbol{q}}(t_p;\omega)$ of Fig.~\ref{fig9}(b) through the Fourier-transform of $\chi^\prime_{\boldsymbol{q}}(t_p+\tau,t_p)$.
Without this, $\bar{A}_{\boldsymbol{q}}(t_p;\omega)$ is just of symmetric shape, as shown in Fig.~\ref{fig9}(c).
The alteration of $\vartheta_{\boldsymbol{q}}(\tau)$ in the case of Fig.~\ref{fig9}(b) is originated from the fact that the effective coupling constant in Si, namely, $M^D_{\boldsymbol{q}\bar{\alpha}_2}$, is real.
It is remarked that the spectral-peak frequency in Fig.~\ref{fig9}(c) is slightly different from $\omega^{(LO)}_{\boldsymbol{q}}$ equal to $63.0\:{\rm meV}$, since adiabatic energy $\mathcal{E}^{(r)}_{\boldsymbol{q}\alpha_2}(t)$ is employed in the calculation by extrapolation to the given asymptotic value of $\mathcal{E}^{(r)}_{\boldsymbol{q}\alpha_2}(t_D)$ equal to $64.3\;{\rm meV}$.
This difference is considered to be just an artifact, not due to a physical effect such as self-energy renormalization.

Figure~\ref{fig10} shows $\bar{A}_{\boldsymbol{q}}(t_p;\omega)$ of  GaAs by a red line, corresponding to Fig.~\ref{fig9} for Si.
As seen in Fig.~\ref{fig10}(a),   a discernible peak of $\tilde{\chi}^\prime _{\boldsymbol{q}}(t_p;\omega)$  -- indicated by a green line --   is superimposed with a structureless continuum background.
Here, this background is  composed of a background contribution of  $\tilde{\chi}^\prime _{\boldsymbol{q}}(t_p;\omega)$, and $\tilde{\chi}_{\boldsymbol{q}}(t_p;\omega)$  -- indicated by a blue line -- with comparable order, since both are in proportion to $|\boldsymbol{q}|^2$.
The spectra  at $t_p=$ 65 fs shown in Fig.~\ref{fig10}(b), which are dominated by 
$\tilde{\chi}^\prime _{\boldsymbol{q}}(t_p;\omega)$, differ a lot  from the counterpart spectra of Fig.~\ref{fig9}(b) of Si in that the former spectra are of symmetric shape.
Here, the contribution of 
$\tilde{\chi}^\prime _{\boldsymbol{q}}(t_p;\omega)$ is dominated by 
the term $\Pi^{(r)}_{\boldsymbol{q}\alpha_2}$  in Eq.~(\ref{chiprime2}).
Actually, manifestation of such symmetric spectra is brought to light by inspecting the combined phase of Eq.~(\ref{vartheta}).
That is,  $\vartheta_{\boldsymbol{q}}(\tau)$ remains almost zero over the whole range of $\tau$  when $t_p=65$ fs; to be precise, $\xi_{\boldsymbol{q}}(t)\approx 0$ for all $t$-region.\cite{SMmathcalD}
This leads to the symmetric profile through the Fourier-transform of $\chi^\prime_{\boldsymbol{q}}(t_p+\tau,t_p)$.
The same situation as that in Fig.~\ref{fig10}(b) holds correctly in the spectra  at $t_p=$ 100 fs shown in Fig.~\ref{fig10}(c), aside from  smaller-peak intensity in the latter due to the fact that $|\mathcal{D}_{\boldsymbol{q}\alpha_2}(t_p=100\:{\rm fs},t_D)| <|\mathcal{D}_{\boldsymbol{q}\alpha_2}(t_p=65\:{\rm fs},t_D)| $.
Similarly to the above-mentioned discussion of Si, the quantity of $\vartheta_{\boldsymbol{q}}(\tau)$ is closely related to an effective coupling constant.
In this case of GaAs, the obtained result of $\vartheta_{\boldsymbol{q}}(\tau)\approx 0$ is due to
the fact that $M^F_{\boldsymbol{q}\bar{\alpha}_2}$ is pure-imaginary.
In addition, the spectral-peak frequency in Fig.~\ref{fig10} is slightly different from $\omega^{(LO)}_{\boldsymbol{q}}$ equal to $35.0\:{\rm meV}$ for the same reason as that in Fig.~\ref{fig9}(c).

\begin{figure}[tb]
\begin{center}
\includegraphics[width=7.0cm,clip]{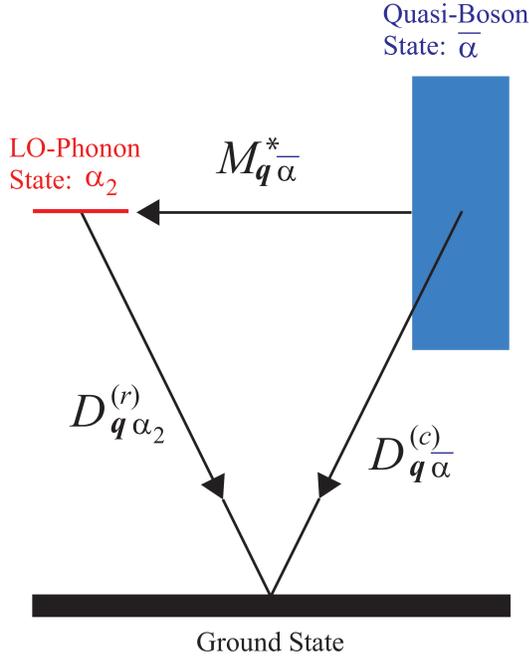}
\caption{(Color online)  Schematic diagram of FR dynamics of LO-phonon state $\alpha_2$ with energy $\omega^{(LO)}_{\boldsymbol{q}}$ embedded in quasi-boson state $\bar{\alpha}$ accompanied by induced photoemission process.
$D^{(r)}_{\boldsymbol{q}\alpha_2}$ and $D^{(c)}_{\boldsymbol{q}\bar{\alpha}}$ represent transition matrices of photoemission from $\alpha_2$ and $\bar{\alpha}$ to the ground state, respectively.  $M^*_{\boldsymbol{q}\bar{\alpha}}$ represents a coupling matrix between $\bar{\alpha}$ and $\alpha_2$, where
this is a complex  matrix in general. 
}
\label{fig11}
\end{center}
\end{figure}

Here, the spectra of Figs.~\ref{fig9}(b) and \ref{fig10}(b) are compared each other in more detail from the viewpoint of
Shore's spectral formula given by Eq.~(\ref{shore0}).
The degree of asymmetry is characterized by the parameter
$\mathcal{A}_{\boldsymbol{q}\alpha_2}(t_p)$ pertinent to 
Fano's $q$-parameter $q_{\boldsymbol{q}\alpha_2}(t_p)$ through Eq.~(\ref{q0}).
Further, the origin of the asymmetry is traced back to Eq.~(\ref{RemathcalA0}) in terms of
$\Psi^{(r)}_{\boldsymbol{q}\bar{\alpha}_2}(t_p+\tau,t_p)$ of Eq.~(\ref{Psir0}).
It is evident that the character of this function  is closely related to that of 
$\mathfrak{D}_{\boldsymbol{q}\alpha_2}(\tau,t_D)$.
In the spectra of Si in Fig.~\ref{fig9}(b), $\Psi^{(r)}_{\boldsymbol{q}\bar{\alpha}_2}(t_p,t_p)$ is undoubtedly  complex and  ${\rm Im}\/\Psi^{(r)\prime}_{\boldsymbol{q}\bar{\alpha}_2}(t_p,t_p) \not= 0$, leading to $\mathcal{A}_{\boldsymbol{q}\alpha_2}(t_p) \not= 0$ and $q_{\boldsymbol{q}\alpha_2}(t_p)$ being finite, whereas in the 
spectra of GaAs in Fig.~\ref{fig10}(b), this is almost real and  ${\rm Im}\/\Psi^{(r)\prime}_{\boldsymbol{q}\bar{\alpha}_2}(t_p,t_p) \approx 0$, leading to  $\mathcal{A}_{\boldsymbol{q}\alpha_2}(t_p) \approx 0$ and $|q_{\boldsymbol{q}\alpha_2}(t_p)| \approx \infty$.
By consulting the above discussion on the correlation of an effective coupling constant with $\mathfrak{D}_{\boldsymbol{q}\alpha_2}(\tau,t_D)$, it would be concluded that the $q$-parameter is exclusively determined by the argument of the effective coupling constant, as long as ${\rm Im}\/\mathfrak{I}^{(C)}_{\boldsymbol{q}\bar{\alpha}_2\bar{\alpha}_2}=0$ [see Eq.~(\ref{upsilonapp})].

Such phase-dependent FR dynamics would be understood in a qualitative manner by consulting the diagram shown in Fig.~\ref{fig11}.
This shows the transient FR dynamics at time $t_p$ accompanied by induced photoemission process.
It is seen that there are two transition paths:
one is a direct process mediated by an optical transition matrix $D^{(c)}_{\boldsymbol{q}\bar{\alpha}}$ from quasi-boson state $\bar{\alpha}$ to the ground state, and the other is 
a two-step resonant process mediated by $M^*_{\boldsymbol{q}\bar{\alpha}}$ from $\bar{\alpha}$ to 
LO-phonon state $\alpha_2$, followed by a deexcited process mediated by an optical transition matrix $D^{(r)}_{\boldsymbol{q}\alpha_2}$ from $\alpha_2$ to the ground state.
Here, $t_p$-dependence of $D^{(c)}_{\boldsymbol{q}\bar{\alpha}}$, $D^{(r)}_{\boldsymbol{q}\alpha_2}$, and 
 $M^*_{\boldsymbol{q}\bar{\alpha}}$ are omitted for the sake of simplicity, and $M_{\boldsymbol{q}\bar{\alpha}}$
is given by Eq.~(\ref{M2DF}).
Thus, the whole transition matrix $D_{\boldsymbol{q}\bar{\alpha}}(t_p;\omega)$ is given by\cite{shore}
\begin{equation}
D_{\boldsymbol{q}\bar{\alpha}}(t_p;\omega)
=D^{(c)}_{\boldsymbol{q}\bar{\alpha}}
+\frac{
D^{(r)}_{\boldsymbol{q}\alpha_2}M^*_{\boldsymbol{q}\bar{\alpha}}
}{
\omega-\omega^{(LO)}_{\boldsymbol{q}}+i\Gamma_{\boldsymbol{q}\alpha_2}/2
},
\label{FanoScheme}
\end{equation}
and the induced photoemission spectra $\bar{A}_{\boldsymbol{q}}(t_p,\omega)$ is approximated by
$\bar{A}_{\boldsymbol{q}}(t_p;\omega)\approx|D_{\boldsymbol{q}\bar{\alpha}}(t_p;\omega)|^2$.
Comparing the obtained result with Eq.~(\ref{shore0}), the three Shore's parameters are provided as
\begin{equation}
\mathcal{A}_{\boldsymbol{q}\alpha_2}(t_p)=2D^{(c)}_{\boldsymbol{q}\bar{\alpha}}D^{(r)}_{\boldsymbol{q}\alpha_2}
|M_{\boldsymbol{q}\bar{\alpha}}|\cos{\phi_{\boldsymbol{q}}},
\label{Aapp}
\end{equation}
\begin{eqnarray}
\mathcal{B}_{\boldsymbol{q}\alpha_2}(t_p)
&=&-2D^{(c)}_{\boldsymbol{q}\bar{\alpha}}D^{(r)}_{\boldsymbol{q}\alpha_2}
|M_{\boldsymbol{q}\bar{\alpha}}|\sin{\phi_{\boldsymbol{q}}}
\nonumber\\
&&
+{D^{(r)2}_{\boldsymbol{q}\alpha_2}
|M_{\boldsymbol{q}\bar{\alpha}}|^2 \over \Gamma_{\boldsymbol{q}\alpha_2}/2},
\label{Bapp}
\end{eqnarray}
and
\begin{equation}
\mathcal{C}_{\boldsymbol{q}\alpha_2}(t_p)=|D^{(c)}_{\boldsymbol{q}\bar{\alpha}}|^2,
\label{Capp}
\end{equation}
where both of $D^{(c)}_{\boldsymbol{q}\bar{\alpha}}$ and $D^{(r)}_{\boldsymbol{q}\alpha_2}$ are considered as real, and
$M_{\boldsymbol{q}\bar{\alpha}}$ is a complex number given by
$M_{\boldsymbol{q}\bar{\alpha}}=|M_{\boldsymbol{q}\bar{\alpha}}|e^{i\phi_{\boldsymbol{q}}}$.

It is readily seen that $\mathcal{A}_{\boldsymbol{q}\alpha_2}(t_p)$
depends on the argument of $M^*_{\boldsymbol{q}\bar{\alpha}}$.
This result shows that  $\mathcal{A}_{\boldsymbol{q}\alpha_2}(t_p)$ vanishes for $M^*_{\boldsymbol{q}\bar{\alpha}}$ being a pure-imaginary number as in the case of GaAs, and the observed spectra  $\bar{A}_{\boldsymbol{q}}(t_p;\omega)$ becomes of symmetric shape.
This implies that despite of the symmetric spectral-profile in GaAs,
the associated spectral-peak is formed by a FR state similarly to that in Si, because the natural spectral-width $\Gamma_{\boldsymbol{q}\alpha_2}$ is still  finite.  
Actually, according to the calculated results, 
 $\Gamma_{\boldsymbol{q}\alpha_2}$ of GaAs is 0.46 and  $9.0\times 10^{-3}$ meV at
 $t_p=t_D(=7.5$ fs) and 65 fs, respectively.
 In passing, $\Gamma_{\boldsymbol{q}\alpha_2}$ of Si is $1.6\times10^{-3}$ and  $1.4\times10^{-5}$ meV  at
 $t_p=t_D$ and 65 fs, respectively.
As referred to in Ref.~\onlinecite{comment1}, the function of $e^{-\tau/T_{ph}}$ is implicitly convoluted with the integrand of Eq.~(\ref{tildechitot}) with $2/T_{ph}=0.27$ meV.
Hence, the spectral widths of  Figs.~\ref{fig9}(b) and \ref{fig10}(b) are determined by 
$2/T_{ph}$ rather than  $\Gamma_{\boldsymbol{q}\alpha_2}$.
In the meantime, for $t_p=100$ fs, it is speculated that $D^{(c)}_{\boldsymbol{q}\bar{\alpha}}$ vanishes, and
$\bar{A}_{\boldsymbol{q}}(t_p;\omega)$ is  governed by a discrete state with $\Gamma_{\boldsymbol{q}\alpha_2}=0$.
Here, the spectral profile becomes symmetric due to $\mathcal{A}_{\boldsymbol{q}\alpha_2}(t_p)=0$,
as seen in Figs.~\ref{fig9}(c) and \ref{fig10}(c). 

\begin{figure}[tb]
\begin{center}
\includegraphics[width=7cm,clip]{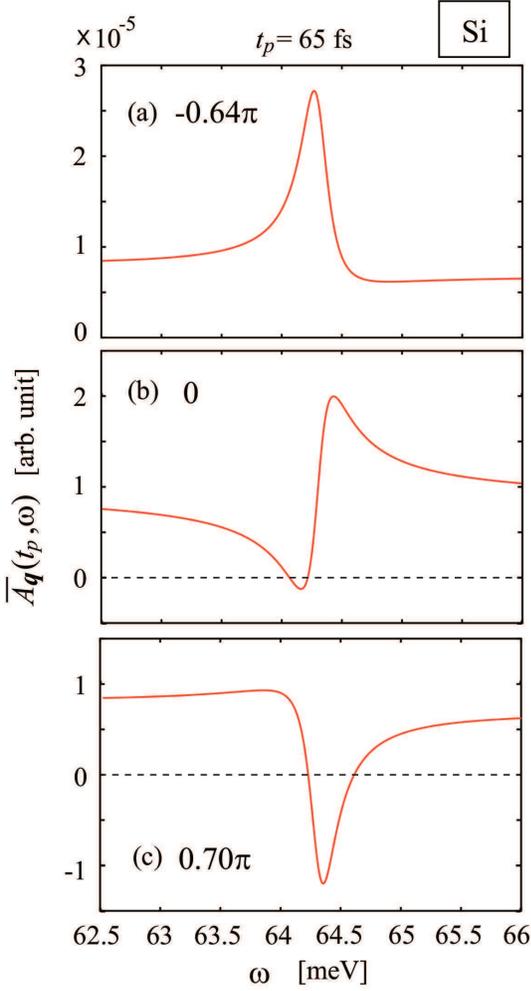}
\caption{(Color online) Transient photoemission spectra $\bar{A}_{\boldsymbol{q}}(t_p,\omega)$ of Si (red line) as a function of frequency $\omega$ (in the unit of meV) at probe time $t_p= 65$ with ${\rm Im}\/\mathfrak{I}^{(C)}_{\boldsymbol{q}\bar{\alpha}_2\bar{\alpha}_2}$ of
(a) $-0.64\pi$, (b) $0$, and (c) $0.70\pi$. 
}
\label{fig12}
\end{center}
\end{figure}

\begin{figure}[tb]
\begin{center}
\includegraphics[width=7cm,clip]{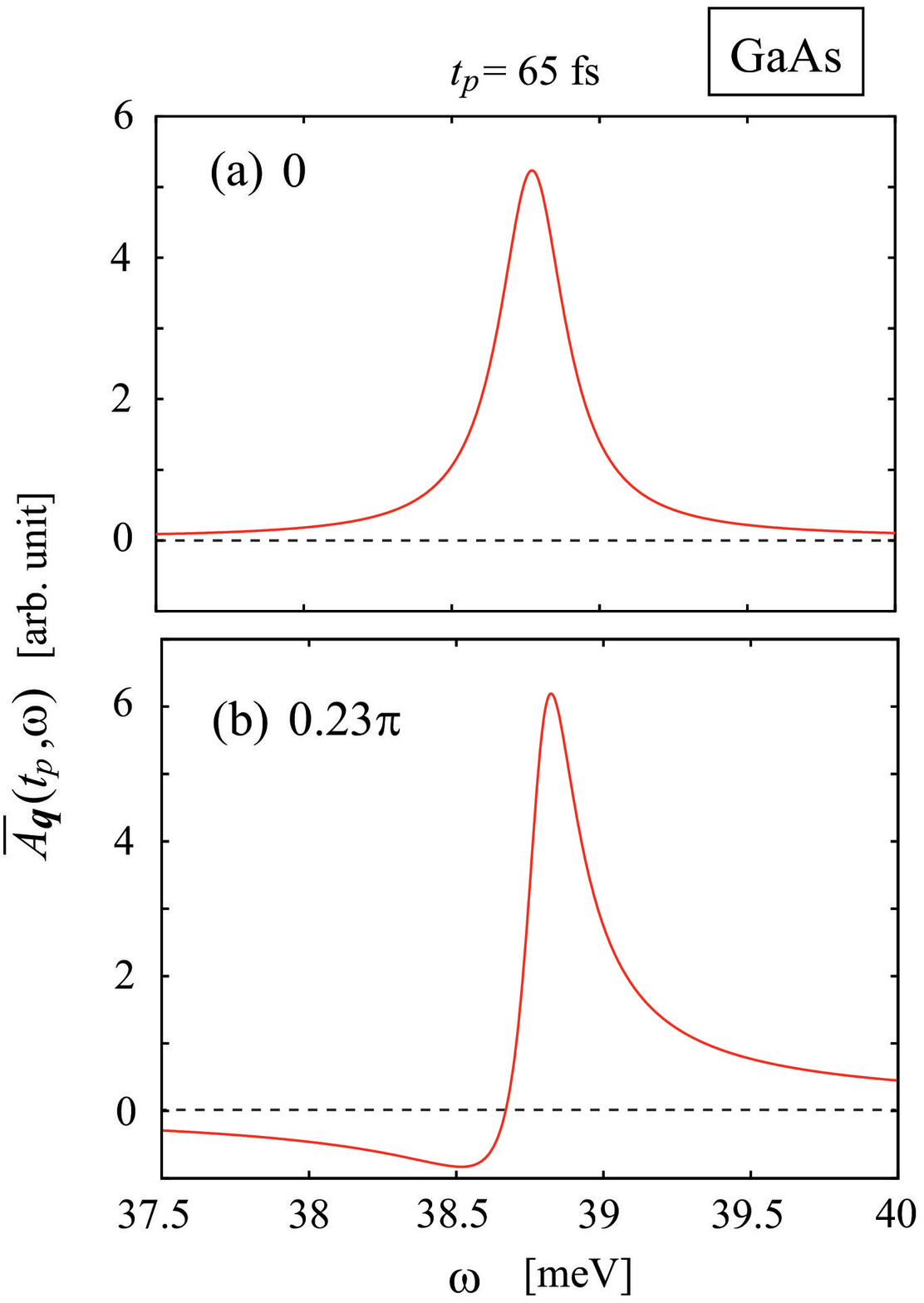}
\caption{(Color online) The same as Fig.~\ref{fig12} but for GaAs with ${\rm Im}\/\mathfrak{I}^{(C)}_{\boldsymbol{q}\bar{\alpha}_2\bar{\alpha}_2}$ of
(a) 0 and (b) $0.23\pi$. 
}
\label{fig13}
\end{center}
\end{figure}

Next, the effect of ${\rm Im}\/\mathfrak{I}^{(C)}_{\boldsymbol{q}\bar{\alpha}_2\bar{\alpha}_2}$
on $\bar{A}_{\boldsymbol{q}}(t_p;\omega)$ is examined, where this is fully neglected in the calculations of Figs.~\ref{fig9} and \ref{fig10}.
Figures~\ref{fig12} and \ref{fig13} show $\bar{A}_{\boldsymbol{q}}(t_p;\omega)$ of Si and GaAs, respectively, at $t_p=$ 65 fs with given numerical values of  ${\rm Im}\/\mathfrak{I}^{(C)}_{\boldsymbol{q}\bar{\alpha}_2\bar{\alpha}_2}$.
It is noted that these spectra are affected by the non-adiabatic coupling of concern in the limited region of $t_p < t_C$.
Thus, both  spectra of  Figs.~\ref{fig9}(c) and \ref{fig10}(c) are free from this effect, sustaining the symmetric profiles.
First, as regards Si shown in Fig.~\ref{fig12}, the profile of $\bar{A}_{\boldsymbol{q}}(t_p;\omega)$ depends definitely on the value of ${\rm Im}\/\mathfrak{I}^{(C)}_{\boldsymbol{q}\bar{\alpha}_2\bar{\alpha}_2}$.
The profile of Fig.~\ref{fig12}(b) -- the same as Fig.~\ref{fig9}(b) -- changes to the shape with a peak followed by a dip, as shown in  Fig.~\ref{fig12}(a), when ${\rm Im}\/\mathfrak{I}^{(C)}_{\boldsymbol{q}\bar{\alpha}_2\bar{\alpha}_2}=-0.64\pi$.
Thus this profile is
characterized by $q_{\boldsymbol{q}\alpha_2}(t_p) < 0$.
Further, even a window-resonance-shaped profile with   $q_{\boldsymbol{q}\alpha_2}(t_p) \approx 0$ is revealed in the spectra shown in  Fig.~\ref{fig12}(c), when ${\rm Im}\/\mathfrak{I}^{(C)}_{\boldsymbol{q}\bar{\alpha}_2\bar{\alpha}_2}=0.70\pi$.
The similar changes of spectra with   $q_{\boldsymbol{q}\alpha_2}(t_p) \ge 0 $ are also recognized in GaAs, as shown in Fig.~\ref{fig13}, where
the profile of Fig.~\ref{fig13}(a) is the same as Fig.~\ref{fig10}(b).
The spectra with ${\rm Im}\/\mathfrak{I}^{(C)}_{\boldsymbol{q}\bar{\alpha}_2\bar{\alpha}_2} < 0$
 is of the shape with $q_{\boldsymbol{q}\alpha_2}(t_p) \le 0 $, though not shown here.

Prior to closing this subsection, some remarks are made on the allocation of various time-constants shown in Fig.~\ref{fig8}.
Asymmetry in the FR spectra is successfully explained  under such a condition that these  time-constants line up in this order, as demonstrated above.
It is necessary to bear in mind that  there is an additive restriction on it that $T^\prime_{\boldsymbol{q}12}  \ll T^\prime_{12}$.
In the temporal region of  $t_p< T^\prime_{\boldsymbol{q}12}$,
the contribution of 
$\tilde{\chi}_{\boldsymbol{q}}(t_p;\omega)$  to $\bar{A}_{\boldsymbol{q}}(t_p,\omega)$ is dominant over 
that of $\tilde{\chi}^\prime_{\boldsymbol{q}}(t_p;\omega)$ until $t_p $ approximately equal to  $T^\prime_{\boldsymbol{q}12}$, and the observed spectra would just show structureless continuum, as shown in Fig.~\ref{fig9}(a).
After this, in the temporal region of  $T^\prime_{\boldsymbol{q}12} <t_p \ll T^\prime_{12}$, it is likely that the asymmetry is observed even without non-adiabatic couplings,  as shown in Fig.~\ref{fig9}(b) for Si,
whereas this asymmetry vanishes in the region of  $t_p \gtrsim T^\prime_{12}$, as shown in Fig.~\ref{fig9}(c).
Thus, the above restriction seems crucial to the manifestation of asymmetry in the FR profile.
On the other hand,   in the case of $T^\prime_{\boldsymbol{q}12}  \sim T^\prime_{12}$, $\bar{A}_{\boldsymbol{q}}(t_p,\omega)$ would no longer show asymmetry in the whole region of $t_p$.

\begin{figure}[tb]
\begin{center}
\includegraphics[width=7.0cm,clip]{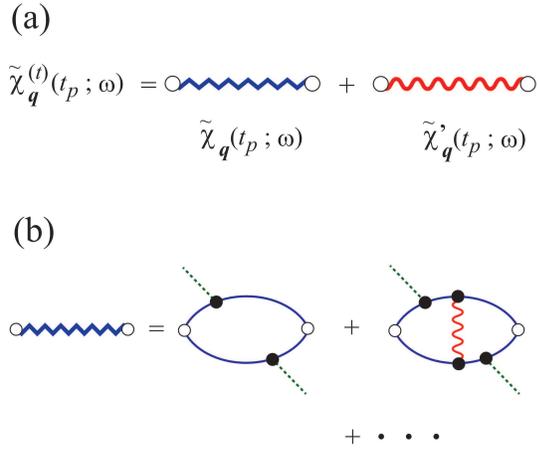}
\caption{(Color online) 
A diagrammatic representation of the retarded longitudinal susceptibility.
(a) $\tilde{\chi}^{(t)}_{\boldsymbol{q}}(t_p;\omega)$ of Eq.~(\ref{tildechitot2}) represented 
in a diagrammatic manner as a sum of $\tilde{\chi}_{\boldsymbol{q}}(t_p;\omega)$ and $\tilde{\chi}^\prime_{\boldsymbol{q}}(t_p;\omega)$.
(b) A many-body perturbation expansion of $\tilde{\chi}_{\boldsymbol{q}}(t_p;\omega)$ with respect to the number of internal phonon lines represented by a red wavy thin-line.
}
\label{fig14}
\end{center}
\end{figure}

\subsection{Interpretation of CP generation dynamics}
\label{sec3C}

The expression of Eq.~(\ref{tildechitot2}) is depicted in a diagrammatic manner in Fig.~\ref{fig14}(a).
Here, a blue  angled-wavy thick-line and a red wavy thick-line represent 
$\tilde{\chi}_{\boldsymbol{q}}(t_p;\omega)$ and $\tilde{\chi}^\prime_{\boldsymbol{q}}(t_p;\omega)$, respectively, and 
an open circle represents a vertex attributed to an interaction with a probe laser.
As shown in Eqs.~(\ref{chi}) and (\ref{chiprime}), these blue and red thick-lines correspond to a retarded density-density correlation function and a retarded phonon Green function, respectively.
Figure~\ref{fig14}(b) shows the many-body perturbation expansion of the diagram of $\tilde{\chi}_{\boldsymbol{q}}(t_p;\omega)$ with respect to the number of internal phonon lines represented by a red wavy thin-line, which corresponds to a retarded free phonon-Green function.
Here, a loop depicted by a blue solid thin-line represents a polarization function, where
this line represents a retarded Hartree-Fock Green function of electron with an effect of Coulomb exchange,\cite{fetter}
a dotted green line represents an external line relevant to  a pump laser, and
a filled circle represents a vertex attributed  to either an interaction with the pump laser  or an electron-LO-phonon interaction.
It is understood  that both of propagators of electron and hole are depicted  just by  this single blue  solid thin-line without distinction for the sake of simplicity.
As regards the second diagram in the right-hand side of the expression of Fig.~\ref{fig14}(a) for $\tilde{\chi}^\prime_{\boldsymbol{q}}(t_p;\omega)$, the many-body perturbation expansion of it does not seem straightforward, since the  retarded phonon Green function of concern is pertinent to 
a retarded density-density correlation function with respect to charge density of ionic core.
Below, the understandings of underlying physics of both of $\tilde{\chi}_{\boldsymbol{q}}(t_p;\omega)$ and $\tilde{\chi}^\prime_{\boldsymbol{q}}(t_p;\omega)$ are deepened based on the many-body perturbation theory for the former,\cite{mahan} and on the group theory for the latter,\cite{inui} respectively.

\begin{figure}[tb]
\begin{center}
\includegraphics[width=7.0cm,clip]{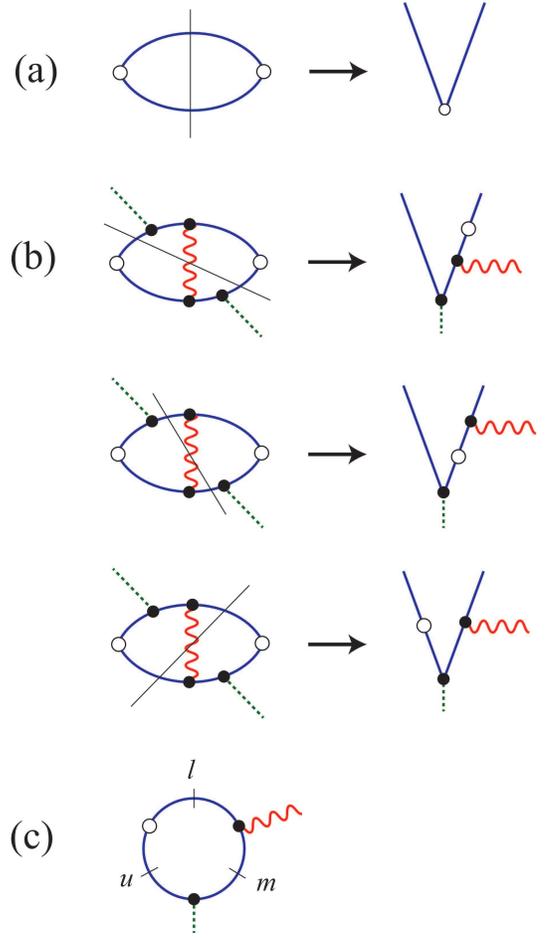}
\caption{(Color online) Diagrammatic representation of the partition-procedure of taking a transition process out of a polarization function.
The meanings of  every line and circle are the same as in Fig.~\ref{fig14}(b).
(a)  An example of this procedure  in the case of a transition process of interband photoabsorption generated by a probe laser within linear optical response.
(b) The procedure in the case of the transient induced photoemission process of concern.
(c) A diagram of a vibrational Raman process. The meanings of labels of $u$, $m$, and $l$ correspond to the diagrams of panel (b) lined up in a sequential order from the upper diagram to the lower one.
}
\label{fig15}
\end{center}
\end{figure}

Bearing in mind that ${\rm Im}\/\tilde{\chi}_{\boldsymbol{q}}(t_p;\omega)$ is in proportion to ${\rm Im}\/\tilde{D}^R_{\boldsymbol{q}}(t_p;\omega)$, as shown in Eq.~(\ref{chi}),
the former quantity is provided by replacing each propagator  incorporated in $\tilde{D}^R_{\boldsymbol{q}}(t_p;\omega)$ by the associated imaginary part that corresponds to a spectral function,
where $\tilde{D}^R_{\boldsymbol{q}}(t_p;\omega)$ represents the Fourier-transform of $D^R_{\boldsymbol{q}}(t_p+\tau,t_p)$ with respect to $\tau$.
One of the most familiar  problems  exemplifying this statement would be posed by 
the procedure of obtaining interband linear-absorption spectra from a retarded longitudinal susceptibility obtained within the linear optical response.
In the procedure of concern,  an energy-denominator of electron-hole propagator is replaced by an energy-on-shell contribution of it, which is equivalent to a delta-function providing  energy-conservation in a final state of  an electron-hole pair produced by photoabsorption.\cite{belitsky}
This procedure is simply shown  in a diagrammatic manner by Fig.~\ref{fig15}(a), in which the associated lowest-order polarization function, namely, a retarded Hartree-Fock Green function of electron,
 is partitioned to provide a diagram corresponding to a transition process of an electron-hole excitation generated  by a probe laser; the resulting diagram is right-arrowed from the original diagram relevant to the polarization function.
 Here, a black solid line represents the partition-procedure of taking an energy on-shell contribution out of the polarization function, and this function is represented by a loop depicted by a blue solid lines with a vertex represented by an open circle, similarly to Fig.~\ref{fig14}.

The above partition-procedure is applied to the second diagram in the right-hand side of  Fig.~\ref{fig14}(b) for $\tilde{\chi}_{\boldsymbol{q}}(t_p;\omega)$ in order to understand more details of the transient induced photoemission process of concern,\cite{belitsky} where this diagram is referred to as the original diagram.
In passing,  the first diagram in the right-hand side of  the same figure does not come into play because of no participation of a phonon.
As shown in Fig.~\ref{fig15}(b), there are  three ways of partitioning the original diagram; where the meanings of  every line and circle are the same as in Fig.~\ref{fig14}(b).
The resulting three diagrams that are right-arrowed are the leading contribution to the  transition processes under consideration in the sense of a perturbation expansion, where every diagram includes 
three vertexes -- indicated by one open circle and two filled circles ---  attributed to a distinct interaction each.
The upper diagram of these three ones shows the transition of electron-hole excitation resulting from
an interaction of the ground state with a pump laser, followed by an interaction of electron (hole) with an LO-phonon, and then an interaction of electron (hole) with a probe laser in this chronological order.
The middle diagram is the same as the upper one but with the inverse of the chronological order between the interactions with an LO-phonon and a probe laser.
The lower diagram shows the transition of electron-hole excitation resulting from
an interaction of the ground state with a pump laser, followed by both of  the interaction of electron (hole) with an LO-phonon and that of hole  (electron)  with a probe laser.

Figure~\ref{fig15}(c) shows a diagram of a vibrational Raman process, which is generated by irradiation of a pump laser, accompanying photoemission induced by a probe laser and emission of an LO-phonon.\cite{yu}
It is seen that applying the partition-procedure mentioned above to the propagator of electron labeled as $u$ leads straightforward to the right-arrowed diagram shown in the upper panel of Fig.~\ref{fig15}(b).
In the same manner as this, the partition-procedures with respect to the propagators labeled as $m$ and $l$ result in the right-arrowed diagrams shown in the middle and lower panels of Fig.~\ref{fig15}(b).
Accordingly, it would be concluded  that the transition process regarding $\tilde{\chi}_{\boldsymbol{q}}(t_p;\omega)$ is caused by the vibrational Raman process accompanying real electronic excitation, differing from a usual vibrational Raman process in that the latter process involves just virtual, namely, non-resonant  electronic-excitation.
Such real electronic excitation is introduced to the vibrational Raman process by taking just the energy-on-shell contribution from a propagator of electron.

\begin{figure}[tb]
\begin{center}
\includegraphics[width=7.0cm,clip]{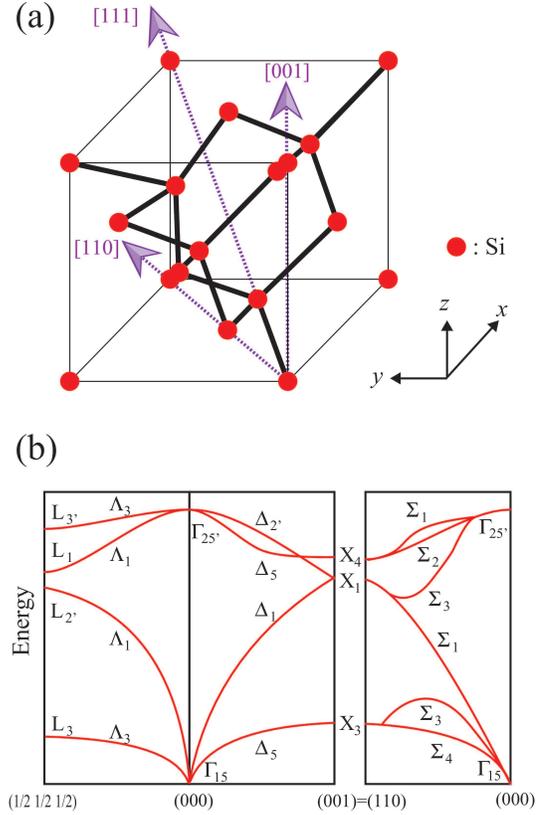}
\caption{(Color online) (a) The crystal structure of Si (the diamond structure).
Here, each Si atom is depicted by a red filled circle, and an arrow with a purple dotted line indicates the direction of electric field of a pump laser; the representative three directions of $[001], [111]$, and $[110]$ are chosen.
(b) Schematic phonon energy-dispersion curve in Si along high-symmetry axes.
The number in the abscissa represents Bloch momentum  in the unit of $2\pi/d$ with $d$ as a lattice constant.
}
\label{fig16}
\end{center}
\end{figure}

Next, the transition process relevant to $\tilde{\chi}^\prime_{\boldsymbol{q}}(t_p;\omega)$ depicted in Fig.~\ref{fig14}(a) is examined based on the group theory.
As well known,  a $\Gamma_4$ optical phonon of GaAs -- classified to the space group ${\rm T_d^2\:(F\bar{4}3m)}$ --  can be directly excited by an infrared photon via an electric dipole transition,
while a $\Gamma_{25^\prime}$ optical phonon of Si -- classified  to the space group ${\rm O_h^7\:(Fd3m)}$ -- is not infrared-active because of the presence of inversion symmetry.
This is the reason why in $\bar{A}_{\boldsymbol{q}}(t_p; \omega)$ of GaAs, the background contribution from $\tilde{\chi}^\prime_{\boldsymbol{q}}(t_p;\omega)$ is of the comparable order with the contribution  from $\tilde{\chi}_{\boldsymbol{q}}(t_p;\omega)$ even in the case of $t_p=15$ fs, as shown in Fig.~\ref{fig10}(a).
On the other hand, in the case of Si, the contribution from $\tilde{\chi}^\prime_{\boldsymbol{q}}(t_p;\omega)$ is negligibly small compared to that from $\tilde{\chi}_{\boldsymbol{q}}(t_p;\omega)$, as  shown in Fig.~\ref{fig9}(a).
However, the former contribution to $\bar{A}_{\boldsymbol{q}}(t_p; \omega)$ never vanishes, and  with the increase of  $t_p$, this becomes more dominant, leading to the key result of the manifestation of transient FR.

In what follows, the transition mechanism of $\bar{A}_{\boldsymbol{q}}(t_p; \omega)$ attributed to $\tilde{\chi}^\prime_{\boldsymbol{q}}(t_p;\omega)$ is scrutinized exclusively in Si.
Despite the space group ${\rm O_h^7}$ being non-symmorphic, it is enough to consider a factor group ${\rm O_h^7}/T$ with respect to the translational group $T$, since no attention is paid to Brillouin-zone edge-points hereafter.
The physical system to be concerned is a crystal with application of a pump laser in a certain direction, as schematically represented in Fig.~\ref{fig16}(a), rather than the free crystal of Si belonging to the above  factor group;\cite{yu} the former crystal is termed as a {\it dressed} crystal.
On the occasion that an electric field of the laser is applied in the three representative directions of  $[001]$, $[111]$, and $[110]$, symmetry of the {\it dressed} crystal is lowered from the original point group ${\rm O_h}$ -- equivalent to ${\rm O_h^7}/T$ -- into point groups of ${\rm C_{4v}}$, ${\rm C_{3v}}$, and ${\rm C_{2v}}$, respectively.
Irreducible representations subduced  from the most concerned irreducible representation $\Gamma_{25^\prime}$ of ${\rm O_h}$ to the respective point groups are 
readily obtained as follows:\cite{inui}
\begin{equation}
\Gamma_{25^\prime} \downarrow {\rm C_{4v}} = \Delta_{2^\prime}+\Delta_5,
\label{c4v}
\end{equation}
\begin{equation}
\Gamma_{25^\prime} \downarrow {\rm C_{3v}} = \Lambda_1+\Lambda_3,
\label{c3v}
\end{equation}
and
\begin{equation}
\Gamma_{25^\prime} \downarrow {\rm C_{2v}} = \Sigma_1+\Sigma_2 +\Sigma_3.
\label{c2v}
\end{equation}
Among the subduced  representations thus obtained, the irreducible representations of 
$\Lambda_1$, $\Lambda_3$, and $\Sigma_3$ are consistent with the symmetry of an ionic momentum operator; that is, $\Lambda_1:\{z\}$, $\Lambda_3:\{x\pm iy\}$, and $\Sigma_3:\{z\}$, where $\Lambda_3$ is a double-valued representation, and the others are single-valued ones.
This result implies that the {\it dressed} crystal can be infrared-active and  optically deexcited through an emission process induced by an infrared laser, differing a lot from the free crystal being infrared-inactive.

Equations~(\ref{c4v})-(\ref{c2v}) are interpreted by consulting a phonon energy-dispersion diagram of Si shown schematically in Fig.~\ref{fig16}(b).\cite{yu}
The subduced representations obtained here are in harmony with the compatibility relations with respect to  $\Gamma_{25^\prime}$ point.\cite{inui}
To be specific, for instance, the $\boldsymbol{k}$-group $\mathcal{G}_\Lambda$ relevant to  $\Lambda$ point ($\boldsymbol{k}_\Lambda$) along the  $(111)$-axis of Bloch momentum, namely, the direction of  ${\rm L}$ point, is a subgroup of the $\boldsymbol{k}$-group $\mathcal{G}_\Gamma$ relevant to  $\Gamma$ point ($\boldsymbol{k}_\Gamma=\boldsymbol{0}$) at the zone center.
Here, the symmetry lowering from $\boldsymbol{k}_\Gamma$ to $\boldsymbol{k}_\Lambda$ along the  $(111)$-axis
makes a threefold degenerate level  $\Gamma_{25^\prime}$ lifted into non-degenerate level $\Lambda_1$ and a twofold degenerate level $\Lambda_3$.
The similar statement also holds for the subgroups of  $\mathcal{G}_\Delta$ and $\mathcal{G}_\Sigma$.

According to the above discussion of the symmetry of the dresses crystal in Si, the degree of magnitude of the symmetry lowering from $\Gamma_{25^\prime}$, for instance, to $\Lambda_1$ is of the order of the momentum change of $|\boldsymbol{q}|$ with $\boldsymbol{q}\equiv\boldsymbol{k}_\Lambda-\boldsymbol{k}_\Gamma$.
In reality, the change of momentum results from spatial inhomogeneity entailed by the formation of polarized charge induced by  the pump laser.\cite{pfeifer2,scholz1,sabbah1,hase1}
That is, the generation of such polarization leads to the breaking of spatial inversion symmetry.
For this reason, it is speculated that the intensity of $\bar{A}_{\boldsymbol{q}}(t_p; \omega)$ due to the contribution from $\tilde{\chi}^\prime_{\boldsymbol{q}}(t_p;\omega)$ in Si is by the order of $|\boldsymbol{q}|^2$ smaller than
that in GaAs, as seen from Figs.~\ref{fig9} and \ref{fig10}, though the {\it dressed} crystal of Si is infrared-active.
Therefore, as regards a transition process governing $\tilde{\chi}^\prime_{\boldsymbol{q}}(t_p;\omega)$,
this is considered to be an electric-dipole transition in the {\it dressed} crystal with emission of, for instance, a $\Lambda_1$ optical phonon.
In other words, this is also considered to be an electric quadrupole transition in the original crystal  with emission of a $\Gamma_{25^\prime}$ optical phonon, as long as $\boldsymbol{q} \approx \boldsymbol{0}$;
actually, the irreducible representation is consistent with the symmetry $\{xy, yz, zx \}$.\cite{yu,inui}
A significant role of  the electric quadrupole transition is also addressed in optical second-harmonic generation from Si, in which the associated reflection with anisotropic response is governed by optically nonlinear susceptibility of  both of bulk and surface layer.\cite{tom}
Further,  it is reported that high density of excited carriers by a strong fs-laser-pulse induces lattice instability of Si and GaAs due to LO-phonon distortions in addition to transverse acoustic-phonon distortions.\cite{stampfli}
It is remarked that the results thus obtained from the group-theoretical point of view remain unaltered in the case of a cubic crystal modeled in the actual calculations of spectra, as mentioned in
Sec.~\ref{sec2A1}.

\subsection{Power spectra of LO-phonon-displacement function}
\label{sec3D}

Both of
Fourier-transforms of a phonon displacement function $\tilde{Q}_{\boldsymbol{q}}(\omega)$   and the related power spectra $S_{\boldsymbol{q}}(\omega)$ of Si and GaAs as  functions of frequency $\omega$ are taken into account.
By comparing Eqs.~(\ref{Pr}) and (\ref{P}), it is readily recognized that both functions of 
$\Pi ^{(r)}_{\boldsymbol{q}\alpha_2}(t_p+\tau,t_p)$ and
$P_{\boldsymbol{q}\alpha_2}(t,t_D)$
are akin each other, and thus,  basic properties of $\tilde{Q}_{\boldsymbol{q}}(\omega)$ would be similar to those of $\tilde{\chi}^\prime_{\boldsymbol{q}}(t_p;\omega)$.
To make this statement more specific, Eq.~(\ref{tildeQ}) is represented as
\(\displaystyle
\tilde{Q}_{\boldsymbol{q}}(\omega)
\approx
\int^\infty_{t_D} e^{-i\omega t}  Q_{\boldsymbol{q}}(t) dt
\equiv
e^{i{\rm Im}\/\mathfrak{I}^{(D)}_{\boldsymbol{q}\alpha_2\alpha_2}}\:\tilde{Q}^\prime_{\boldsymbol{q}}(\omega),
\)
where the small contribution of an interval $[0, t_D]$ to the Fourier integral is neglected, and
$\tilde{Q}^\prime_{\boldsymbol{q}}(\omega)$ is defined in view of
Eq.~(\ref{upsilonapp}).
Thus, it can be shown that ${\rm Re}\/\tilde{Q}^\prime_{\boldsymbol{q}}(\omega)$ and ${\rm Im}\/\tilde{Q}^\prime_{\boldsymbol{q}}(\omega)$ correspond to 
 ${\rm Im}\/\tilde{\chi}^\prime_{\boldsymbol{q}}(t_p;\omega) \propto \bar{A}_{\boldsymbol{q}}(t_p,\omega)$
 and ${\rm Re}\/\tilde{\chi}^\prime_{\boldsymbol{q}}(t_p;\omega)$, respectively, and $S_{\boldsymbol{q}}(\omega)$ is approximated as proportional to $\left|\tilde{Q}^\prime_{\boldsymbol{q}}(\omega)\right|^2$ in place of Eq.~(\ref{spectraS2}).
It is remarked that
such correspondence is no longer correct for $t_p \gg T_{\boldsymbol{q}12},\:t_C$.
Therefore, the transient FR dynamics, which is accompanied by the CP generation occurring at relatively  early time,  is properly reflected on  ${\rm Re}\/\tilde{Q}^\prime_{\boldsymbol{q}}(\omega)$.

\begin{figure}[tb]
\begin{center}
\includegraphics[width=7cm,clip]{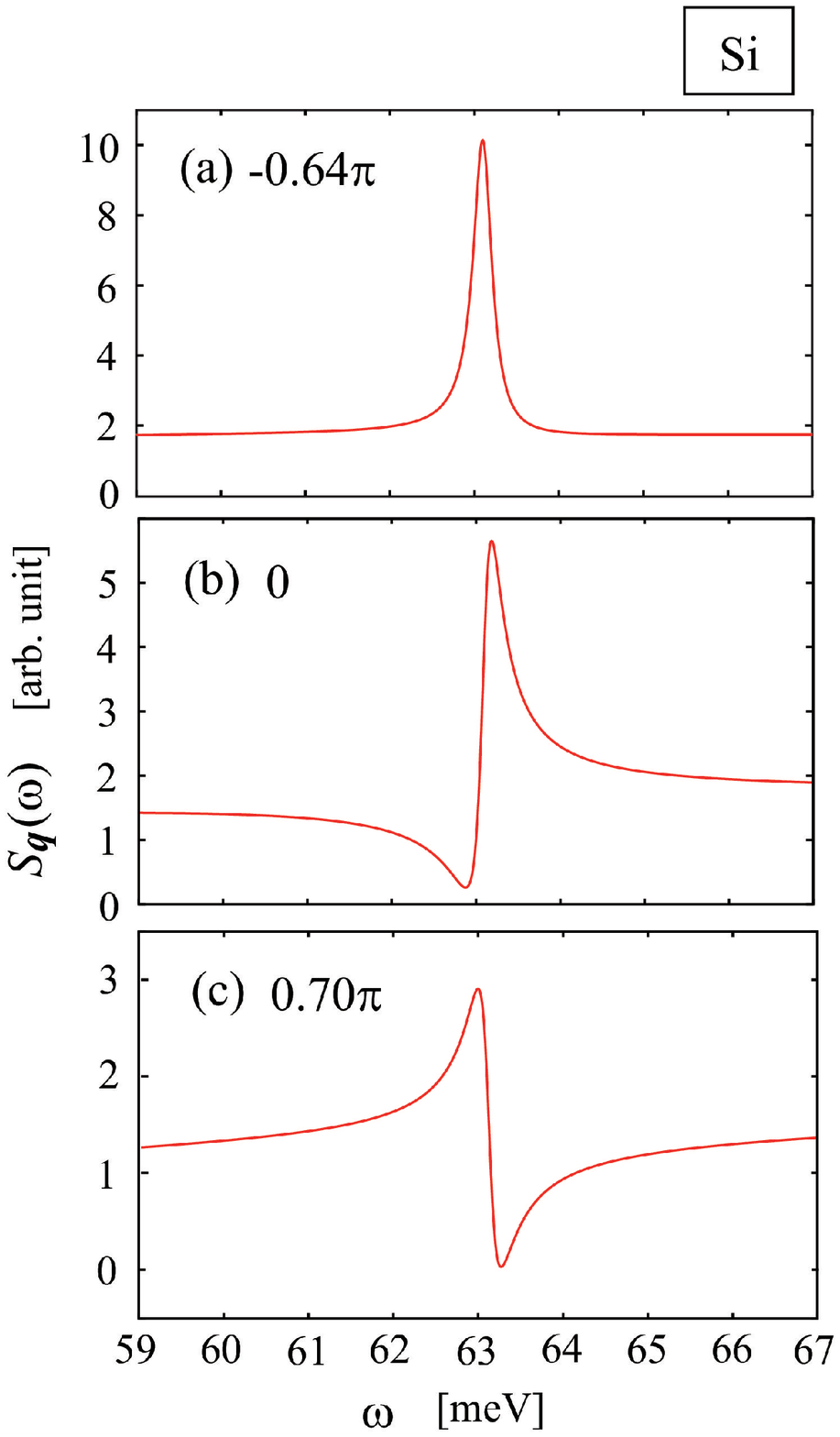}
\caption{(Color online) Power spectra $S_{\boldsymbol{q}}(\omega)$ of Si as a function of frequency $\omega$ (in the unit of meV)  with  ${\rm Im}\/\mathfrak{I}^{(C)}_{\boldsymbol{q}\bar{\alpha}_2\bar{\alpha}_2}$ of
(a) $-0.64\pi$, (b) 0, and (c) $0.70\pi$. 
}
\label{fig18}
\end{center}
\end{figure}

\begin{figure}[tb]
\begin{center}
\includegraphics[width=7.0cm,clip]{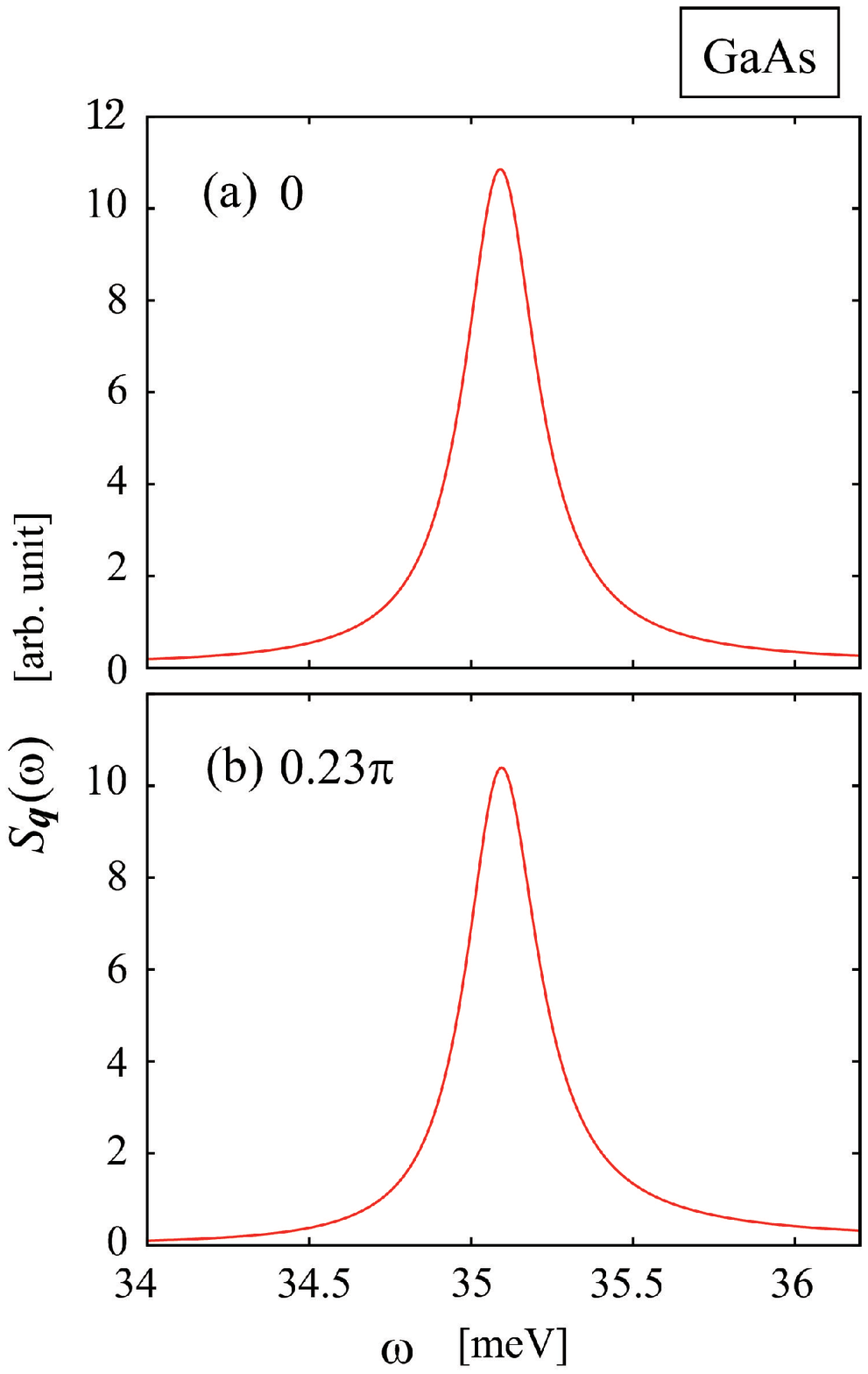}
\caption{(Color online) The same as Fig.~\ref{fig18} but for GaAs with ${\rm Im}\/\mathfrak{I}^{(C)}_{\boldsymbol{q}\bar{\alpha}_2\bar{\alpha}_2}$ of
(a) 0  and (b) $0.23\pi$. 
}
\label{fig20}
\end{center}
\end{figure}

Figure~\ref{fig18} shows the power spectra $S_{\boldsymbol{q}}(\omega)$ as a function of $\omega$.
Here, the asymmetric  spectral shape shown in each panel is reminiscent of FR profile, and this  would be characterized by another asymmetry parameter $\bar{q}_{\boldsymbol{q}\alpha_2}$ with mind to the fact that
this does not always match the Fano $q$-parameter $q_{\boldsymbol{q}\alpha_2}(t_p)$ of $\bar{A}_{\boldsymbol{q}}(t_p,\omega)$.
For instance, with ${\rm Im}\/\mathfrak{I}^{(C)}_{\boldsymbol{q}\bar{\alpha}_2\bar{\alpha}_2}=0.70\pi $, $S_{\boldsymbol{q}}(\omega)$ shown in Fig.~\ref{fig18}(c) has a peak followed by a dip with  $\bar{q}_{\boldsymbol{q}\alpha_2} < 0$, whereas 
$\bar{A}_{\boldsymbol{q}}(t_p,\omega)$ shown in Fig.~\ref{fig12}(c) indicates almost window resonance with  $q_{\boldsymbol{q}\alpha_2}(t_p) \approx 0$.
Unlike with this,  with ${\rm Im}\/\mathfrak{I}^{(C)}_{\boldsymbol{q}\bar{\alpha}_2\bar{\alpha}_2}=0$, both of spectra profiles shown in Figs.~\ref{fig12}(b) and  \ref{fig18}(b) look identical each other, namely, 
$\bar{q}_{\boldsymbol{q}\alpha_2} \approx q_{\boldsymbol{q}\alpha_2}(t_p) >0$.
This implies that differing from the close resemblance between ${\rm Re}\/\tilde{Q}^\prime_{\boldsymbol{q}}(\omega)$ and $\bar{A}_{\boldsymbol{q}}(t_p,\omega)$, the detail of the transient FR dynamics is not necessarily reflected on $S_{\boldsymbol{q}}(\omega)$ due to the relatively large contribution from 
${\rm Im}\/\tilde{Q}^\prime_{\boldsymbol{q}}(\omega)$
to $S_{\boldsymbol{q}}(\omega)$.
To be more specific, just the analysis of $S_{\boldsymbol{q}}(\omega)$ is not  enough to precisely understand the transient FR dynamics.
The spectral pattern at each $t_p$ is characterized by 
$q_{\boldsymbol{q}\alpha_2}(t_p)$ with a firm physical meaning of the way of interaction between both states of  LO-phonon and quasi-boson [see Fig.~\ref{fig11}].
However, $\bar{q}_{\boldsymbol{q}\alpha_2}$ is nothing but a fitting parameter of $S_{\boldsymbol{q}}(\omega)$ irrelevant to this physical meaning.

Figure~\ref{fig20} shows $S_{\boldsymbol{q}}(\omega)$ of GaAs as a function of $\omega$.
In marked contrast with that of Si shown in Fig.~\ref{fig18}, all spectra look almost symmetric, namely,  $1/\bar{q}_{\boldsymbol{q}\alpha_2} \approx 0$, irrelevant to the change of parameters of ${\rm Im}\/\mathfrak{I}^{(C)}_{\boldsymbol{q}\bar{\alpha}_2\bar{\alpha}_2}$; this tendency holds correctly for other values of ${\rm Im}\/\mathfrak{I}^{(C)}_{\boldsymbol{q}\bar{\alpha}_2\bar{\alpha}_2}$, though not shown here.
This does not imply that the transient FR is absent from the concerned system of GaAs, since
$\bar{A}_{\boldsymbol{q}}(t_p,\omega)$ at $t_p=$ 65 fs can be asymmetric with non-vanishing values of $1/q_{\boldsymbol{q}\alpha_2}(t_p)$ [see Fig.~\ref{fig13}(b)].

Next, discussion is made on the initial phase $\theta_{\boldsymbol{q}}(t_L)$.
According to Eq.~(\ref{thetatL}), this is composed of three phases with different origins as follows:
$\pi/2-\xi_{\boldsymbol{q}}(t_L)$, $\Delta\alpha_{\boldsymbol{q}}\approx A_L$, and $\upsilon_{\boldsymbol{q}\bar{\alpha}_2}(t_L,t_0)$.
The first one, represented as $\theta^{(0)}_{\boldsymbol{q}}(t_L)\equiv \pi/2-\xi_{\boldsymbol{q}}(t_L)$,
arises exclusively from
$\mathcal{D}_{\boldsymbol{q}\alpha_2}(t_L,t_D)$ given by 
Eq.~(\ref{calDtL}).
As stated in Sec.~\ref{sec2C3},
$\theta^{(0)}_{\boldsymbol{q}}(t_L)$ approximately equals either
zero or $\pi/2$, depending on the type of electron-phonon interaction.
The second one,  given by Eq.~(\ref{Deltaalpha}), is relevant to properties of pump pulse-laser.\cite{hasex,nakamura}
The third one, given by Eq.~(\ref{Upsilonsum}), equals an imaginary part of all diagonal-components of non-adiabatic coupling integrated over the whole $t$-region of $(t_0,t_L)$.

Under the limited conditions that $A_L\approx 0$ and $\upsilon_{\boldsymbol{q}\bar{\alpha}_2}(t_L,t_0) \approx 0$,
$\theta_{\boldsymbol{q}}(t_L)$, which  is approximated as $\theta^{(0)}_{\boldsymbol{q}}(t_L)$,
almost vanishes for Si, where this is governed by the deformation potential interaction.
This result is reminiscent of the ISRS model of CP generation dynamics (see Sec.~\ref{sec1}).
On the other hand, $\theta^{(0)}_{\boldsymbol{q}}(t_L)$ almost equals $\pi/2$ modulus $\pi$ for GaAs, where this  is mostly governed by the Fr\"{o}hlich interaction.
This result is readily reminiscent of the DECP model of CP generation dynamics (see Sec.~\ref{sec1}).
It seems that $\theta^{(0)}_{\boldsymbol{q}}(t_L)$ is irrelevant to the manifestation of the transient FR
 in contrast with  the aforementioned result that  the temporal variation of $\xi_{\boldsymbol{q}}(t)$ gives rise to the asymmetry in $S_{\boldsymbol{q}}(\omega)$, as shown in  Fig.~\ref{fig18}(b).

The numerical values of $A_L$  employed here for Si and GaAs are set to be $0.12\pi(=21.6^\circ)$ and $0.20\pi(=35.1^\circ)$, respectively.\cite{SMparameter}
As regards the effect of $\upsilon_{\boldsymbol{q}\bar{\alpha}_2}(t_L,t_0)$, unfortunately, the complete evaluation of this phase is beyond the scope of the numerical method employed in the actual calculations,
though this would be enabled by more sophisticated numerical methods suitable for  the present PQ theory [see Sec.~\ref{sec4}].
Although throughout Sec.~\ref{sec3},
$\upsilon_{\boldsymbol{q}\bar{\alpha}_2}(t_L,t_0)$ is simply approximated by
Eq.~(\ref{upsilonapp}) consisting of  the two terms, namely, ${\rm Im}\/\mathfrak{I}^{(D)}_{\boldsymbol{q}\bar{\alpha}_2\bar{\alpha}_2}$ and ${\rm Im}\/\mathfrak{I}^{(C)}_{\boldsymbol{q}\bar{\alpha}_2\bar{\alpha}_2}$,
just by the second term are crucially affected the FR profiles of both $\bar{A}_{\boldsymbol{q}}(t_p,\omega)$ and $S_{\boldsymbol{q}}(\omega)$, as stated above.
Nevertheless, the first term has to be also incorporated with $\theta_{\boldsymbol{q}}(t_L)$, differing from
 $\bar{A}_{\boldsymbol{q}}(t_p,\omega)$ and $S_{\boldsymbol{q}}(\omega)$.
Given the reported value of $\theta_{\boldsymbol{q}}(t_L)$  for Si is roughly $0.3\pi$,\cite{hase1}$\upsilon_{\boldsymbol{q}\bar{\alpha}_2}(t_L,t_0)$ is estimated to $0.2\pi$.
Aside from the above-mentioned difficulty of the {\it ab initio} evaluation of $\upsilon_{\boldsymbol{q}\bar{\alpha}_2}(t_L,t_0)$, Eq.~(\ref{thetatL}) provides a well-defined microscopic description of the initial phase without a phenomenological ansatz such as an artifact which hybridizes the two extreme models of ISRS and DECP.\cite{garrett,merlin1,stevens,bragas,riffe1}
The explicit expressions 
for the Fr\"{o}hlich interaction and the deformation potential interaction are given by
$\theta^F_{\boldsymbol{q}}(t_L)$ and $\theta^D_{\boldsymbol{q}}(t_L)$ in Eqs.~(\ref{thetatL1}) and (\ref{thetatL2}), respectively.
It is seen from these expressions that the asymmetry parameter $\bar{q}_{\boldsymbol{q}\alpha_2}$ is not incorporated in the initial phase obtained here in an explicit fashion, differing from the claims made in Refs.~\onlinecite{misochko7} and \onlinecite{riffe2};
recall that $\bar{q}_{\boldsymbol{q}\alpha_2}$ does not always equal Fano's asymmetry parameter.

\subsection{Comparison with other studies}
\label{sec3E}

The present work is compared with the three studies by Hase, {\it et al.},\cite{hase1} Lee, {\it et al.},\cite{lee} and Riffe.\cite{riffe2}
One begins with  a comparison  with the pioneering work by Hase, {\it et al.}, 
 in which transient electro-optic reflectivity signals due to the CP generation were measured in a lightly $n$-doped Si crystal.\cite{hase1}
In the associated continuous-wavelet-transformed spectra, an asymmetric profile  was found around 50 fs immediately after the onset of the irradiation of pump pulse.
This conspicuous phenomenon was suggestive of the transient manifestation of  quantum-interference  between excited electrons and an LO-phonon leading to FR.
Further, given the situation that this coupling was such a strong interaction that the electron and phonon were considered as a composite particle, the authors supposed the birth of PQ.
Actually, the present theoretical model is absolutely based on this supposition, though still naive.
As illustrated in Sec.~\ref{sec3}, the PQ picture thus developed can
successfully explain the manifestation of the FR just under the condition that various time-constants line up in the order shown in Fig.~\ref{fig8}.

The transient reflectivity signals accompanying  CP generation were observed in the experimentally setup $\Gamma_{25^\prime}$-geometry.
This geometry would correspond to the $\Lambda_1$ irreducible representation in the
symmetry $\Gamma_{25^\prime} \downarrow {\rm C_{3v}}$ given in  Eq.~(\ref{c3v}).
This vibrational state of {\it dressed} crystal
can be optically deexcited through an emission process induced by an infrared laser polarized in the $[001]$-direction.
In the actual measurements, CP signals are caused by a lot of carrier transitions along the $\Lambda$ and $\Delta$ directions in the Brillouin zone as well as near $\Gamma$ point. 
Since the energy dispersions of conduction bands along  the $\Lambda$ (L-valley) direction and the $\Delta$ (X-valley) direction are almost parallel to the energy dispersion of the heavy-hole valence band, 
relative energy of an electron-hole pair, given by 
$w_{cv\boldsymbol{kq}}$ of Eq.~(\ref{w}),  is almost dispersionless, namely, independent of $\boldsymbol{k}$.
Thus, it is speculated that just the electron excitation near $\Gamma$ point forms broad band spectra due to  electron-hole continuum states with $\boldsymbol{k}$-dispersion, giving rise to transient FR; this
differs  a lot from the transitions along  the $\Lambda$ and $\Delta$ directions.

The phonon displacement function $Q_{\boldsymbol{q}}(t)$  of Eq.~(\ref{Q})  is considered as equivalent to the transient reflectivity signal of concern.
The parameter  $\bar{q}_{\boldsymbol{q}\alpha_2}$ obtained by a fitting procedure in this experiment was a negative value.
As stated in Sec.~\ref{sec1}, in FR spectra generated by incoherent Raman scattering, a sign of Fano's $q$-parameter  $q^{(F)}$ is indicative of the way of carrier-LO-phonon interaction.
On the other hand, in the present CP generation, 
there is no contribution of excited electron at $\Gamma$-point to an electron-LO-phonon deformation potential interaction because of restriction of symmetry, namely, $g_{c\boldsymbol{0}}^D=0$.\cite{SMparameter}
Thus, the phonon is exclusively coupled with holes at $\Gamma$-point in this transient system that is, so to speak, an optically $n$-doped Si.
For the following reason, it does not imply that
the sign of $\bar{q}_{\boldsymbol{q}\alpha_2}$ corresponds to that of  $q^{(F)}$.
The sign  of $\bar{q}_{\boldsymbol{q}\alpha_2}$ as well as that 
of $q_{\boldsymbol{q}\alpha_2}(t_p)$ is not simply determined by a sign of carrier density, differing a lot from $q^{(F)}$;
the former sign strongly depends on both of carrier-density-dependent phase factor $\vartheta_{\boldsymbol{q}}(\tau)$
of Eq.~(\ref{vartheta}) 
and  the non-adiabatic coupling ${\rm Im}\/\mathfrak{I}^{(C)}_{\boldsymbol{q}\alpha_2\alpha_2}$
independent of it.

As described above, the calculated results based on the transient PQ picture employed here seem in harmony to a certain extent with the experimental results of Hase, {\it  et al.}
Nevertheless, as it stands, this fact does not necessarily imply that the PQ introduced here is a real entity.
Unfortunately, this issue is beyond the scope of the present study, though quite challenging.

Lee, {\it  et al.} presented the first theoretical approach to the FR problem associated with the CP generation.\cite{lee}
They calculated a phonon displacement function similar to $Q_{\boldsymbol{q}}(t)$ by solving time-dependent Schr\"{o}dinger equations for the system of  GaAs.
An asymmetric line-shape was obtained  in the continues-wavelet-transformed spectra of the associated time signals.
Further, they interpreted the underlying dynamics causing the FR as quantum interference between two diagrams corresponding to  the one-phonon Raman process [see the diagram of Fig.~\ref{fig15}(c)].
These results differ from those obtained here in that in the latter,
the transient FR never takes place in $S_{\boldsymbol{q}}(\omega)$ relevant to $Q_{\boldsymbol{q}}(t)$ in GaAs and that the origin of FR observed in Si is not the interference between the Raman diagrams; thus far, FR spectra have not been observed in GaAs in the existing experiments.\cite{misochko1,Hase09}

Riffe proposed a classical Fano oscillator model based
on the Fano-Anderson Hamiltonian, in which a set of coupled oscillators are driven by as many  fictitious  external-forces introduced a posteriori.\cite{riffe2}
According to the present study, this Hamiltonian is read as 
\begin{eqnarray}
\hat{H}_{\boldsymbol{q}}^{(FA)}
&=&
\omega_{\boldsymbol{q}}^{(LO)}c^\dagger_{\boldsymbol{q}}c_{\boldsymbol{q}}
+\sum_\alpha \mathcal{E}_{\boldsymbol{q}\alpha}B^\dagger_{\boldsymbol{q}\alpha}
B_{\boldsymbol{q}\alpha}
\nonumber\\
&&
+
\sum_{\alpha}
\left(
M_{\boldsymbol{q}\alpha}
c_{\boldsymbol{q}} B^\dagger_{\boldsymbol{q}\alpha}
+M^*_{\boldsymbol{q}\alpha}c^\dagger_{\boldsymbol{q}} B_{\boldsymbol{q}\alpha}
\right),
\label{FAH}
\end{eqnarray}
where the quasi-boson is considered as a real boson with real eigenenergy $\mathcal{E}_{\boldsymbol{q}\alpha}$.
Thus, defining displacement functions relevant to quasi-boson and LO-phonon operators as
\(
X^{(+)}_{\boldsymbol{q}}(t)=
\bigl\langle
 c_{\boldsymbol{q}}(t)+ c^\dagger_{-\boldsymbol{q}}(t)
\bigr\rangle/2
\label{Xpm}
\)
and
\(
x^{(+)}_{\boldsymbol{q}\alpha}(t)=
\bigl\langle
 B_{\boldsymbol{q}\alpha}(t)+ B^\dagger_{-\boldsymbol{q}\alpha}(t)
\bigr\rangle/2,
\)
respectively,
the associated equations of motion are readily obtained from Eq.~(\ref{FAH}).
Here, external forces associated with $X^{(+)}_{\boldsymbol{q}}(t)$ and $x^{(+)}_{\boldsymbol{q}\alpha}(t)$
are denoted as $\mathfrak{F}_{\boldsymbol{q}}(t)$ and $\mathfrak{f}_{\boldsymbol{q}\alpha}(t)$, respectively.
These forces just correspond to the fictitious forces introduced by Riffe, which resulted in FR.
In Riffe's work, quasi-boson energy $\mathcal{E}_{\boldsymbol{q}\alpha}$ and an effective coupling 
$M_{\boldsymbol{q}\alpha}$ were assumed independent of time $t$
in addition to $X^{(+)}_{\boldsymbol{q}}(t)$ and $x^{(+)}_{\boldsymbol{q}\alpha}(t)$ being regarded as expectation values with respect to a coherent state.
This assumption is brought to the result that
both of $W_{\boldsymbol{q}\alpha\alpha^\prime}(t)$ of Eq.~(\ref{W}) and
$dM_{\boldsymbol{q}\alpha}(t)/dt$ vanish simultaneously.
Thus, it  is revealed that $\mathfrak{F}_{\boldsymbol{q}}(t)$ and $\mathfrak{f}_{\boldsymbol{q}\alpha}(t)$
 are identified as 
frictional forces due to an effective  damping factor $\gamma^{(B)}_{\boldsymbol{q}\alpha\alpha^\prime}(t)$
of Eq.~(\ref{gammaB}), that is,
\(
\mathfrak{F}_{\boldsymbol{q}}(t)
\approx -\sum_{\alpha,\alpha^\prime} x^{(+)}_{\boldsymbol{q}\alpha}(t)
\/{\rm Im}\/\left[\gamma^{(B)}_{\boldsymbol{q}\alpha\alpha^\prime}(t)M_{\boldsymbol{q}\alpha^\prime}\right]/2
\)
and 
\(
\mathfrak{f}_{\boldsymbol{q}\alpha}(t)
\approx - \mathcal{E}_{\boldsymbol{q}\alpha}\sum_{\alpha^\prime} x^{(+)}_{\boldsymbol{q}\alpha^\prime}(t)
\/{\rm Im}\/\gamma^{(B)}_{\boldsymbol{q}\alpha^\prime\alpha}(t)/2.
\)
In particular, in Si,  given the fact that $M_{\boldsymbol{q}\alpha^\prime}$ and $\gamma^{(B)}_{\boldsymbol{q}\alpha\alpha^\prime}(t)$ are real, 
both forces end up with \(
\mathfrak{F}_{\boldsymbol{q}}(t)
\approx 0
\)
and 
\(
\mathfrak{f}_{\boldsymbol{q}\alpha}(t)
\approx 0.
\)
Therefore, Riffe's results seem incompatible with those based on the PQ model concerned here.

\section{Conclusions}
\label{sec4}

The PQ model is established to apply to CP generation dynamics of semiconductors -- Si and GaAs -- accompanying transient and optically non-linear FR.
Based on this model, a transient dielectric function $\tilde{\epsilon}_{\boldsymbol{q}}(t_p;\omega)$ at probe time $t_p$ is derived and 
the associated induced photoemission spectra $\bar{A}_{\boldsymbol{q}}(t_p;\omega)$ from strongly-excited electron-hole continuum coupled with LO-phonon  are calculated.
The present model succeeds in illustrating  the FR appearing in a flash
immediately after the completion of pump-laser irradiation, where
the resulting spectral profile is determined by $M_{\boldsymbol{q}\alpha}(t)$ and
a diagonal part of  non-adiabatic coupling matrix $I_{\boldsymbol{q}\beta\beta^\prime}(t)$  of Eq.~(\ref{I}).
 It is likely that asymmetric spectra due to FR appear both in Si and GaAs just in the temporal region of 
$ T^\prime_{\boldsymbol{q}12} < t_p  \ll T^\prime_{\boldsymbol{q}12}$.
In particular, in Si, the FR takes place even without $I_{\boldsymbol{q}\beta\beta}(t)$.
This model is also applied to a LO-phonon displacement function to evaluate the power spectra
$S_{\boldsymbol{q}}(\omega)$.
Differing a lot from an asymmetric spectral profile of $S_{\boldsymbol{q}}(\omega)$ in Si, however, the profile  in GaAs is always symmetric.
The present model is nicely in agreement with the experimental results obtained by Hase, {\it  et al.},\cite{hase1} though not consistent with the existing theoretical results.\cite{lee,riffe2}
This can  be  straightforward applicable to CP generation dynamics in heavily $n$-doped and $p$-doped semiconductor systems as well, irrespective of  whether FR manifests itself nor not.

Despite the success attained to some extent here, there is still room of improving the present PQ model, since this does not suit for fully quantitative calculations, as it is.
Actually, an imaginary part of $I_{\boldsymbol{q}\beta\beta}(t)$ is dealt with just as given parameters throughout, though this would be important for determining FR spectra.
There is numerical difficulty of evaluating  $I_{\boldsymbol{q}\beta\beta^\prime}(t)$ due to its spike-like behavior in the vicinity of crossing region between states of $\beta$ and  $\beta^\prime$ [see Eq.~(\ref{I4-1})].
This is the price to be paid for an adiabatic expansion method employed here, which makes it possible to interpret the transient PQ as adiabatic state in a transparent manner.
In order to remedy this difficulty and to develop the theoretical framework so as to bear more quantitative investigation,
one would require a more sophisticated numerical recipe to be substituted for the adiabatic expansion method, for instance, a diabatic-by-sector method\cite{tang,hino6} and a R-matrix propagation method,\cite{hino3,hino4,hino5,burke} aside from heavy numerical-burden
possibly  incurred by these recipes.
Without such improvement, the present theoretical framework would be inadequate for understandings of details of experimental results immediately after the onset of pulse excitation of carriers.\cite{hase9}

There would be another drawback to be pointed out   in the present PQ model.
As shown in Eq.~(\ref{H'}), a laser field is assumed as a classical external field, not a quantized photon field.
This assumption results in  difficulty that contributions of Raman scattering\cite{belitsky} are unable to be incorporated in 
$\tilde{\chi}_{\boldsymbol{q}}(t_p;\omega)$ that is shown diagrammatically in Fig.~\ref{fig14}(b), since an internal photon line -- a retarded photon propagator -- cannot be described in terms of the classical external field.
Therefore, both of real excitation of carriers that is shown diagrammatically in Fig.~\ref{fig15}(b) and virtual excitation of carries due to the Raman process are not dealt with on an equal footing as it is.
To overcome this difficulty, one has to take account of a more expanded composite-particle operator  consisting of photon operator as well as quasi-boson and LO-phonon operators, rather than the PQ operator  of Eq.~(\ref{F-}), though
the resulting problem to be solved would be demanding.

\begin{acknowledgments}
This work was supported by Grants-in-Aids for Scientific Research (C) (Grant No. 23540360 and Grant No. 15K05121)
of the Ministry of Education, Culture, Sports, Science and Technology (MEXT), Japan.  
\end{acknowledgments}




\bibliographystyle{elsarticle-num}


\begin{thebibliography}{00}
\bibitem{ultrafastXIX}K.~Yamanouchi, S.~Cundiff, R.~de~Vivie-Riedle, M.~Kuwata-Gonokami, and L.~DiMauro (Editors),  {\it Ultrafast Phenomena XIX: Proceedings of the 19th International Conference}, Springer Proceedings in Physics {\bf 162},
(Springer-Verlag, Berlin, 2015).

\bibitem{lightscatteringVIII}T.~Dekorsy, G.~C.~Cho, and H.~Kurz, in {\it Light Scattering in Solids VIII: Fullerens, Semiconductor Surfaces, Coherent Phonons}, Topics in Applied Physics {\bf  76}, ed. by M.~Cardona and G.~G\"{u}ntherodt
 (Springer-Verlag, Berlin, 2000) Chap. 4.
 
\bibitem{kuznetsov1}A.~V.~Kuznetsov and C.~J.~Stanton, in {\it Ultrafast Phenomena in Semiconductors}, ed. by K.~T.~Tsen (Springer-Verlag, Berlin, 2001) Chap. 7. 

\bibitem{kuznetsov2}A.~V.~Kuznetsov and C.~J.~Stanton, Phys. Rev. Lett. {\bf 73}, 3243 (1994).

\bibitem{cho1}G.~C.~Cho, W.~K\"{u}tt, and H.~Kurz,  Phys. Rev. Lett. {\bf 65}, 764 (1990).

\bibitem{pfeifer1}T.~Pfeifer, T.~Derkosy, W.~K\"{u}tt, and H.~Kurz,  Appl. Phys. A: Solids Surf.  {\bf 55}, 482 (1992).

\bibitem{dekorsy1}T.~Dekorsy, T.~Pfeifer,  W.~K\"{u}tt, and H.~Kurz,  Phys. Rev. B {\bf 47}, 3842 (1993).

\bibitem{dekorsy2}T.~Dekorsy, H.~Kurz, X.~Q.~Zhou, and K.~Ploog, Appl. Phys. Lett. {\bf 63}, 2899 (1993).

\bibitem{cho2}G.~C.~Cho, H.~J.~Bakker, T.~Dekorsy,  and H.~Kurz,  Phys. Rev. B {\bf 53}, 6904 (1996).

\bibitem{chang}Y.~M.~Chang, L.~Xu, and H.~W.~K.~Tom,  Phys. Rev. Lett. {\bf 78}, 4649 (1997).

\bibitem{misochko1}O.~V.~Misochko, JETP {\bf 92}, 246 (2001).

\bibitem{sabbah1}A.~J.~Sabbah and D.~M.~Riffe, Phys. Rev. B {\bf 66}, 165217 (2002).

\bibitem{hase1}M.~Hase, M.~Kitajima, A.~M.~Constantinescu, and H.~Petek, Nature (London)  {\bf 426}, 51 (2003).

\bibitem{riffe1}D.~M.~Riffe and A.~J.~Sabbah, Phys. Rev. B {\bf 76}, 085207 (2007).

\bibitem{kato}K.~Kato, A.~Ishizawa, K.~Oguri, K.~Tateno, T.~Tawara, H.~Gotoh, M.~Kitajima,  and H.~Nakano, Jpn. J. Appl. Phys. {\bf 48}, 100205 (2009).

\bibitem{nelson}K.~A.~Nelson, D.~D.~Dlott, and M.~D.~Fayer, Chem. Phys. Lett. {\bf  64}, 88 (1979).

\bibitem{bakker1}H.~J.~Bakker, S.~Hunsche, and H.~Kurz, Phys. Rev. Lett. {\bf  69}, 2823 (1992).

\bibitem{bakker2}H.~J.~Bakker, S.~Hunsche, and H.~Kurz, Phys. Rev. B {\bf  50}, 914 (1994).

\bibitem{bakker3}H.~J.~Bakker, S.~Hunsche, and H.~Kurz, Rev. Mod. Phys. {\bf  70}, 523 (1998).

\bibitem{cheng1}T.~K.~Cheng, S.~D.~Brorson, A.~S.~Kazeroonian, J.~S.~Moodera, G.~Dresselhaus, M.~S.~Dresselhaus, and E.~P.~Ippen, Appl. Phys. Lett. {\bf 57}, 1004 (1990).

\bibitem{cheng2}T.~K.~Cheng, J.~Vidal, H.~J.~Zeiger, G.~Dresselhaus, M.~S.~Dresselhaus, E.~P.~Ippen, Appl. Phys. Lett. {\bf 59}, 1923 (1991).

\bibitem{dekorsy3}T.~Dekorsy, H.~Auer, C.~Waschke, H.~J.~Bakker, H.~G.~Roskos, H.~Kurz, V.~Wagner, and P.~Grosse, Phys. Rev. Lett. {\bf 74}, 738 (1995).

\bibitem{hunsche}S.~Hunsche, K.~Wienecke, T.~Dekorsy, and H.~Kurz,  Phys. Rev. Lett. {\bf 75}, 1815 (1995).

\bibitem{hase2}M.~Hase, K.~Mizoguchi, H.~Harima, S.~Nakashima, M.~Tani, K.~Sakai, and M.~Hangyo, Appl. Phys. Lett. {\bf 69}, 2474 (1996).

\bibitem{hase6}M.~Hase, K.~Mizoguchi, H.~Harima,  S.~I.~Nakashima, and K.~Sakai, Phys. Rev. B {\bf 58}, 5448 (1998).

\bibitem{hase3}M.~Hase, K.~Ishioka, M.~Kitajima, K.~Ushida, and S.~Hishita, Appl. Phys. Lett. {\bf 76}, 1258 (2000).

\bibitem{misochko3}O.~V.~Misochko, K.~Sakai, and S.~I.~Nakashima, Phys. Rev. B {\bf 61}, 11225 (2000).

\bibitem{decamp}M.~F.~DeCamp, D.~A.~Reis, P.~H.~Bucksbaum, and R.~Merlin, Phys. Rev. B {\bf 64}, 092301 (2001).

\bibitem{hase4}M.~Hase, M.~Kitajima, S.~I.~Nakashima, and K.~Mizoguchi, Phys. Rev. Lett. {\bf 88}, 067401 (2002).

\bibitem{misochko2}O.~V.~Misochko, M.~Hase, K.~Ishioka, and M.~Kitajima, Phys. Rev. Lett. {\bf 92}, 197401 (2004).

\bibitem{murray}\'{E}.~D.~Murray, D.~M.~Fritz, J.~K.~Wahlstrand, S.~Fahy, and D.~A.~Reis, Phys. Rev. B {\bf 72}, 060301(R)  (2005).

\bibitem{misochko6}O.~V.~Misochko, K.~Ishioka, M.~Hase, and M.~Kitajima, J. Phys.: Condens. Matter {\bf 19}, 156227 (2007).

\bibitem{misochko7}O.~V.~Misochko and M.~V.~Lebedeva, JETP {\bf 120}, 651 (2015).

\bibitem{hase5}M.~Hase, K.~Ishioka, J.~Demsar, K.~Ushida, and M.~Kitajima, Phys. Rev. B {\bf 71}, 184301 (2005).

\bibitem{hase7}M.~Hase, J.~Demsar, and M.~Kitajima, Phys. Rev. B {\bf 74}, 212301 (2006).

\bibitem{li}J.~J.~Li, J.~Chen, D.~A.~Reis, S.~Fahy, and R.~Merlin, Phys. Rev. Lett. {\bf  110}, 047401 (2013).

\bibitem{chwalek1}J.~M.~Chwalek, C.~Uher, J.~F.~Whitaker, G.~Mourou, J.~Agostinelli, and. M.~Lelental, Appl. Phys. Lett. {\bf 57}, 1696 (1990).

\bibitem{chwalek2}J.~M.~Chwalek, C.~Uher, J.~F.~Whitaker, G.~A.~Mourou, and J.~Agostinelli, Appl. Phys. Lett. {\bf 58}, 980 (1991).

\bibitem{albrecht}W.~Albrecht, Th.~Kruse, and H.~Kurz, Phys. Rev. Lett. {\bf 69}, 1451 (1992).

\bibitem{misochko4}O.~V.~Misochko, Phys. Lett. A {\bf 269}, 97 (2000).

\bibitem{misochko5}O.~V.~Misochko, K.~Kisoda, K.~Sakai, and S.~Nakashima, Phys. Rev. B {\bf 61}, 4305 (2000).

\bibitem{bozovic}I.~Bozovic, M.~Schneider, Y.~Xu, R.~Sobolewski, Y.~H.~Ren, G.~L\"{u}pke, J.~Demsar, A.~J.~Taylor, and M.~Onellion, Phys. Rev. B {\bf 69}, 132503 (2004).

\bibitem{mishina}T.~Mishina, K.~Nitta, and Y.~Masumoto, Phys. Rev. B {\bf 62}, 2908 (2000).

\bibitem{ishioka1}K.~Ishioka, M.~Hase, M.~Kitajima, and K.~Ushida, Appl. Phys. Lett. {\bf 78}, 3965 (2001).

\bibitem{watanabe1}K.~Watanabe, N.~Takagi, and Y.~Masumoto, Chem. Phys. Lett. {\bf 366}, 606 (2002).

\bibitem{watanabe2}K.~Watanabe, N.~Takagi, and Y.~Masumoto, Phys. Rev. Lett. {\bf 92}, 57401 (2004).

\bibitem{yan1}Y.~-X.~Yan, E.~B.~Gamble, and K.~Nelson, J. Chem. Phys. {\bf 83}, 5391 (1985).

\bibitem{yan2}Y.~-X.~Yan and K.~Nelson, J. Chem. Phys. {\bf 87}, 6240 (1987).

\bibitem{kutt}W.~K\"{u}tt, W.~Albrecht, and H.~Kurz, IEEE J. Quantum Electron. {\bf 28}, 2434 (1992).

\bibitem{zeiger}H.~J.~Zeiger, J.~Vidal, T.~K.~Cheng, E.~P.~Ippen, G.~Dresselhaus, and M.~S.~Dresselhaus, Phys. Rev. B {\bf 45}, 768 (1992).

\bibitem{garrett}G.~A.~Garrett, T.~F.~Albrecht, J.~F.~Whitaker, and R.~Merlin, Phys. Rev. Lett. {\bf 77}, 3661 (1996).

\bibitem{merlin1}R.~Merlin, Solid State Commun. {\bf 102}, 207 (1977).

\bibitem{hasex}M. Hase, M. Katsuragawa, A. M. Constantinescu, and H. Petek, Nature Photon. {\bf 6},  243 (2012). 

\bibitem{stevens}T.~E.~Stevens, J.~Kuhl, and R.~Merlin, Phys. Rev. B {\bf 65}, 144304 (2002).

\bibitem{bragas}A.~V.~Bragas, C.~Aku-Leh, S.~Costantino, A.~Ingale, J.~ Zhao, R.~Merlin, Phys. Rev. B {\bf 69}, 205306 (2004).

\bibitem{pfeifer2}T.~Pfeifer, W.~K\"{u}tt, H.~Kurz, and R.~Scholz, Phys. Rev. Lett. {\bf 69}, 3248 (1992).

\bibitem{scholz1}R.~Scholz, T.~Pfeifer, and H.~Kurz, Phys. Rev. B {\bf 47}, 16229 (1993).

\bibitem{lee}J.~D.~Lee, J.~Inoue, and M.~Hase, Phys. Rev. Lett. {\bf 97}, 157405 (2006).

\bibitem{shinohara}Y.~Shinohara, K.~Yabana, Y.~Kawashita, J.~I.~Iwata, T.~Otobe, and G.~F.~Bertsch, Phys. Rev. B {\bf 82}, 155110 (2010).

\bibitem{mahan}G.~D.~Mahan, {\it Many-Particle Physics} (Plenum, New York, 1981) Chaps. 4 and 5.

\bibitem{fano1}U. Fano, Phys. Rev. {\bf 124}, 1866 (1961).

\bibitem{riffe2}D.~M.~Riffe, Phys. Rev. B {\bf 84}, 064308 (2011).

\bibitem{nakamura}K.~G.~Nakamura, Y.~Shikano, and Y.~Kayanuma, Phys. Rev. B {\bf 92}, 144304 (2015).

\bibitem{Gaal2007}
P. Gaal, W. Kuehn, K. Reimann, M. Woerner, T. Elsaesser, and R. Hey, Nature {\bf 450}, 1210 (2007).

\bibitem{yoshino}S.~Yoshino, G.~Oohata, and K.~Mizoguchi, Phys.~Rev.~Lett. {\bf 115}, 157402 (2015).

\bibitem{varga}B.~B.~Varga, Phys. Rev. {\bf 137}, A1896 (1965).

\bibitem{mooradian}A.~Mooradian and A.~L.~ McWhorter, Phys. Rev. Lett.  {\bf 19}, 849 (1967).

\bibitem{klein}M.~V.~Klein, in {\it Light Scattering in Solids I: Introductory Concepts}, Topics in Applied Physics {\bf  8}, ed. by M.~Cardona
 (Springer-Verlag, Berlin, 1983) Chap. 4.

\bibitem{kuznetsov3}A.~V.~Kuznetsov and C.~J.~Stanton, Phys. Rev. B {\bf 51}, 7555 (1995).

\bibitem{hase8}M.~Hase, S.~-I.~Nakashima, K.~Mizogichi, H.~Harima, and K.~Sakai, Phys. Rev. B {\bf 60}, 16526 (1999).

\bibitem{ishioka2}K.~Ishioka, A.~K.~Basak, and H.~Petek, Phys. Rev. B {\bf 84}, 235202 (2011).

\bibitem{hu}J.~Hu, O.~V.~Misochoko, A.~Goto, K.~G.~Nakamura, Phys. Rev. B {\bf 86}, 235145 (2012).

\bibitem{Huber05}
R. Huber, C. K\"{u}bler, S. T\"{u}bel, A. Leitenstorfer, Q. T. Vu, H. Haug, F. K\"{o}hler, and M.-C. Amann, Phys. Rev. Lett. {\bf  94}, 027401 (2005).

\bibitem{Chang2010}
Y. -M. Chang, in the Lasers and Electro-Optics (CLEO) and Quantum Electronics and Laser Science Conference (QELS), 2010, paper QThF1.

\bibitem{russell}J.~P.~Russell, Appl. Phys. Lett. {\bf 6}, 223 (1965).

\bibitem{parker}J.~H.~Parker,~Jr., D.~W.~Feldman, and M.~Ashkin, Phys. Rev. {\bf 155}, 712 (1967).

\bibitem{hart}T.~R.~Hart, R.~L.~Aggarwal, and B.~Lax, Phys. Rev. B {\bf 1}, 638 (1970).

\bibitem{cerdeira2}F.~Cerdeira, T.~A.~Fjeldy, and M.~Cardona, Solid State Commum. {\bf 8}, 133 (1970).

\bibitem{cerdeira1}F.~Cerdeira and M.~Cardona, Phys. Rev. B {\bf 5}, 1440 (1972).

\bibitem{cerdeira0}F.~Cerdeira, T.~A.~Fjeldy, and M.~Cardona, Solid State Commum. {\bf 13}, 325 (1973).

\bibitem{temple}P.~A.~Temple and C.~E.~Hathaway, Phys. Rev. B {\bf 7}, 3685 (1973).

\bibitem{cerdeira3}F.~Cerdeira, T.~A.~Fjeldy, and M.~Cardona, Phys. Rev. B {\bf 8}, 4734 (1973).

\bibitem{cerdeira4}F.~Cerdeira, T.~A.~Fjeldy, and M.~Cardona, Phys. Rev. B {\bf 9}, 4344 (1974).

\bibitem{balkanski}M.~Balkanski, K.~P.~Jain, R.~Beserman, and M.~Jouanne, Phys. Rev. B {\bf 12}, 4328 (1975).

\bibitem{chandrasekhar1}M.~Chandrasekhar, J.~B.~Renucci,  and M.~Cardona, Phys. Rev. B {\bf 17}, 1623 (1978).

\bibitem{arya}K.~Arya, M.~A.~Kanehisa, M.~Jouanne, K.~P.~Jain, and M.~Balkanski, J. Phys. C {\bf 12}, 3843 (1979).

\bibitem{chandrasekhar2}M.~Chandrasekhar, H.~R.~Chandrasekhar, M.~Grimsditch, and M.~Cardona, Phys. Rev. B {\bf 22}, 4825  (1980).

\bibitem{menendez}J.~Men\'{e}ndez and M.~Cardona, Phys. Rev. B {\bf 29}, 2051 (1984).

\bibitem{belitsky}V.~I.~Belitsky, A.~Cantarero, M.~Cardona, C.~Trallero-Giner, and S.~T.~Pavlov, J. Phys.: Condens. Matter  {\bf 9}, 5965 (1997).

\bibitem{nunes}L.~A.~O.~Nunes, L.~Ioriatti, L.~T.~Florez, and J.~P.~Harbison, Phys. Rev. B {\bf 47}, R13011 (1993).

\bibitem{pusep}Yu.~A.~Pusep, M.~T.~O.~Silva, J.~C.~Galzerani, S.~W.~da~Silva, L.~M.~R.~Scolfaro, R~.Enderlein, A.~A.~Quivy, A.~P.~Lima, and J.~R.~Leite, Phys. Rev. B {\bf 54}, 13927 (1996).

\bibitem{jin}K.~-J.~Jin, J.~Zhang, Z.~-H.~Chen, G.~-Z.~Yang, Z.~H.~Chen, X.~H.~Shi, and S.~C.~Shen, Phys. Rev. B {\bf  64}, 205203 (2001).
 
\bibitem{miroshnichenko1}A.~E.~Miroshnichenko, S.~Flach, and Y.~S.~Kivshar, Rev. Mod. Phys. {\bf 82}, 2257 (2010).

\bibitem{non-linearFR}
Recently, there are a coupled of reports regarding  experimental observations of non-linear FR in self-assembled semiconductor quantum dots and in the system of semiconductor-metal nanoparticle molecules, where strongly asymmetric spectral shapes are discerned at high cw-light power:
M.~Kroner, A.~O.~Govorov, S.~Remi, B~ Biedermann, S.~Seidl, A.~Badolato, P.~M.~Petroff, W.~Zhang, R.~Barbour, B.~D.~Gerardot, R.~J.~Warburton, and K.~Karrai, Nature {\bf 451}, 311 (2008); W.~Zhang, A.~O.~Govorov, and G.~W.~Bryant, Phys. Rev. Lett. {\bf 97}, 146804 (2006).
 
\bibitem{meier}T.~Meier, A.~Schulze, P.~Thomas, and H~ Vaupel, Phys. Rev. B {\bf 51}, 13977 (1995).

\bibitem{siegner1}U.~Siegner, M.~-A.~Mycek, S.~Glutsch, and D.~S.~Chemla, Phys. Rev. Lett.  {\bf 74}, 470 (1995).

\bibitem{siegner2}U.~Siegner, M~ -A.~Mycek, S.~Glutsch, and D.~S.~Chemla, Phys. Rev. B {\bf 51}, 4953 (1995).

\bibitem{hino2}K.~Hino, K.~Goto, and N~ Toshima, Phys. Rev. B {\bf 69}, 035322 (2004).

\bibitem{SMparameter}See Supplemental Material at http://link.aps.org/
supplemental/10.1103/PhysRevB.xx.yyyyyy for Appendix~A: the parameters of materials and lasers employed in this study.

\bibitem{appA}See Supplemental Material at http://link.aps.org/
supplemental/10.1103/PhysRevB.xx.yyyyyy for Appendix~B: the derivation of Eq.~(\ref{comm}): a factorization approximation.








            


    

  







\bibitem{haug}H.~Haug and S.~W.~Koch, {\it Quantum Theory of the Optical and Electronic Properties of Semiconductors, fifth ed.}, (World Scientific, Singapore, 2009) Chaps.~1 and12.

\bibitem{meystre}P.~Meystre and M.~Sargent III, {\it Elements of  Quantum Optics, third ed.}, (Springer-Verlarg, Berlin,  1999) Chaps.~3 and 15.

\bibitem{morse}P.~M.~Morse and H.~Feshbach, {\it Methods of Theoretical Physics}, (McGraw-Hill,  New York, 1953) Chap.~7.

\bibitem{crossing}N.~Moiseyev, {\it Non-Hermitian Quantum Mechanics}, (Cambridge,  New York, 2011) Chaps.~7-9.

\bibitem{appB}See Supplemental Material at http://link.aps.org/
supplemental/10.1103/PhysRevB.xx.yyyyyy for Appendix~C: the solutions of eigenvalue equations of Eqs.~(\ref{eigenZbarL}) and (\ref{eigenZbarR}).

\bibitem{dyson}According to Eqs.~(\ref{Heisenberg2}) and (\ref{Heisenberg3}), a creation operator of quasi-boson $B^\dagger_{\boldsymbol{q}\alpha}(t)$ is introduced so as to ensure the commutation relation that
$[\hat{\mathcal{H}}^{(eff)}_e(t), B^\dagger_{\boldsymbol{q}\alpha}(t)]=B^\dagger_{\boldsymbol{q}\alpha}(t)
\mathcal{E}_{\boldsymbol{q}\alpha}(t)$, where $\hat{\mathcal{H}}^{(eff)}_e(t)$ is considered to be an effective electronic Hamiltonian with a correction due to a rotational approximation to $\hat{\mathcal{H}}_e(t)$ as seen from the right-hand side of the first equality of  Eq.~(\ref{Heisenberg2}).
Defining the (adiabatic) ground-state of $\hat{\mathcal{H}}^{(eff)}_e(t)$ as $|0\rangle$, one obtains the result that
$[\hat{\mathcal{H}}^{(eff)}_e(t), B^\dagger_{\boldsymbol{q}\alpha}(t)]|0\rangle=[\hat{\mathcal{H}}^{(eff)}_e(t)-\mathcal{E}^{(eff)}_0(t)]|1;\boldsymbol{q}\alpha \rangle =
\mathcal{E}_{\boldsymbol{q}\alpha}(t)|1;\boldsymbol{q}\alpha \rangle$, where $|1;\boldsymbol{q}\alpha \rangle$ is the first (adiabatic) excited-state defined by
$|1;\boldsymbol{q}\alpha \rangle = B^\dagger_{\boldsymbol{q}\alpha}(t)|0\rangle$ and $\mathcal{E}^{(eff)}_0(t)$ represents the ground-state energy.
This leads to the expression that $\hat{\mathcal{H}}^{(eff)}_e(t)|1;\boldsymbol{q}\alpha \rangle =
\epsilon_{\boldsymbol{q}\alpha}(t)|1;\boldsymbol{q}\alpha \rangle$, where
$\epsilon_{\boldsymbol{q}\alpha}(t)=\mathcal{E}_{\boldsymbol{q}\alpha}(t)+\mathcal{E}^{(eff)}_0(t)$, implying energy of a single quasi-boson state $|1;\boldsymbol{q}\alpha \rangle$.
The above procedure of quasi-bosonization is reminiscent of  Dyson's method of bosonization for spin-wave interactions in a ferromagnet; F.~J.~Dyson, Phys. Rev. {\bf 102}, 1217 (1956) and  {\it ibid}., Phys. Rev. {\bf 102}, 1230 (1956).
It should be noted that this quasi-bosonization procedure is correct just for the single quasi-boson state, and would  become more questionable with increasing the number of quasi-bosons.
For instance,  it is readily shown that $\hat{\mathcal{H}}^{(eff)}_e(t)|2;\boldsymbol{q}\alpha \rangle \not=
2\epsilon_{\boldsymbol{q}\alpha}(t)|2;\boldsymbol{q}\alpha \rangle$ with
$|2;\boldsymbol{q}\alpha \rangle$  a two-quasi-boson state defined as $ |2;\boldsymbol{q}\alpha \rangle = \sqrt{2}^{-1}[B^\dagger_{\boldsymbol{q}\alpha}(t)]^2|0\rangle$.


\bibitem{conventionE}According to the convention, eigen energy pertinent to an annihilation operator $B_{\boldsymbol{q}\alpha}(t)$
is defined as  $\mathcal{E}_{\boldsymbol{q}\alpha}(t)$.
Thus, differing from the expression of Eq.~(\ref{B-2}),  a negative value of imaginary part of  it, namely, ${\rm Im}\/\mathcal{E}_{\boldsymbol{q}\alpha}(t) < 0$, gives rise to damping of  quasi-bosonic state $\alpha$ with increasing $t$. 


\bibitem{yu}P.~Y.~Yu and M.~Cardona, {\it Fundamentals of Semiconductors, fourth ed.}, (Springer-Verlarg, Berlin, 2010) Chaps.~3 and 7.

\bibitem{seaton}M.~J.~Seaton, Rep. Prog. Phys. {\bf 46}, 167 (1983).

\bibitem{appC}See Supplemental Material at http://link.aps.org/
supplemental/10.1103/PhysRevB.xx.yyyyyy for Appendix~D: the closed analytic forms of $F^\dagger_{\boldsymbol{q}\beta}$ and $F_{\boldsymbol{q}\beta}$ derived by solving
Eq.~(\ref{HeisenbergF2}).

\bibitem{nikitin}E.~E.~Nikitin and S.~Ya.~Umanskii, {\it Theory of Slow Atomic Collisions},
 Springer Series in Chemical Physics {\bf 30},
(Springer-Verlag, Berlin, 1984) Chap.~7.

\bibitem{schafer}W.~Sch\"{a}fer and M.~Wegener, {\it Semiconductor Optics and Transport Phenomena} (Springer-Verlag, Berlin, 2002) Chaps. 2, 10, and 11.

\bibitem{fetter}A.~L.~Fetter and J.~D.~Walecka, {\it Quantum Theory of Many-Particle Systems},
(McGraw-Hill, Inc., New York, 1971) Chaps.~3-5.

\bibitem{SMchi}See Supplemental Material at http://link.aps.org/
supplemental/10.1103/PhysRevB.xx.yyyyyy for Appendix~E: details of deriving an expression of  total retarded longitudinal-susceptibility $\chi^{(t)}_{\boldsymbol{q}}(t,t^\prime)$.

\bibitem{comment1}It is understood that the factor of $e^{-\tau/T_{ph}}$ due to LO-phonon anharmonicity is implicitly convoluted with the integrand of Eq.~(\ref{tildechitot}), where $T_{ph}$ represents  a phenomenological damping time-constant of  LO-phonon due to the anharmonicity.
This also ensures the convergence of the Fourier transform.

\bibitem{SMappF}See Supplemental Material at http://link.aps.org/
supplemental/10.1103/PhysRevB.xx.yyyyyy for Appendix~F: the derivation of Eq.~(\ref{shore0}): Shore's spectral profile.

\bibitem{shore}B.~W.~Shore, Rev. Mod. Phys. {\bf 39}, 439 (1967).

\bibitem{comment2}The factor of $e^{-t/T_{ph}}$ similar to that in Ref.~\onlinecite{comment1} is implicitly convoluted with the integrand of Eq.~(\ref{tildeQ}).

\bibitem{SMQ}See Supplemental Material at http://link.aps.org/
supplemental/10.1103/PhysRevB.xx.yyyyyy for Appendix~G: the derivation of Eq.~(\ref{Q2}): phonon displacement function $Q_{\boldsymbol{q}}(t)$.

\bibitem{comment3}According to the prsesent theoretical framework developed in Sec.~\ref{sec2}, the implicit assumption is made that the temporal pulse-width $\tau_L$ is small enough to disregard contributions during the period of laser irradiation to the CP generation dynamics.
Hence, it seems more reasonable from the view point of theoretical consistency to approximate $\Delta\alpha_{\boldsymbol{q}}$ as $A_L$, as given in the second equality of Eq.~(\ref{Deltaalpha}).

\bibitem{SMmathcalD}See Supplemental Material at http://link.aps.org/
supplemental/10.1103/PhysRevB.xx.yyyyyy for Appendix~H: the properties of dimensionless function  $\mathcal{D}_{\boldsymbol{q}\alpha_p}(t,t_D)$.

\bibitem{inui}T.~Inui, Y.~Tanabe, and Y.~Onodera, {\it Group Theory and Its Applications in Physics},
 Springer Series in Solid-State Sciences {\bf 78},
(Springer-Verlag, Berlin, 1990) Chaps.~11 and 12.

\bibitem{tom}H.~W.~K.~Tom, T~.F.~Heinz, and Y.~R.~Shen, Phys. Rev. Lett. {\bf 51}, 1983 (1983).

\bibitem{stampfli}P.~Stampfli and K.~H.~Bennemann, Phys. Rev. B {\bf 49}, 7299 (1994).

\bibitem{Hase09}
M. Hase, Appl. Phys. Lett. {\bf 94}, 112111 (2009).


\bibitem{tang}J.~Z.~Tang, S.~Watanabe, and M.~Matsuzawa, Phys. Rev. A {\bf 46}, 2437 (1992).

\bibitem{hino6}K.~Hino, A.~Igarashi, and J.~H.~Macek, Phys. Rev. A {\bf 56}, 1038 (1997).

\bibitem{hino3}K. Hino, Phys. Rev. B {\bf 62}, R10626 (2000).

\bibitem{hino4}K. Hino, Phys. Rev. B {\bf 64}, 075318 (2001).

\bibitem{hino5}K. Hino and N. Toshima, Phys. Rev. B {\bf 71}, 205326  (2005).

\bibitem{burke}P.~G. Burke, {\it R-Matrix Theory of Atomic Collisions: Application to Atomic, Molecular and Optical Processes}  (Springer-Verlag, Berlin, 2011) Chap.~9.

\bibitem{hase9}M.~Hase (unpublished work).








\end{thebibliography}

%

%
%

\end{document}